\documentclass[11pt]{article}
\usepackage{amssymb,amsmath,amsfonts}
\usepackage{graphicx}
\usepackage{graphics}
\usepackage{eepic,epsfig}
\usepackage{dcolumn}
\usepackage{color}

\@addtoreset{equation}{section}

\makeatother

\begin{document}

\title{Casimir densities induced by a sphere in the hyperbolic \\
vacuum of de Sitter spacetime }
\author{A. A. Saharian$^{1,2}$, T. A. Petrosyan$^{1,2}$ \\
\\
\textit{$^1$ Department of Physics, Yerevan State University,}\\
\textit{1 Alex Manoogian Street, 0025 Yerevan, Armenia}\vspace{0.3cm}\\
\textit{$^2$ Institute of Applied Problems in Physics NAS RA,}\\
\textit{25 Nersessian Street, 0014 Yerevan, Armenia}}
\maketitle

\begin{abstract}
Complete set of modes and the Hadamard function are constructed for a scalar
field inside and outside a sphere in $(D+1)$-dimensional de Sitter spacetime
foliated by negative constant curvature spaces. We assume that the field
obeys Robin boundary condition on the sphere. The contributions in the
Hadamard function induced by the sphere are explicitly separated and the
vacuum expectation values (VEVs) of the field squared and energy-momentum
tensor are investigated for the hyperbolic vacuum. In the flat spacetime
limit the latter is reduced to the conformal vacuum in the Milne universe
and is different from the maximally symmetric Bunch-Davies vacuum state. The
vacuum energy-momentum tensor has a nonzero off-diagonal component that
describes the energy flux in the radial direction. The latter is a purely
sphere-induced effect and is absent in the boundary-free geometry. Depending
on the constant in Robin boundary condition and also on the radial
coordinate, the energy flux can be directed either from the sphere or
towards the sphere. At early stages of the cosmological expansion the
effects of the spacetime curvature on the sphere-induced VEVs are weak and
the leading terms in the corresponding expansions coincide with those for a
sphere in the Milne universe. The influence of the gravitational field is
essential at late stages of the expansion. Depending on the field mass and
the curvature coupling parameter, the decay of the sphere-induced VEVs, as
functions of the time coordinate, is monotonic or damping oscillatory. At
large distances from the sphere the fall-off of the sphere-induced VEVs, as
functions of the geodesic distance, is exponential for both massless and
massive fields.
\end{abstract}

\section{Introduction}

The quantum field-theoretical effects in background of de Sitter (dS)
spacetime (for geometrical properties and coordinate systems see, for
instance, \cite{Hawk94,Grif09}) continue to be the subject of active
research. There are several motivations for that. First of all, the high
symmetry of dS spacetime allows to obtain closed analytic solutions in
numerous physical problems with important applications in cosmology of the
early Universe. On the basis of this, one can reveal the features of the
influence of gravitational fields on quantum effects in more complicated
geometries, including those describing more general class of cosmological
models and black hole physics. The most inflationary models for the
expansion of the early Universe are based on an approximately dS geometry
sourced by slowly evolving scalar fields. The short period of the
corresponding quasi-exponential expansion provides a natural solution to a
number of problems in Big Bang cosmology \cite{Bass06,Mart14}. An important
effect of the rapid expansion during the inflation is the magnification of
quantum fluctuations of fields, including those for inflaton, to macroscopic
scales. The related inhomogeneities in the distribution of the energy
density act as seeds for subsequent large-scale structure formation in the
Universe. This mechanism for the galaxy formation has been supported by the
observational data on the temperature anisotropies of the cosmic microwave
background radiation. Another important discovery based on those data, in
combination with observations of high redshift supernovae and galaxy
clusters, is the accelerated expansion of the Universe at the present epoch.
The observational data are well approximated by the Lambda-CDM model with a
positive cosmological constant responsible for the accelerated expansion.
The dS spacetime is the future attractor of this model. In addition to the
above, interesting topics related to the physics in dS geometry are the
string-theoretical models of dS inflation and the holographic duality
between quantum gravity on dS spacetime and a quantum field theory living on
its timelike infinity (dS/CFT correspondence, see \cite{Stro01,Noji02,Anni17}
and references therein).

In the present paper we consider the effect of a spherical boundary on dS
bulk, foliated by negative constant curvature spaces, on the local
properties of quantum vacuum for a massive scalar field with general
curvature coupling parameter. The influence of the sphere originates from
the modification of the spectrum for the vacuum fluctuations, induced by the
boundary condition on the field operator. This type of boundary-induced
effects are widely investigated in the literature for different bulk and
boundary geometries and are known under the general name of the Casimir
effect (see for reviews \cite{Most97}-\cite{Casi11}). For quantum fields in
a given curved background, closed analytic expressions for the
characteristics of the vacuum, such as the vacuum energy, the Casimir forces
and vacuum expectation values (VEVs) of the energy-momentum tensor, are
obtained for geometries with high symmetry. In particular, motivated by
radion stabilization and generation of the cosmological constant on branes,
the investigation of boundary-induced quantum effects in anti-de Sitter
(AdS) spacetime has attracted a great deal of attention (see references
given in \cite{Saha20AdS,Saha20AdSb}).

The Casimir effect for planar boundaries in dS spacetime has been discussed
in \cite{Eliz03dS}-\cite{Kota15} for scalar and electromagnetic fields. It
has been shown that the influence of the gravitational field on the local
characteristics of the vacuum state is essential at distance from the
boundaries larger than the curvature radius of the background geometry. The
VEVs of the field squared and energy-momentum tensor for scalar and
electromagnetic fields induced by a cylindrical boundary in dS bulk have
been investigated in \cite{Saha15cyl,Saha16cyl}. Another class of exactly
solvable problems correspond to spherical boundaries. The corresponding
Casimir densities were discussed in \cite{Seta01a,Seta01b} for a conformally
coupled massless scalar field and in \cite{Milt12} for a massive field with
general coupling to the curvature. In the conformally coupled massless case
the VEVs in the dS spacetime are obtained from the corresponding results for
a spherical boundary in the Minkowski bulk by a conformal transformation. By
using the conformal relation between dS (described in static coordinates)
and Rindler spacetimes, the vacuum densities for a more complicated boundary
have been studied in \cite{Saha04dSbr}. The VEVs in geometries with
spherical dS bubbles have been investigated in \cite{Bell14}. The
topological Casimir effect induced by toroidal compactification of a part of
spatial dimensions and by the presence of topological defects in locally dS
spacetime was discussed in \cite{Saha08Compds}-\cite{Saha19Izv}.

An important step to quantize fields in curved spacetimes is the choice of a
coordinate system and related complete set of mode functions being solutions
of the classical field equations. In general, the different sets of mode
functions will lead to different Fock spaces, in particular, to inequivalent
vacuum states. A well known example of this kind in flat spacetime is the
quantization of fields in Cartesian coordinates, relevant for inertial
observers, and in Rindler coordinates, adapted for uniformly accelerated
observers. These two ways of quantization give rise to different vacuum
states, the Minkowski and Fulling-Rindler vacua for inertial and uniformly
accelerated observers, respectively. In dS spacetime, depending on the
specific physical problem, different coordinate systems have been used. The
global coordinates, with spatial sections being spheres, cover the whole dS
spacetime. In planar (or inflationary) coordinates the spatial sections are
flat and they only cover half of dS spacetime. These coordinates are the
most suitable for cosmological applications, in particular, in models of
inflation. Though the dS spacetime has timelike isometries, the metric
tensor in both the global and planar coordinates is time-dependent. The
existence of time isometries is explicit in static coordinates with
time-independent metric tensor. These coordinates are analogue of the
Schwarzschild coordinates for black holes and cover the region in dS
spacetime accessible to a single observer. They are well-adapted for
discussions of thermal aspects of dS spacetime. Another coordinate system
with spatial sections having constant negative curvature has been employed
in recent investigations of the entanglement entropy in dS spacetime (see 
\cite{Mald13}-\cite{Bhat19} and references therein). These hyperbolic
coordinates provide a natural setup to discuss long range quantum
correlations between causally disconnected regions (L and R regions in the
discussion below) separated by another finite region (region C below).

In the present paper we investigate the influence of a spherical boundary on
the vacuum fluctuations of a massive scalar field in background of $(D+1)$%
-dimensional dS spacetime with negative curvature spatial foliation for the
general curvature coupling. The paper is organized as follows. In Section %
\ref{sec:Setup} we describe the bulk and boundary geometries and the
boundary condition imposed on the scalar field operator. The general form of
the mode functions is obtained by solving the field equation. In Section \ref%
{sec:Vacuum} the mode functions are specified for the special case of the
hyperbolic vacuum. It is shown that the latter coincides with the conformal
vacuum. In Section \ref{sec:Had} the Hadamard functions for the
boundary-free geometry and for the regions outside and inside a spherical
boundary are evaluated. The eigenvalues of the radial quantum number are
specified inside the spherical shell. The sphere-induced contributions in
the Hadamard function are separated explicitly for both the exterior and
interior regions. In the case of the hyperbolic vacuum, representations for
those contributions, well-adapted for the investigation of local VEVs, are
provided. The VEVs of the field squared inside and outside a spherical shell
are studied in Section \ref{sec:phi2i}. The results of numerical analysis
are presented. The corresponding investigations for the VEVs of the
energy-momentum tensor are presented in Section {\ref{sec:EMT}}. The main
results of the paper are summarized in Section \ref{sec:Conc}. In Appendix %
\ref{sec:ApCoord} the coordinates in different regions of the dS spacetime,
foliated by negative curvature spaces, and their relations to the global and
inflationary coordinates are discussed. In Appendix \ref{sec:VEVbf} the
expression for the Hadamard function in the boundary-free dS spacetime with
negative curvature spatial foliation is presented without specifying the
vacuum state.

\section{Problem setup and the scalar modes}

\label{sec:Setup}

\begin{figure}[tbph]
\begin{center}
\epsfig{figure=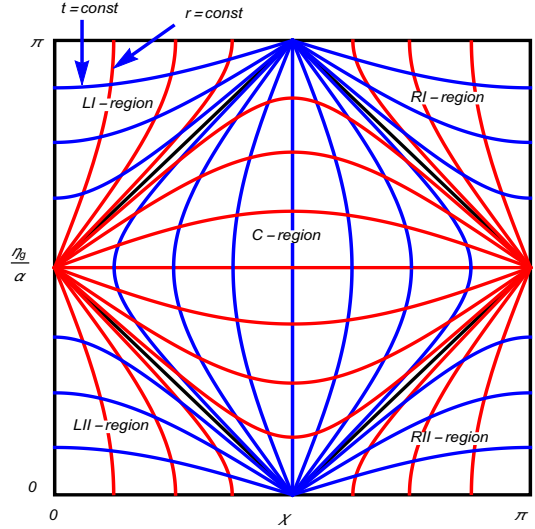,width=7cm,height=7cm}
\end{center}
\caption{The Penrose diagram for dS spacetime covered by the coordinates
corresponding to the negative curvature spatial foliation. }
\label{fig1}
\end{figure}

We consider $(D+1)$-dimensional dS spacetime with negative curvature spatial
foliation. The relations between the coordinates realizing the foliation and
the global conformal coordinates $(\eta _{g},\chi ,\vartheta ,\phi )=(\eta
_{g},\chi ,\theta _{1},\ldots \theta _{n},\phi )$, $n=D-2$, are discussed in
Appendix \ref{sec:ApCoord}. Here, $(\vartheta ,\phi )=(\theta _{1},\ldots
\theta _{n},\phi )$ are the angular coordinates on a sphere $S^{D-1}$. The
corresponding Penrose diagram, mapped on the square $(0\leqslant \eta
_{g}/\alpha \leqslant \pi ,0\leqslant \chi \leqslant \pi )$, is presented in
Figure \ref{fig1}. The five regions designated by LI, LII, RI, RII and C are
separated by the line segments $\eta _{g}/\alpha =\pi /2\pm \chi $, $\eta
_{g}/\alpha =3\pi /2-\chi $, $\eta _{g}/\alpha =\chi -\pi /2$. In what
follows the discussion will be presented for the LI-region defined by (\ref%
{LI}). The corresponding line element reads%
\begin{equation}
ds^{2}=dt^{2}-\alpha ^{2}\sinh ^{2}\left( t/\alpha \right) (dr^{2}+\sinh
^{2}rd\Omega _{D-1}^{2}),  \label{LE}
\end{equation}%
where $0\leq t<\infty $, $0\leq r<\infty $ and $d\Omega _{D-1}^{2}$ is the
line element on a sphere $S^{D-1}$ with unit radius. The metric tensors in
the regions LII, RI, RII have similar forms. Note that the radial coordinate 
$r$ is dimensionless. The line element (\ref{LE}) is conformally related to
the line element of static spacetime with negative constant curvature space.
In order to see that we introduce a new time coordinate $\eta $, $-\infty
<\eta \leq 0$, in accordance with%
\begin{equation}
e^{\eta /\alpha }=\tanh \left( t/2\alpha \right) .  \label{etac}
\end{equation}%
The line element takes the form%
\begin{equation}
ds^{2}=\frac{d\eta ^{2}-\alpha ^{2}\left( dr^{2}+\sinh ^{2}rd\Omega
_{D-1}^{2}\right) }{\sinh ^{2}\left( \eta /\alpha \right) }.  \label{ds2c}
\end{equation}%
Note that we have the relations $\sinh \left( \eta /\alpha \right) =-1/\sinh
\left( t/\alpha \right) $ and $\coth \left( \eta /\alpha \right) =-\cosh
\left( t/\alpha \right) $ between the conformal and synchronous time
coordinates.

We are interested in effects of a spherical boundary with radius $r=r_{0}$
on the local characteristics of the vacuum state for a scalar field $\varphi
(x)$ with curvature coupling parameter $\xi $. The corresponding field
equation has the form%
\begin{equation}
\left( \nabla _{\mu }\nabla ^{\mu }+m^{2}+\xi R\right) \varphi =0,
\label{Feq}
\end{equation}%
where $\nabla _{\mu }$ is the covariant derivative operator and the Ricci
scalar is given by $R=D\left( D+1\right) /\alpha ^{2}$. On the sphere the
field obeys the Robin boundary condition 
\begin{equation}
\left( A-\delta _{\mathrm{(j)}}B\partial _{r}\right) \varphi (x)=0,\;r=r_{0},
\label{Rbc}
\end{equation}%
where $\mathrm{j}=\mathrm{i}$ and $\mathrm{j}=\mathrm{e}$ correspond to the
interior ($r\leq r_{0}$) and exterior ($r\geq r_{0}$) regions with $\delta _{%
\mathrm{(i)}}=1$ and $\delta _{\mathrm{(e)}}=-1$. It is of interest to have
the radius of the sphere $r_{\mathrm{I}0}$ in inflationary coordinates. By
using the relations (\ref{rInf}), we can see that%
\begin{equation}
r_{\mathrm{I}0}/\alpha =\coth r_{0}-\sqrt{\sinh ^{-2}r_{0}+e^{-2t_{\mathrm{I}%
}/\alpha }}.  \label{rI0}
\end{equation}%
As seen, in inflationary coordinates the radius of the sphere is
time-dependent. One has $r_{\mathrm{I}0}=0$ for $t_{\mathrm{I}}=0$. With
increasing $t_{\mathrm{I}}$ the radius $r_{\mathrm{I}0}$ increases and in
the limit $t_{\mathrm{I}}\rightarrow \infty $ it tends to the value $%
\lim_{t_{\mathrm{I}}\rightarrow \infty }r_{\mathrm{I}0}=\alpha \tanh
(r_{0}/2)$.

The VEVs of the physical quantities bilinear in the field operator are
obtained from the two-point functions or their derivatives in the
coincidence limit of the arguments. As a two-point function we will consider
the Hadamard function $G(x,x^{\prime })=\langle 0|\varphi (x)\varphi
(x^{\prime })+\varphi (x^{\prime })\varphi (x)|0\rangle $, where $|0\rangle $
stands for the vacuum state and $x=\left( t,r,\vartheta ,\phi \right) $. For
the evaluation of the Hadamard function we will employ the mode sum formula%
\begin{equation}
G(x,x^{\prime })=\sum_{\sigma }\left[ \varphi _{\sigma }\left( x\right)
\varphi _{\sigma }^{\ast }\left( x^{\prime }\right) +\varphi _{\sigma
}\left( x^{\prime }\right) \varphi _{\sigma }^{\ast }\left( x\right) \right]
,  \label{WF}
\end{equation}%
where $\{\varphi _{\sigma }\left( x\right) ,\varphi _{\sigma }^{\ast }\left(
x^{\prime }\right) \}$ is a complete set of solutions to the classical field
equation obeying the boundary condition and the collective index $\sigma $
specifies the quantum numbers. The symbol $\sum_{\sigma }$ includes the
summation over discrete quantum numbers and the integration over the
continuous ones. Given the Hadamard function, the VEVs of the field squared, 
$\langle 0|\varphi ^{2}(x)|0\rangle \equiv \left\langle \varphi
^{2}(x)\right\rangle $, and of the energy-momentum tensor, $\langle
0|T_{ik}(x)|0\rangle \equiv \left\langle T_{ik}(x)\right\rangle $, are found
in the coincidence limit of the arguments as follows:%
\begin{eqnarray}
\left\langle \varphi ^{2}(x)\right\rangle  &=&\frac{1}{2}\underset{x^{\prime
}\rightarrow x}{\lim }G\left( x,x^{\prime }\right) ,  \notag \\
\left\langle T_{ik}(x)\right\rangle  &=&\frac{1}{2}\underset{x^{\prime
}\rightarrow x}{\lim }\partial _{i^{\prime }}\partial _{k}G\left(
x,x^{\prime }\right) +\left( \xi -\frac{1}{4}\right) g_{ik}\nabla _{p}\nabla
^{p}\left\langle \varphi ^{2}\right\rangle -\xi \nabla _{i}\nabla
_{k}\left\langle \varphi ^{2}\right\rangle -\xi R_{ik}\left\langle \varphi
^{2}\right\rangle ,  \label{TikVev}
\end{eqnarray}%
where $R_{ik}$ is the Ricci tensor. Of course, the expressions in the
right-hand sides diverge and a renormalization is required. Here we are
interested in the contributions to the VEVs induced by the spherical
boundary. In the discussion below the corresponding contribution in the
Hadamard function will be extracted explicitly. The divergences are
determined by the local geometry and for points away from the sphere they
are the same in the problems without and with spherical boundary. This means
that for those points the renormalization in (\ref{TikVev}) is reduced to
the one in the problem where the spherical boundary is absent.

As the first step we need to specify the mode functions $\varphi _{\sigma
}\left( x\right) $. In accordance with the symmetry of the problem the
solution of the field equation (\ref{Feq}) can be presented in the form (for
a discussion of the scalar field mode function in $D=3$ dS spacetime with
negative curvature spatial sections see also \cite{Sasa95,Dimi15}) 
\begin{equation}
\varphi \left( x\right) =f\left( t/\alpha \right) g\left( r\right) Y\left(
m_{p};\vartheta ,\phi \right) ,  \label{GenSol}
\end{equation}%
where $Y(m_{p};\vartheta ,\phi )$ are hyperspherical harmonics of degree $%
l=0,1,2,\ldots $. For the set of quantum numbers $m_{p}$ one has $%
m_{p}=(m_{0}\equiv l,m_{1},\ldots ,m_{n})$, with $m_{1},m_{2},\ldots ,m_{n}$
being integers such that $-m_{n-1}\leqslant m_{n}\leqslant m_{n-1}$ and 
\begin{equation}
0\leqslant m_{n-1}\leqslant m_{n-2}\leqslant \cdots \leqslant m_{1}\leqslant
l.  \label{mn-1}
\end{equation}%
The angular part in (\ref{GenSol}) obeys the equation 
\begin{equation}
\Delta _{\vartheta ,\phi }Y\left( m_{p};\vartheta ,\phi \right) =-l\left(
l+n\right) Y\left( m_{p};\vartheta ,\phi \right) ,  \label{Yeq}
\end{equation}%
with $\Delta _{\vartheta ,\phi }$ being the Laplace operator on a unit
sphere. Substituting (\ref{GenSol}) into the field equation we get separate
equations for the functions $f(t/\alpha )$ and $g(r)$:%
\begin{eqnarray}
\frac{\partial _{\tau }\left[ \sinh ^{D}\tau \partial _{\tau }f\left( \tau
\right) \right] }{\sinh ^{D}\tau }+\left[ m^{2}\alpha ^{2}+\xi D\left(
D+1\right) +\frac{\gamma ^{2}}{\sinh ^{2}\tau }\right] f\left( \tau \right) 
&=&0,  \notag \\
\frac{\partial _{r}\left[ \sinh ^{D-1}r\partial _{r}g\left( r\right) \right] 
}{\sinh ^{D-1}r}+\left[ \gamma ^{2}-\frac{l\left( l+n\right) }{\sinh ^{2}r}%
\right] g\left( r\right)  &=&0,  \label{fgeq}
\end{eqnarray}%
where $\tau =t/\alpha $ and $\gamma ^{2}$ is the separation constant.

The equations (\ref{fgeq}) have the same structure and the corresponding
solutions are expressed in terms of the associated Legendre functions $%
P_{\nu }^{\mu }\left( u\right) $ and $Q_{\nu }^{\mu }\left( u\right) $ (for
the properties of the associated Legendre functions see \cite{Abra72,Nist10}%
). The solutions are presented in the form%
\begin{eqnarray}
f\left( \tau \right)  &=&\frac{X_{\nu }^{iz}\left( \cosh \tau \right) }{%
\sinh ^{(D-1)/2}\tau },  \notag \\
g\left( r\right)  &=&\frac{Z_{iz-1/2}^{-\mu }\left( \cosh r\right) }{\sinh
^{D/2-1}r},  \label{gr}
\end{eqnarray}%
with the functions%
\begin{eqnarray}
X_{\nu }^{iz}\left( y\right)  &=&d_{1}P_{\nu -1/2}^{iz}\left( y\right)
+d_{2}Q_{\nu -1/2}^{iz}\left( y\right) ,  \notag \\
Z_{iz-1/2}^{-\mu }\left( u\right)  &=&c_{1}P_{iz-1/2}^{-\mu }\left( u\right)
+c_{2}Q_{iz-1/2}^{-\mu }\left( u\right) ,  \label{XZ}
\end{eqnarray}%
and notations%
\begin{eqnarray}
\mu  &=&l+\frac{D}{2}-1,  \notag \\
\nu  &=&\sqrt{\frac{D^{2}}{4}-\xi D\left( D+1\right) -m^{2}\alpha ^{2}}.
\label{nu}
\end{eqnarray}%
The separation constant is expressed in terms of $z$ as $\gamma
^{2}=z^{2}+\left( D-1\right) ^{2}/4$. The parameter $\nu $ can be either
real or purely imaginary.

On the base of (\ref{gr}), the mode functions are presented in the form%
\begin{equation}
\varphi _{\sigma }\left( x\right) =\frac{X_{\nu }^{iz}\left( \cosh (t/\alpha
)\right) }{\sinh ^{(D-1)/2}(t/\alpha )}\frac{Z_{iz-1/2}^{-\mu }\left( \cosh
r\right) }{\sinh ^{D/2-1}r}Y\left( m_{p};\vartheta ,\phi \right) ,
\label{phisig}
\end{equation}%
where the set of quantum numbers is specified by $\sigma =(z,m_{p})$. The
coefficients $c_{1},c_{2},d_{1},d_{2}$ in the linear combinations of the
associated Legendre functions are determined by the choice of the vacuum
state and by the boundary and normalization conditions. The latter is given
by%
\begin{equation}
\int d^{D}x\sqrt{\left\vert g\right\vert }\varphi _{\sigma }\left( x\right) 
\overleftrightarrow{\partial _{t}}\varphi _{\sigma ^{\prime }}^{\ast }\left(
x\right) =i\delta _{\sigma \sigma ^{\prime }},  \label{NC}
\end{equation}%
where $\delta _{\sigma \sigma ^{\prime }}$ is understood as Kronecker delta
for discrete quantum numbers and Dirac delta function for continuous ones.
Note that we can also present the function $X_{\nu }^{iz}\left( y\right) $
as a linear combination of the functions $P_{\nu -1/2}^{\pm iz}\left(
y\right) $: 
\begin{equation}
X_{\nu }^{iz}\left( y\right) =\sum_{j=+,-}b_{j}P_{\nu -1/2}^{jiz}\left(
y\right) .  \label{X2}
\end{equation}%
By using the relation between the functions $Q_{\nu -1/2}^{iz}\left(
y\right) $ and $P_{\nu -1/2}^{\pm iz}\left( y\right) $, for the
corresponding coefficients one gets 
\begin{eqnarray}
b_{+} &=&d_{1}-\frac{i\pi e^{-\pi z}}{2\sinh \left( \pi z\right) }d_{2}, 
\notag \\
b_{-} &=&\frac{i\pi e^{-\pi z}}{2\sinh \left( \pi z\right) }\frac{\Gamma
\left( \nu +iz+1/2\right) }{\Gamma \left( \nu -iz+1/2\right) }d_{2},
\label{b12}
\end{eqnarray}%
where $\Gamma (x)$ is the gamma function. Note that one has the relation 
\begin{equation}
P_{\nu ^{\ast }-1/2}^{-iz}\left( y\right) =P_{\nu -1/2}^{-iz}\left( y\right)
,  \label{Prel}
\end{equation}%
for both real and purely imaginary $\nu $.

We impose an additional condition 
\begin{equation}
W\left\{ f\left( \tau \right) ,f^{\ast }\left( \tau \right) \right\}
=f\left( \tau \right) \partial _{\tau }f^{\ast }\left( \tau \right)
-\partial _{\tau }f\left( \tau \right) f^{\ast }\left( \tau \right) =\frac{i%
}{\sinh ^{D}\tau },  \label{fcond}
\end{equation}%
on the function $f\left( \tau \right) $ in (\ref{gr}), where $W\left\{
f_{1}\left( \tau \right) ,f_{2}\left( \tau \right) \right\} $ stands for the
Wronskian. This imposes a constraint on the coefficients of the linear
combination of the associated Legendre functions in the expression for the
function $X_{\nu }^{iz}\left( y\right) $. In order to obtain that constraint
it is convenient to employ the representation (\ref{X2}). By using (\ref%
{Prel}) and the Wronskian 
\begin{equation}
W\left\{ P_{\nu -1/2}^{iz}\left( y\right) ,P_{\nu -1/2}^{-iz}\left( y\right)
\right\} =\frac{2i}{\pi }\frac{\sinh \left( \pi z\right) }{y^{2}-1},
\label{WP}
\end{equation}%
the following relation is obtained for the coefficients in (\ref{X2}):%
\begin{equation}
\left\vert b_{+}\right\vert ^{2}-\left\vert b_{-}\right\vert ^{2}=\frac{\pi 
}{2\sinh \left( \pi z\right) }.  \label{brel}
\end{equation}%
In deriving this relation we have assumed that $z$ is real. In the case of
purely imaginary $z$ the condition (\ref{fcond}) is reduced to 
\begin{equation}
b_{+}b_{-}^{\ast }-b_{+}^{\ast }b_{-}=\frac{\pi }{2\sinh \left( \pi z\right) 
}.  \label{brelIm}
\end{equation}

Having the condition (\ref{fcond}) and by taking into account the formula 
\begin{equation}
\int d\Omega \,\,Y(m_{p};\vartheta ,\phi )Y^{\ast }(m_{p}^{\prime
};\vartheta ,\phi )=N(m_{p})\delta _{m_{p}m_{p}^{\prime }},  \label{Yint}
\end{equation}%
for the integral over the angular coordinates, the normalization condition (%
\ref{NC}) is written in terms of the radial functions:%
\begin{equation}
\int du\,Z_{iz-1/2}^{-\mu }\left( u\right) [Z_{iz^{\prime }-1/2}^{-\mu
}\left( u\right) ]^{\ast }=\frac{\alpha ^{1-D}}{N\left( m_{p}\right) }\delta
_{zz^{\prime }}.  \label{NCZ}
\end{equation}%
Here, the integration goes over the region $[1,\cosh r_{0}]$ for the mode
functions inside the sphere and over the region $[\cosh r_{0},\infty )$ for
the exterior modes. The explicit expression for $N(m_{p})$ is not required
in the following discussion and can be found, for example, in \cite{Erd53V2}.

An alternative representation of the time-dependent part in the mode
functions is obtained by using the relation%
\begin{equation}
P_{\nu -1/2}^{\pm iz}\left( \cosh (t/\alpha )\right) =\frac{\sqrt{2/\pi }%
e^{i\nu \pi }Q_{\mp iz-1/2}^{-\nu }\left( \coth (t/\alpha )\right) }{\Gamma
\left( 1/2-\nu \mp iz\right) \sqrt{\sinh (t/\alpha )}},  \label{LegRel2}
\end{equation}%
between the associated Legendre functions. For the function in (\ref{phisig}%
) this gives%
\begin{equation}
X_{\nu }^{iz}\left( \cosh (t/\alpha )\right) =\frac{\sqrt{2/\pi }e^{i\nu \pi
}}{\sqrt{\sinh (t/\alpha )}}\sum_{j=+,-}\frac{b_{j}Q_{-jiz-1/2}^{-\nu
}\left( \coth (t/\alpha )\right) }{\Gamma \left( 1/2-\nu -jiz\right) }.
\label{X3}
\end{equation}%
Equivalently, we can use the formula%
\begin{equation}
e^{i\pi \nu }Q_{\pm iz-1/2}^{-\nu }(y)=-\frac{\pi }{2\sin \left( \pi \nu
\right) }\left[ P_{iz-1/2}^{-\nu }(y)-\frac{\Gamma \left( \pm iz-\nu
+1/2\right) }{\Gamma \left( \pm iz+\nu +1/2\right) }P_{iz-1/2}^{\nu }(y)%
\right] ,  \label{LegRel3}
\end{equation}%
in order to express the modes in terms of the functions $P_{iz-1/2}^{\pm \nu
}(y)$:%
\begin{equation}
X_{\nu }^{iz}\left( \cosh (t/\alpha )\right) =\frac{\sqrt{\pi /2}}{\sin
\left( \pi \nu \right) }\sum_{j=+,-}jc_{j}\frac{P_{iz-1/2}^{j\nu }(\coth
(t/\alpha ))}{\sqrt{\sinh (t/\alpha )}},  \label{X4}
\end{equation}%
with the coefficients 
\begin{equation}
c_{\pm }=\sum_{j=+,-}\frac{b_{j}}{\Gamma \left( 1/2\pm \nu -jiz\right) }.
\label{cj}
\end{equation}

\section{Vacuum states}

\label{sec:Vacuum}

The coefficients in the linear combination (\ref{X2}) are related by (\ref%
{brel}) and (\ref{brelIm}) for modes with real and purely imaginary $z$,
respectively. The remaining degree of freedom is fixed by the choice of the
vacuum state. In order to discuss the vacuum states let us consider special
and limiting cases. For a conformally coupled massless scalar field one has $%
\xi =\xi _{D}=(D-1)/(4D)$ and $\nu =1/2$. For the associated Legendre
functions in the expressions of the scalar modes we get%
\begin{equation}
P_{0}^{\pm iz}\left( \cosh \left( t/\alpha \right) \right) =\frac{e^{\mp
iz\eta /\alpha }}{\Gamma \left( 1\mp iz\right) },  \label{P0}
\end{equation}%
and, hence, 
\begin{equation}
X_{1/2}^{iz}\left( \cosh \left( t/\alpha \right) \right) =\sum_{j=+,-}\frac{%
b_{j}e^{-jiz\eta /\alpha }}{\Gamma \left( 1-jiz\right) },  \label{X12}
\end{equation}%
where the conformal time $\eta $ is defined by (\ref{etac}). The modes
realizing the conformal vacuum are related to the mode functions $\varphi
_{\sigma }^{\mathrm{(st)}}\left( x\right) $ in static spacetime with a
negative constant curvature space (with the line element given by the
expression in the square brackets of (\ref{ds2c})) by the formula $\varphi
_{\sigma }\left( x\right) =\Omega ^{\left( 1-D\right) /2}\varphi _{\sigma }^{%
\mathrm{(st)}}\left( x\right) $ with the conformal factor $\Omega ^{2}=\sinh
^{-2}\left( \eta /\alpha \right) $.

For a conformally coupled massless field the static spacetime positive
energy mode functions are expressed as%
\begin{equation}
\varphi _{\sigma }^{\mathrm{(st)}}(x)=\frac{Z_{iz-1/2}^{-\mu }\left( \cosh
r\right) }{\sinh ^{D/2-1}r}Y\left( m_{k};\vartheta ,\phi \right) e^{-iz\eta
/\alpha },  \label{phist}
\end{equation}%
with the energy $E=z/\alpha \geq 0$. Comparing (\ref{phist}) with (\ref{X12}%
), we see that for the conformal vacuum the quantum number $z$ is real and $%
b_{-}=0$. The other coefficient $b_{+}$ is found from the relation (\ref%
{brel}):%
\begin{equation}
\left\vert b_{+}\right\vert ^{2}=\frac{\pi }{2\sinh \left( z\pi \right) }.
\label{b1cc}
\end{equation}%
The corresponding mode functions in dS spacetime are given by 
\begin{equation}
\varphi _{\sigma }\left( t,r,\vartheta ,\phi \right) =\sinh ^{\frac{D-1}{2}%
}\left( |\eta |/\alpha \right) e^{-iz\eta /\alpha }\frac{Z_{iz-1/2}^{-\mu
}\left( \cosh r\right) }{\sinh ^{D/2-1}r}Y\left( m_{p};\vartheta ,\phi
\right) ,  \label{psigc}
\end{equation}%
where we have used $1/\sinh (t/\alpha )=-\sinh \left( \eta /\alpha \right) $
and have excluded the factor $b_{+}/\Gamma \left( 1-iz\right) $ by
redefining the coefficients $c_{1}$ and $c_{2}$ in (\ref{XZ}). For a massive
field with general curvature coupling the mode functions corresponding to
the conformal vacuum are obtained from (\ref{phisig}) with $b_{-}=0$ and $%
b_{+}$ given by (\ref{b1cc}). The corresponding eigenvalues of the quantum
number $z$ are real.

In order to discuss the adiabatic vacuum we introduce the function $h(\eta )$
of the conformal time in accordance with $h(\eta )=$ $\sinh
^{(D-1)/2}(t/\alpha )f\left( t/\alpha \right) $, where the function $%
t=t(\eta )$ is given by (\ref{etac}). This function obeys the equation 
\begin{equation}
\partial _{\eta }^{2}h\left( \eta \right) +\omega ^{2}\left( z,\eta \right)
h\left( \eta \right) =0,  \label{heq}
\end{equation}%
with time-dependent frequency%
\begin{equation}
\omega \left( z,\eta \right) =\frac{1}{\alpha }\left[ z^{2}-\frac{\nu
^{2}-1/4}{\sinh ^{2}\left( \eta /\alpha \right) }\right] ^{1/2}.  \label{om}
\end{equation}%
From here it follows that the limit $\eta \rightarrow -\infty $ corresponds
to asymptotically static region (static in-region). In terms of the proper
time $t$ this corresponds to the region $t/\alpha \ll 1$. In the zeroth
adiabatic order, for the modes realizing the in-vacuum one has $h^{(0)}(\eta
)\sim e^{-iz\eta /\alpha }$, $\eta \rightarrow -\infty $. Let us consider
the behavior of the mode functions (\ref{phisig}) in that region. By using
the asymptotics for the associated Legendre functions one gets%
\begin{equation}
X_{\nu }^{iz}\left( \cosh (t/\alpha )\right) \approx \sum_{j=+,-}\frac{%
b_{j}e^{-jiz\eta /\alpha }}{\Gamma \left( 1-jiz\right) },\;t/\alpha \ll 1.
\label{XasAd}
\end{equation}%
From here it follows that for the mode functions that are reduced to the
positive energy modes in static spacetime we should take $b_{-}=0$ and,
hence, the conformal and adiabatic vacua coincide. The corresponding state
is also known as hyperbolic vacuum. Note that the latter is different from
the maximally symmetric Bunch-Davies vacuum state (for the relation between
the hyperbolic and Bunch-Davies vacua in the special case of $D=3$
boundary-free dS spacetime see also \cite{Sasa95,Dimi15}).

Now let us consider the flat spacetime limit $\alpha \rightarrow \infty $.
The line element takes the form 
\begin{equation}
ds^{2}=dt^{2}-t^{2}(dr^{2}+\sinh ^{2}rd\Omega _{D-1}^{2}),  \label{ds2Milne}
\end{equation}%
which corresponds to the Milne universe. In order to find the limiting form
of the scalar mode functions (\ref{phisig}) we note that in the limit under
consideration $\nu \approx im\alpha $ and $|\nu |$ is large. We can use the
relation%
\begin{equation}
\underset{\alpha \rightarrow \infty }{\lim }\left[ \left( m\alpha \right)
^{\pm iz}P_{im\alpha -1/2}^{\mp iz}\left( \cosh (t/\alpha )\right) \right]
=J_{\pm iz}\left( mt\right) ,  \label{AsMilne}
\end{equation}%
where $J_{\nu }(x)$ is the Bessel function. For the scalar modes one gets $%
\lim_{\alpha \rightarrow \infty }\varphi _{\sigma }\left( x\right) =\varphi
_{\sigma }^{\mathrm{(Milne)}}\left( x\right) $, where the mode functions in
the Milne universe are given by 
\begin{equation}
\varphi _{\sigma }^{\mathrm{(Milne)}}\left( x\right) =c\frac{\tilde{b}%
_{+}J_{-iz}(mt)+\tilde{b}_{-}J_{iz}(mt)}{t^{\left( D-1\right) /2}}\frac{%
P_{iz-1/2}^{-\mu }\left( \cosh r\right) }{\sinh ^{D/2-1}r}Y(m_{p};\vartheta
,\phi ),  \label{phiMilne}
\end{equation}%
with $|\tilde{b}_{j}|=|b_{j}|$ and 
\begin{equation}
|c|^{2}=\frac{z\sinh (\pi z)}{\pi N\left( m_{p}\right) }|\Gamma (iz+\mu
+1/2)|.  \label{cMilne}
\end{equation}%
These mode functions have been discussed in \cite{Saha20}. In the Milne
universe the conformal and adiabatic vacua are different. The conformal
vacuum corresponds to the special case $\tilde{b}_{-}=0$ with $|\tilde{b}%
_{+}|^{2}=\pi /[2\sinh \left( \pi z\right) ]$ and for the adiabatic vacuum
in the Milne universe 
\begin{equation}
\tilde{b}_{+}=\frac{\sqrt{\pi }e^{\pi z/2}}{2\sinh \left( \pi z\right) },\;%
\tilde{b}_{-}=-\tilde{b}_{+}e^{-\pi z}.  \label{c12Milne}
\end{equation}%
For the adiabatic vacuum the time-dependence in the corresponding mode
function (\ref{phiMilne}) is expressed in terms of the function $t^{\left(
1-D\right) /2}H_{iz}^{\left( 2\right) }\left( mt\right) $ with the Hankel
function $H_{iz}^{\left( 2\right) }\left( x\right) $.

\section{Hadamard function}

\label{sec:Had}

Having specified the general structure of the mode functions we turn to the
construction of the Hadamard function in accordance with (\ref{WF}). The
boundary-free, exterior and interior geometries will be considered
separately.

\subsection{Boundary-free geometry}

We start with the problem where the sphere is absent. For the modes regular
at the origin $r=0$ one should take $c_{2}=0$ in (\ref{XZ}) and the
corresponding mode functions take the form%
\begin{equation}
\varphi _{\sigma }^{(0)}\left( x\right) =C_{0}\frac{X_{\nu }^{iz}\left(
\cosh (t/\alpha )\right) }{\sinh ^{(D-1)/2}(t/\alpha )}\frac{%
P_{iz-1/2}^{-\mu }\left( \cosh r\right) }{\sinh ^{D/2-1}r}Y\left(
m_{p};\vartheta ,\phi \right) ,  \label{phi0}
\end{equation}%
with $0\leq r<\infty $. The spectrum of the quantum number $z$ is
continuous, $0\leq z<\infty $, and in the right-hand side of the
normalization condition (\ref{NCZ}) we take $\delta _{zz^{\prime }}=\delta
(z-z^{\prime })$ with the integration range $u\in \lbrack 1,\infty )$. By
using the result%
\begin{equation}
\int_{1}^{\infty }duP_{iz-1/2}^{-\mu }\left( u\right) P_{iz^{\prime
}-1/2}^{-\mu }\left( u\right) =\frac{\pi \delta \left( z-z^{\prime }\right) 
}{z\sinh \left( \pi z\right) \left\vert \Gamma \left( iz+\mu +1/2\right)
\right\vert ^{2}},  \label{Pint0}
\end{equation}%
for the normalization coefficient one gets%
\begin{equation}
\left\vert C_{0}\right\vert ^{2}=\frac{z\sinh \left( \pi z\right) }{\pi
N\left( m_{p}\right) }\frac{\left\vert \Gamma \left( \mu +iz+1/2\right)
\right\vert ^{2}}{\alpha ^{D-1}}.  \label{C0}
\end{equation}

Substituting the mode functions (\ref{phi0}) into the corresponding mode sum
formula (\ref{WF}) and by using the addition theorem%
\begin{equation}
\sum_{m_{k}}\frac{Y\left( m_{p};\vartheta ,\phi \right) }{N\left(
m_{p}\right) }Y^{\ast }\left( m_{p};\vartheta ^{\prime },\phi ^{\prime
}\right) =\frac{2l+n}{nS_{D}}C_{l}^{n/2}\left( \cos \theta \right) ,
\label{AdY}
\end{equation}%
for spherical harmonics, for the Hadamard function in the boundary-free
geometry we find%
\begin{eqnarray}
G_{0}(x,x^{\prime }) &=&\frac{2\alpha ^{1-D}}{\pi nS_{D}}\overset{\infty }{%
\sum_{l=0}}\mu C_{l}^{n/2}\left( \cos \theta \right) \int_{0}^{\infty }dz%
\text{\thinspace }z\sinh \left( \pi z\right) \left\vert \Gamma \left( \mu
+iz+1/2\right) \right\vert ^{2}  \notag \\
&&\times \frac{X_{\nu }^{iz}\left( y\right) \left[ X_{\nu }^{iz}\left(
y^{\prime }\right) \right] ^{\ast }+X_{\nu }^{iz}\left( y^{\prime }\right) %
\left[ X_{\nu }^{iz}\left( y\right) \right] ^{\ast }}{\left[ \sinh (t/\alpha
)\sinh (t^{\prime }/\alpha )\right] ^{\frac{D-1}{2}}}\frac{P_{iz-1/2}^{-\mu
}\left( u\right) P_{iz-1/2}^{-\mu }\left( u^{\prime }\right) }{\left( \sinh
r\sinh r^{\prime }\right) ^{\frac{D}{2}-1}},  \label{W0g}
\end{eqnarray}%
with $\mu $ defined in (\ref{nu}) and 
\begin{eqnarray}
y &=&\cosh (t/\alpha ),\;y^{\prime }=\cosh (t^{\prime }/\alpha ),  \notag \\
u &=&\cosh r,\;u^{\prime }=\cosh r^{\prime }.  \label{uup}
\end{eqnarray}%
In this expression, $S_{D}=2\pi ^{D/2}/\Gamma (D/2)$ is the surface area of
the unit sphere in $D$-dimensional space, $C_{l}^{n/2}(\cos \theta )$ is the
Gegenbauer polynomial and $\theta $ is the angle between the directions
determined by $(\vartheta ,\phi )$ and $(\vartheta ^{\prime },\phi ^{\prime
})$.

For the hyperbolic vacuum 
\begin{equation}
X_{\nu }^{iz}\left( y\right) =\sqrt{\frac{\pi }{2\sinh \left( \pi z\right) }}%
P_{\nu -1/2}^{iz}\left( y\right) ,  \label{Xad}
\end{equation}%
and the function (\ref{W0g}) is reduced to 
\begin{eqnarray}
G_{0}\left( x,x^{\prime }\right)  &=&\frac{\alpha ^{1-D}}{nS_{D}}%
\sum_{l=0}^{\infty }\mu C_{l}^{n/2}\left( \cos \theta \right)
\int_{0}^{\infty }dz\,z\left\vert \Gamma \left( \mu +iz+1/2\right)
\right\vert ^{2}  \notag \\
&&\times \frac{P_{iz-1/2}^{-\mu }\left( u\right) P_{iz-1/2}^{-\mu }\left(
u^{\prime }\right) }{\left( \sinh r\sinh r^{\prime }\right) ^{\frac{D}{2}-1}}%
\frac{\sum_{j=+,-}P_{\nu -1/2}^{jiz}\left( y\right) P_{\nu
-1/2}^{-jiz}\left( y^{\prime }\right) }{\left[ \sinh (t/\alpha )\sinh
(t^{\prime }/\alpha )\right] ^{\frac{D-1}{2}}}.  \label{W0}
\end{eqnarray}%
The further transformation of the Hadamard function in the boundary-free
geometry is presented in Appendix \ref{sec:VEVbf}. In particular, the
corresponding expression for the hyperbolic vacuum is obtained from (\ref%
{W0g1}) with the function $X_{\nu }^{ix}\left( y\right) $ from (\ref{Xad}):%
\begin{eqnarray}
G_{0}\left( x,x^{\prime }\right)  &=&\frac{\alpha ^{1-D}}{2(2\pi )^{D/2}}%
\int_{0}^{\infty }dz\,z\left\vert \Gamma \left( \frac{D-1}{2}+iz\right)
\right\vert ^{2}  \notag \\
&&\times \frac{\sum_{j=+,-}P_{\nu -1/2}^{jiz}\left( y\right) P_{\nu
-1/2}^{-jiz}\left( y^{\prime }\right) }{\left[ \sinh (t/\alpha )\sinh
(t^{\prime }/\alpha )\right] ^{\frac{D-1}{2}}}\frac{P_{iz-1/2}^{1-D/2}\left( 
\bar{u}\right) }{\left( \bar{u}^{2}-1\right) ^{\frac{D-2}{4}}},  \label{W01}
\end{eqnarray}%
where%
\begin{equation}
\bar{u}=\cosh r\cosh r^{\prime }-\sinh r\sinh r^{\prime }\cos \theta .
\label{u}
\end{equation}%
In the limit $\alpha \rightarrow \infty $, by using the relation (\ref%
{AsMilne}), from (\ref{W01}) we obtain the Hadamard function for the
conformal vacuum in the Milne universe:%
\begin{eqnarray}
G_{0}^{\mathrm{(Milne)}}\left( x,x^{\prime }\right)  &=&\frac{(tt^{\prime
})^{(1-D)/2}}{2(2\pi )^{D/2}}\int_{0}^{\infty }dz\,z\left\vert \Gamma \left( 
\frac{D-1}{2}+iz\right) \right\vert ^{2}  \notag \\
&&\times \frac{P_{iz-1/2}^{1-D/2}\left( \bar{u}\right) }{\left( \bar{u}%
^{2}-1\right) ^{\frac{D-2}{4}}}\sum_{j=+,-}J_{jiz}\left( mt\right)
J_{-jiz}\left( mt^{\prime }\right) .  \label{W01M}
\end{eqnarray}%
It can be checked that this formula is obtained from the corresponding
expression in \cite{Saha20} by making use of the addition theorem (\ref{AdP2}%
).

\subsection{Region outside the sphere}

In the region outside the sphere, $r>r_{0}$, the mode functions have the
form (\ref{phisig}) where the function $Z_{iz-1/2}^{-\mu }\left( u\right) $
is given by (\ref{XZ}). For the exterior region it is more convenient to
take the linear combination of the functions $Q_{iz-1/2}^{-\mu }\left(
u\right) $ and $Q_{-iz-1/2}^{-\mu }\left( u\right) $ by using the relation%
\begin{equation}
P_{iz-1/2}^{-\mu }\left( u\right) =\frac{ie^{i\mu \pi }}{\pi \sinh \left(
\pi z\right) }\sum_{j=+,-}j\cos \left[ \pi \left( \mu -jiz\right) \right]
Q_{jiz-1/2}^{-\mu }\left( u\right) .  \label{PQQ}
\end{equation}%
For the modes with real values of $z$ the ratio of the coefficients in that
combination is determined by the boundary condition (\ref{Rbc}) and is
expressed as $-\bar{Q}_{-iz-1/2}^{-\mu }\left( u_{0}\right) /\bar{Q}%
_{iz-1/2}^{-\mu }\left( u_{0}\right) $ with 
\begin{equation}
u_{0}=\cosh r_{0}.  \label{u0}
\end{equation}%
Here and below, for a given function $F(u)$, the notation with bar is
defined as%
\begin{equation}
\bar{F}\left( u\right) =\left[ B\left( u\right) \partial _{u}+A\left(
u\right) \right] F\left( u\right) ,  \label{Fbar}
\end{equation}%
where the functions $A(u)$ and $B(u)$ are expressed in terms of the Robin
coefficients: 
\begin{eqnarray}
A\left( u\right)  &=&A\sqrt{u^{2}-1}+\left( \frac{D}{2}-1\right) \delta _{%
\mathrm{(j)}}Bu,  \notag \\
B\left( u\right)  &=&-\delta _{\mathrm{(j)}}B\left( u^{2}-1\right) .
\label{ABu}
\end{eqnarray}%
Here, $\mathrm{j=e}$ and $\mathrm{j=i}$ for the regions outside and inside
the sphere, respectively, with $\delta _{\mathrm{(e)}}=-1$ and $\delta _{%
\mathrm{(i)}}=1$. For the corresponding scalar modes one obtains 
\begin{equation}
\varphi _{\sigma }^{\mathrm{(e)}}\left( x\right) =C_{\mathrm{(e)}}\frac{%
X_{\nu }^{iz}\left( \cosh (t/\alpha )\right) }{\sinh ^{(D-1)/2}(t/\alpha )}%
\frac{W_{iz}^{-\mu }\left( \cosh r\right) }{\sinh ^{D/2-1}r}Y\left(
m_{p};\vartheta ,\phi \right) ,  \label{phiExt}
\end{equation}%
where we have defined the function%
\begin{equation}
W_{iz}^{-\mu }\left( u\right) =\bar{Q}_{iz-1/2}^{-\mu }\left( u_{0}\right)
Q_{-iz-1/2}^{-\mu }\left( u\right) -\bar{Q}_{-iz-1/2}^{-\mu }\left(
u_{0}\right) Q_{iz-1/2}^{-\mu }\left( u\right) .  \label{Wf}
\end{equation}%
Similar to the case of the boundary-free geometry, in the exterior region
the eigenvalues for $z$ are continuous.

From (\ref{NCZ}) the following orthonormalization condition is obtained in
terms of the function (\ref{Wf}): 
\begin{equation}
\left\vert C_{\mathrm{(e)}}\right\vert ^{2}\int_{u_{0}}^{\infty
}duW_{iz}^{-\mu }\left( u\right) [W_{iz^{\prime }}^{-\mu }\left( u\right)
]^{\ast }=\frac{\delta \left( z-z^{\prime }\right) }{\alpha ^{D-1}N\left(
m_{p}\right) }.  \label{NCW}
\end{equation}%
The $u$-integral diverges in the upper limit for $z=z^{\prime }$ and, hence,
the contribution from the integration range with large $u$ dominates. So, in
order to evaluate this integral, we can replace the associated Legendre
functions $Q_{\pm iz-1/2}^{-\mu }\left( u\right) $ in (\ref{Wf}) by their
asymptotic expressions for large values of the argument:%
\begin{equation}
Q_{\pm iz-1/2}^{-\mu }\left( u\right) \approx \sqrt{\pi }e^{-i\mu \pi }\frac{%
\Gamma \left( 1/2\pm iz-\mu \right) }{\Gamma \left( 1\pm iz\right) }\frac{%
e^{\mp iz\ln \left( 2u\right) }}{\sqrt{2u}}.  \label{Qas}
\end{equation}%
This leads to the following result%
\begin{equation}
\left\vert C_{\mathrm{(e)}}\right\vert ^{2}=\frac{z\left\vert \Gamma \left(
1/2+iz-\mu \right) \bar{Q}_{iz-1/2}^{-\mu }\left( u_{0}\right) \right\vert
^{-2}}{\pi \alpha ^{D-1}N\left( m_{p}\right) \sinh \left( \pi z\right) },
\label{Ce}
\end{equation}%
for the normalization coefficient.

With the mode functions (\ref{phiExt}), from the mode sum formula (\ref{WF}%
), by using the addition theorem (\ref{AdY}), the following representation
is obtained for the Hadamard function in the exterior region: 
\begin{eqnarray}
G(x,x^{\prime }) &=&\frac{2\alpha ^{1-D}}{\pi nS_{D}}\frac{\left( \sinh
r\sinh r^{\prime }\right) ^{1-\frac{D}{2}}}{\left[ \sinh (t/\alpha )\sinh
(t^{\prime }/\alpha )\right] ^{\frac{D-1}{2}}}\sum_{l=0}^{\infty }\mu
C_{l}^{n/2}\left( \cos \theta \right) \int_{0}^{\infty }dz\,z  \notag \\
&&\times \frac{X_{\nu }^{iz}\left( y\right) W_{iz}^{-\mu }\left( u\right)
[X_{\nu }^{iz}\left( y^{\prime }\right) W_{iz}^{-\mu }\left( u^{\prime
}\right) ]^{\ast }+\left\{ (y,u)\rightleftarrows (y^{\prime },u^{\prime
})\right\} }{\sinh \left( \pi z\right) \left\vert \Gamma \left( 1/2+iz-\mu
\right) \right\vert ^{2}\left\vert \bar{Q}_{iz-1/2}^{-\mu }\left(
u_{0}\right) \right\vert ^{2}}.  \label{WFe}
\end{eqnarray}%
We can also write this expression in terms of the function%
\begin{equation}
Y_{iz-1/2}^{-\mu }\left( u\right) =\bar{Q}_{iz-1/2}^{-\mu }\left(
u_{0}\right) P_{iz-1/2}^{-\mu }\left( u\right) -\bar{P}_{iz-1/2}^{-\mu
}\left( u_{0}\right) Q_{iz-1/2}^{-\mu }\left( u\right) ,  \label{Y}
\end{equation}%
by using the relation%
\begin{equation}
W_{iz}^{-\mu }\left( u\right) =\frac{i\pi e^{-i\mu \pi }\sinh \left( \pi
z\right) }{\cos \left[ \pi \left( \mu +iz\right) \right] }Y_{iz-1/2}^{-\mu
}\left( u\right) .  \label{WY}
\end{equation}%
The corresponding expression takes the form%
\begin{eqnarray}
G(x,x^{\prime }) &=&\frac{2\alpha ^{1-D}}{\pi nS_{D}}\overset{\infty }{%
\sum_{l=0}}\mu C_{l}^{n/2}\left( \cos \theta \right) \int_{0}^{\infty
}dz\,z\sinh \left( \pi z\right) \frac{\left\vert \Gamma \left( \mu
+iz+1/2\right) \right\vert ^{2}}{\left\vert \bar{Q}_{iz-1/2}^{-\mu }\left(
u_{0}\right) \right\vert ^{2}}  \notag \\
&&\times \frac{X_{\nu }^{iz}\left( y\right) Y_{iz-1/2}^{-\mu }\left(
u\right) [X_{\nu }^{iz}\left( y^{\prime }\right) Y_{iz-1/2}^{-\mu }\left(
u^{\prime }\right) ]^{\ast }+\left\{ (y,u)\rightleftarrows (y^{\prime
},u^{\prime })\right\} }{\left[ \sinh (t/\alpha )\sinh (t^{\prime }/\alpha )%
\right] ^{\frac{D-1}{2}}\left( \sinh r\sinh r^{\prime }\right) ^{\frac{D}{2}%
-1}}.  \label{WFe2}
\end{eqnarray}

Depending on the ratio of the coefficients in the Robin boundary condition,
in addition to the modes with real $z$, one can have exterior modes with
purely imaginary $z=\pm i\chi $, $\chi >0$. The radial dependence of the
corresponding normalizable mode functions is expressed in terms of the
function $Q_{\chi -1/2}^{-\mu }(u)/\sinh ^{D/2-1}r$ and they correspond to
bound states. From the boundary condition (\ref{Rbc}) we get the equation $%
\bar{Q}_{\chi -1/2}^{-\mu }(u_{0})=0$ that determines the eigenvalues for $%
\chi $. As it has been already discussed in \cite{Saha14}, one has a
critical value $\beta _{l}^{\mathrm{(e)}}(u_{0})$ for the ratio $\beta =A/B$
such that there are no roots for this equation in the range $\beta \leq
\beta _{l}^{\mathrm{(e)}}(u_{0})$ and a single root exists in the region $%
\beta >\beta _{l}^{\mathrm{(e)}}(u_{0})$. In \cite{Saha14} it has been shown
that $\beta _{l}^{\mathrm{(e)}}(u_{0})$ is an increasing function of $l$ and
a decreasing function of $u_{0}$. In addition, we have $\beta _{l}^{\mathrm{%
(e)}}(u_{0})\geq (D-1)/2$. For the hyperbolic vacuum the corresponding
function $X_{\nu }^{iz}\left( y\right) $ is given by (\ref{Xad}) and the
allowed values for $z$ are real. In order to exclude the modes with $z=\pm
i\chi $, it will be assumed that $\beta \leq \beta _{0}^{\mathrm{(e)}%
}(u_{0}) $. Note that the Neumann boundary condition ($\beta =0$) belongs to
this range. In the special case $D=3$, by using 
\begin{equation}
Q_{\chi -1/2}^{-1/2}(u_{0})=-i\frac{\sqrt{\pi /2}e^{-\chi r_{0}}}{\chi \sqrt{%
\sinh r_{0}}},  \label{Q12}
\end{equation}%
for the bound state corresponding to the mode $l=0$ we get $\chi =\beta
-\coth r_{0}$. From here it follows that $\beta _{0}^{\mathrm{(e)}%
}(u_{0})=\coth r_{0}$ for $D=3$.

We are interested in the effects of the sphere on the properties of the
hyperbolic vacuum. The corresponding Hadamard function outside the sphere is
given by (\ref{WFe2}) with the function (\ref{Xad}). By taking into account
the expression (\ref{W0}) for the boundary-free geometry, the corresponding
sphere-induced contribution $G_{\mathrm{s}}\left( x,x^{\prime }\right)
=G\left( x,x^{\prime }\right) -G_{0}\left( x,x^{\prime }\right) $ is
presented as 
\begin{eqnarray}
G_{\mathrm{s}}\left( x,x^{\prime }\right) &=&\frac{\alpha ^{1-D}}{nS_{D}}%
\overset{\infty }{\sum_{l=0}}\mu C_{l}^{n/2}\left( \cos \theta \right)
\int_{0}^{\infty }dz\,z\left\vert \Gamma \left( \mu +iz+1/2\right)
\right\vert ^{2}  \notag \\
&&\times \frac{P_{\nu -1/2}^{iz}\left( y\right) P_{\nu -1/2}^{-iz}\left(
y^{\prime }\right) U_{iz-1/2}^{-\mu }(u,u^{\prime })+\left\{
(y,u)\rightleftarrows (y^{\prime },u^{\prime })\right\} }{\left[ \sinh
(t/\alpha )\sinh (t^{\prime }/\alpha )\right] ^{\frac{D-1}{2}}\left( \sinh
r\sinh r^{\prime }\right) ^{\frac{D}{2}-1}},  \label{Wse}
\end{eqnarray}%
with the notation%
\begin{equation}
U_{iz-1/2}^{-\mu }(u,u^{\prime })=\frac{Y_{iz-1/2}^{-\mu }\left( u\right)
[Y_{iz-1/2}^{-\mu }\left( u^{\prime }\right) ]^{\ast }}{\left\vert \bar{Q}%
_{iz-1/2}^{-\mu }\left( u_{0}\right) \right\vert ^{2}}-P_{iz-1/2}^{-\mu
}\left( u\right) P_{iz-1/2}^{-\mu }\left( u^{\prime }\right) .  \label{Uf}
\end{equation}%
For the further transformation of the function (\ref{Wse}) it is convenient
to use the relation%
\begin{equation}
U_{iz-1/2}^{-\mu }(u,u^{\prime })=\frac{-ie^{i\mu \pi }}{\pi \sinh \left(
\pi z\right) }\underset{j=+,-}{\sum }j\cos \left[ \pi \left( \mu -jiz\right) %
\right] \frac{\bar{P}_{iz-1/2}^{-\mu }\left( u_{0}\right) }{\bar{Q}%
_{jiz-1/2}^{-\mu }\left( u_{0}\right) }Q_{jiz-1/2}^{-\mu }\left( u\right)
Q_{jiz-1/2}^{-\mu }\left( u^{\prime }\right) .  \label{RelU}
\end{equation}%
The term with $j=-$ ($j=+$) exponentially decreases in the upper (lower)
half-plane of the complex variable $z$ in the limit $\mathrm{Im}%
\,z\rightarrow +\infty $ ($\mathrm{Im}\,z\rightarrow -\infty $). On the base
of these properties, in (\ref{Wse}), with substitution (\ref{RelU}), we can
rotate the contour of the integration in the complex plane $z$ by the angles 
$\pi /2$ and $-\pi /2$ for the terms with $j=-$ and $j=+$, respectively.
This leads to the representation 
\begin{eqnarray}
G_{\mathrm{s}}\left( x,x^{\prime }\right) &=&-\frac{2\alpha ^{1-D}}{nS_{D}}%
\overset{\infty }{\sum_{l=0}}\mu C_{l}^{n/2}\left( \cos \theta \right)
\int_{0}^{\infty }dz\,\frac{ze^{-i\mu \pi }}{\sin \left( \pi z\right) }\frac{%
\bar{P}_{z-1/2}^{-\mu }\left( u_{0}\right) }{\bar{Q}_{z-1/2}^{\mu }\left(
u_{0}\right) }  \notag \\
&&\times \frac{Q_{z-1/2}^{\mu }\left( u\right) Q_{z-1/2}^{\mu }\left(
u^{\prime }\right) }{\left( \sinh r\sinh r^{\prime }\right) ^{\frac{D}{2}-1}}%
\frac{\sum_{j=+,-}P_{\nu -1/2}^{-jz}\left( y\right) P_{\nu -1/2}^{jz}\left(
y^{\prime }\right) }{\left[ \sinh (t/\alpha )\sinh (t^{\prime }/\alpha )%
\right] ^{\frac{D-1}{2}}}.  \label{Wse2}
\end{eqnarray}%
Hence, the Hadamard function in the region outside the sphere is presented
as 
\begin{equation}
G\left( x,x^{\prime }\right) =G_{0}\left( x,x^{\prime }\right) +G_{\mathrm{s}%
}\left( x,x^{\prime }\right) ,  \label{WFidec}
\end{equation}%
where the boundary-free contribution for the hyperbolic vacuum is given by (%
\ref{W01}). In the flat spacetime limit, corresponding to $\alpha
\rightarrow \infty $, by using the relation%
\begin{equation}
\lim_{\alpha \rightarrow \infty }\left[ P_{\nu -1/2}^{-jz}\left( y\right)
P_{\nu -1/2}^{jz}\left( y^{\prime }\right) \right] =J_{jz}(mt)J_{-jz}(mt^{%
\prime }),  \label{PPJJp}
\end{equation}%
from (\ref{Wse2}) the boundary-induced Hadamard function is obtained outside
the sphere in the Milne universe \cite{Saha20}.

\subsection{Hadamard function inside the sphere}

For the interior region, $r<r_{0}$, the regularity condition at the sphere
center fixes $c_{2}=0$ in (\ref{XZ}). The corresponding mode functions are
expressed as 
\begin{equation}
\varphi _{\sigma }^{\mathrm{(i)}}\left( t,r,\vartheta ,\phi \right) =C_{%
\mathrm{(i)}}\frac{X_{\nu }^{iz}\left( \cosh (t/\alpha )\right) }{\sinh
^{(D-1)/2}(t/\alpha )}\frac{P_{iz-1/2}^{-\mu }\left( \cosh r\right) }{\sinh
^{D/2-1}r}Y\left( m_{p};\vartheta ,\phi \right) .  \label{phiInt}
\end{equation}%
From the boundary condition (\ref{Rbc}) with $\mathrm{j}=\mathrm{i}$ we
obtain the equation that determines the allowed values of the quantum number 
$z$: 
\begin{equation}
\bar{P}_{iz-1/2}^{-\mu }(u_{0})=0,  \label{Eigeq}
\end{equation}%
with $u_{0}$ defined by (\ref{u0}). For the region under consideration the
notation with bar in (\ref{Eigeq}) is defined by (\ref{Fbar}) where now $%
\delta _{\mathrm{(j)}}=\delta _{\mathrm{(i)}}=1$ in (\ref{ABu}). Hence,
unlike the exterior region, inside the sphere the eigenvalues of $z$ form a
discrete set. The positive solutions of the eigenvalue equation (\ref{Eigeq}%
) we will denote as $z=z_{k}$, $k=1,2,\ldots $, assuming that $z_{k+1}>z_{k}$%
. These solutions do not depend on the curvature coupling parameter and on
the field mass. In the special case $D=3$, by taking into account that 
\begin{equation}
P_{iz-1/2}^{-1/2}\left( u_{0}\right) =\sqrt{\frac{2}{\pi }}\frac{\sin \left(
zr_{0}\right) }{z\sqrt{\sinh r_{0}}},  \label{Pm12}
\end{equation}%
the eigenvalue equation for the mode $l=0$ is simplified to 
\begin{equation}
\left( \beta +\coth r_{0}\right) \sin \left( zr_{0}\right) /z=\cos \left(
zr_{0}\right) ,  \label{Eigeql0}
\end{equation}%
with the notation $\beta =A/B$.

The normalization coefficient $C_{\mathrm{(i)}}$ in (\ref{phiInt}) is
determined by the condition (\ref{NCZ}) with $Z_{iz-1/2}^{-\mu }(u)=C_{%
\mathrm{(i)}}P_{iz-1/2}^{-\mu }\left( u\right) $, $z=z_{k}$, where the
integration goes over the region $[1,u_{0}]$ and in the right-hand side $%
\delta _{zz^{\prime }}=\delta _{kk^{\prime }}$. The corresponding procedure
is similar to that considered in \cite{Saha14}. By using the integral 
\begin{equation}
\int_{1}^{u_{0}}du\,\left[ P_{iz-1/2}^{\mu }\left( u\right) \right] ^{2}=%
\frac{u_{0}^{2}-1}{2z}\left[ \partial _{z}P_{iz-1/2}^{-\mu }\left(
u_{0}\right) \partial _{u}P_{iz-1/2}^{-\mu }\left( u_{0}\right)
-P_{iz-1/2}^{-\mu }\left( u_{0}\right) \partial _{z}\partial
_{u}P_{iz-1/2}^{-\mu }\left( u_{0}\right) \right] ,  \label{Pint}
\end{equation}%
and the eigenvalue equation (\ref{Eigeq}), we can show that%
\begin{equation}
\left\vert C_{\mathrm{(i)}}\right\vert ^{2}=\frac{2\alpha ^{1-D}e^{i\mu \pi
}z}{\pi N\left( m_{p}\right) }\left\vert \Gamma \left( \mu +iz+1/2\right)
\right\vert ^{2}T_{\mu }\left( z,u_{0}\right) ,  \label{Ci}
\end{equation}%
with $z=z_{k}$. Here we have introduced the notation 
\begin{equation}
T_{\mu }\left( z,u\right) =\frac{\bar{Q}_{iz-1/2}^{-\mu }\left( u\right) }{%
\partial _{z}\bar{P}_{iz-1/2}^{-\mu }\left( u\right) }\cos \left[ \pi \left(
\mu -iz\right) \right] .  \label{Tmui}
\end{equation}%
With (\ref{Ci}), the scalar mode functions inside the sphere are completely
determined. Substituting the modes (\ref{phiInt}) into the mode sum (\ref{WF}%
) and making use of (\ref{AdY}), the Hadamard function inside the sphere is
presented in the form%
\begin{eqnarray}
G(x,x^{\prime }) &=&\frac{4\alpha ^{1-D}}{\pi nS_{D}}\sum_{l=0}^{\infty }\mu
C_{l}^{n/2}\left( \cos \theta \right) e^{i\mu \pi }\sum_{k=1}^{\infty
}T_{\mu }\left( z,u_{0}\right) z\left\vert \Gamma \left( \mu +iz+1/2\right)
\right\vert ^{2}  \notag \\
&&\left. \times \frac{X_{\nu }^{iz}\left( y\right) \left[ X_{\nu
}^{iz}\left( y^{\prime }\right) \right] ^{\ast }+X_{\nu }^{iz}\left(
y^{\prime }\right) \left[ X_{\nu }^{iz}\left( y\right) \right] ^{\ast }}{%
\left[ \sinh (t/\alpha )\sinh (t^{\prime }/\alpha )\right] ^{\frac{D-1}{2}}}%
\frac{P_{iz-1/2}^{-\mu }\left( u\right) P_{iz-1/2}^{-\mu }\left( u^{\prime
}\right) }{\left( \sinh r\sinh r^{\prime }\right) ^{\frac{D}{2}-1}}%
\right\vert _{z=z_{k}}.  \label{WFi}
\end{eqnarray}

As in the previous discussion for the exterior region, we will assume that
the field is prepared in the hyperbolic vacuum state with the function (\ref%
{Xad}). It has been already emphasized that for the hyperbolic vacuum the
eigenvalues of the quantum number $z$ are real. However, depending on the
ratio $\beta =A/B$, the eigenvalue equation (\ref{Eigeq}) may have purely
imaginary roots. The conditions for the presence of those roots have been
specified in \cite{Saha14}. For given values of $u_{0}$ and $l$ there exists
a critical value of $\beta $, denoted here by $\beta _{l}^{\mathrm{(i)}%
}(u_{0})$, such that all the roots are real for $\beta \leq \beta _{l}^{%
\mathrm{(i)}}(u_{0})$ and a pair of purely imaginary roots $z_{\pm }=\pm
i|z_{\pm }|$ appears for $\beta >\beta _{l}^{\mathrm{(i)}}(u_{0})$. For the
critical value one has $\beta _{l}^{\mathrm{(i)}}(u_{0})>-(D-1)/2$ and it is
an increasing function of $l$ and a decreasing function of $u_{0}$. For the
critical values of the Robin coefficient in the exterior and interior
regions one has the relation $\beta _{l}^{\mathrm{(e)}}(u_{0})>\beta _{l}^{%
\mathrm{(i)}}(u_{0})$. In the case $D=3$ from (\ref{Eigeql0}) we get%
\begin{equation}
\beta _{0}^{\mathrm{(i)}}(u_{0})=\frac{1}{r_{0}}-\coth r_{0}.  \label{bet0}
\end{equation}%
In the discussion below, for the interior region we will assume the values
of $\beta $ in the range $\beta \leq \beta _{0}^{\mathrm{(i)}}(u_{0})$,
where all the roots of the equation (\ref{Eigeq}) are real.

For the Hadamard function corresponding to the hyperbolic vacuum we get the
representation%
\begin{eqnarray}
G(x,x^{\prime }) &=&\frac{2\alpha ^{1-D}}{nS_{D}}\sum_{l=0}^{\infty }\mu
C_{l}^{n/2}\left( \cos \theta \right) e^{i\mu \pi }\sum_{k=1}^{\infty
}zT_{\mu }\left( z,u_{0}\right) \frac{\left\vert \Gamma \left( \mu
+iz+1/2\right) \right\vert ^{2}}{\sinh \left( \pi z\right) }  \notag \\
&&\left. \times \frac{P_{iz-1/2}^{-\mu }\left( u\right) P_{iz-1/2}^{-\mu
}\left( u^{\prime }\right) }{\left( \sinh r\sinh r^{\prime }\right) ^{\frac{D%
}{2}-1}}\frac{\sum_{j=+,-}P_{\nu -1/2}^{jiz}\left( y\right) P_{\nu
-1/2}^{-jiz}\left( y^{\prime }\right) }{\left[ \sinh (t/\alpha )\sinh
(t/\alpha )^{\prime }\right] ^{\frac{D-1}{2}}}\right\vert _{z=z_{k}},
\label{WFiad}
\end{eqnarray}%
where the relation (\ref{Prel}) has been used. The summation in this formula
goes over the eigenvalues $z_{k}$ that are defined implicitly, as roots of
the equation (\ref{Eigeq}). A more convenient representation is found by
using the formula \cite{Saha20}%
\begin{eqnarray}
\sum_{k=1}^{\infty }T_{\mu }\left( z_{k},w\right) h\left( z_{k}\right)  &=&%
\frac{e^{-i\mu \pi }}{2}\int_{0}^{\infty }dx\,\sinh \left( \pi x\right)
h\left( x\right)   \notag \\
&&+\sum_{k}\cos \left[ \pi \left( \mu -x_{k}\right) \right] \frac{\bar{Q}%
_{x_{k}-1/2}^{-\mu }\left( w\right) }{\bar{P}_{x_{k}-1/2}^{-\mu }\left(
w\right) }\sum_{j=+,-}\mathrm{Res}_{z=jix_{k}}h(z)  \notag \\
&&-\frac{1}{2\pi }\int_{0}^{\infty }dx\,\frac{\bar{Q}_{x-1/2}^{-\mu }\left(
w\right) }{\bar{P}_{x-1/2}^{-\mu }\left( w\right) }\cos \left[ \pi \left(
x-\mu \right) \right] \sum_{j=+,-}h\left( xe^{j\pi i/2}\right) ,  \label{APF}
\end{eqnarray}%
with a function $h(z)$ analytic in the half-plane $\mathrm{Re}\,z>0$. In (%
\ref{APF}), the points $\pm ix_{k}$ are possible poles of the function $h(z)$
on the imaginary axis. In the presence of these poles, it is assumed that
the last integral is convergent in the sense of the principal value. The
corresponding formula in the case when the poles on the imaginary axis are
absent has been derived in \cite{Saha14,Saha08a} by using the generalized
Abel-Plana formula \cite{Saha08b}. Additional conditions on the function $%
h(z)$ can be found in those references. The function $h(z)$ corresponding to
the representation (\ref{WFiad}) of the Hadamard function is real for real
values of $z$ and is expressed as 
\begin{equation}
h\left( z\right) =z\frac{\Gamma \left( \mu +iz+1/2\right) }{\sinh \left( \pi
z\right) }\Gamma \left( \mu -iz+1/2\right) P_{iz-1/2}^{-\mu }\left( u\right)
P_{iz-1/2}^{-\mu }\left( u^{\prime }\right) \sum_{j=+,-}P_{\nu
-1/2}^{jiz}\left( y\right) P_{\nu -1/2}^{-jiz}\left( y^{\prime }\right) .
\label{hzi}
\end{equation}%
It is an even function of $z$ and has simple poles at $z=\pm ix_{k}=\pm i\pi
k$ with $k=1,2,\ldots $. In this special case the residue term in the
right-hand side of (\ref{APF}) vanishes. Note that the poles coming from the
gamma function in the integrand of the last integral are cancelled by the
zeros of the function $\cos \left[ \pi \left( x-\mu \right) \right] $.

The contribution to the Hadamard function coming from the first term in the
right-hand side of (\ref{APF}) gives the corresponding function in the
boundary-free geometry and the Hadamard function inside the sphere is
decomposed as (\ref{WFidec}). The sphere-induced part comes from the last
integral in (\ref{APF}) and is given by the expression 
\begin{eqnarray}
G_{\mathrm{s}}\left( x,x^{\prime }\right)  &=&-\frac{2\alpha ^{1-D}}{nS_{D}}%
\sum_{l=0}^{\infty }\mu C_{l}^{n/2}\left( \cos \theta \right)
\int_{0}^{\infty }dz\frac{ze^{-i\mu \pi }}{\sin \left( \pi z\right) }\frac{%
\bar{Q}_{z-1/2}^{\mu }\left( u_{0}\right) }{\bar{P}_{z-1/2}^{-\mu }\left(
u_{0}\right) }  \notag \\
&&\times \frac{P_{z-1/2}^{-\mu }\left( u\right) P_{z-1/2}^{-\mu }\left(
u^{\prime }\right) }{\left( \sinh r\sinh r^{\prime }\right) ^{\frac{D}{2}-1}}%
\frac{\sum_{j=+,-}P_{\nu -1/2}^{jz}\left( y\right) P_{\nu -1/2}^{-jz}\left(
y^{\prime }\right) }{\left[ \sinh (t/\alpha )\sinh (t/\alpha )^{\prime }%
\right] ^{\frac{D-1}{2}}},  \label{Wsi}
\end{eqnarray}%
with $r,r^{\prime }<r_{0}$. Here, for the transformation of the integrand,
the relation%
\begin{equation}
\Gamma \left( \mu +z+1/2\right) \Gamma \left( \mu -z+1/2\right)
Q_{z-1/2}^{-\mu }\left( u_{0}\right) =\pi \frac{e^{-2i\mu \pi
}Q_{z-1/2}^{\mu }\left( u_{0}\right) }{\cos \left[ \pi \left( \mu -z\right) %
\right] }  \label{RelQQ}
\end{equation}%
has been used. By taking into account the asymptotics of the functions $%
P_{z-1/2}^{-\mu }\left( u\right) $, $Q_{z-1/2}^{\mu }\left( u\right) $, $%
P_{\nu -1/2}^{\pm z}\left( y\right) $ (see, for instance, \cite{Nist10}) it
can be seen that for large values of $z$ the integrand in (\ref{Wsi})
behaves as $e^{z[r+r^{\prime }+\left\vert \eta ^{\prime }-\eta \right\vert
/\alpha -2r_{0}]}/z$. From here it follows that the representation (\ref{Wsi}%
) is valid in the range $r+r^{\prime }+\left\vert \eta ^{\prime }-\eta
\right\vert /\alpha <2r_{0}$. We recall that the integral in (\ref{Wsi}) is
understood in the sense of the principal value. Comparing with (\ref{Wse2}),
we see that the sphere-induced contributions inside and outside the sphere
are obtained from each other by the replacements%
\begin{equation}
Q_{z-1/2}^{\mu }\left( w\right) \rightleftarrows P_{z-1/2}^{-\mu }\left(
w\right) ,\;w=u,u_{0},  \label{ReplPQ}
\end{equation}%
of the associated Legendre functions.

In the limit $r_{0}\rightarrow \infty $ for the associated Legendre
functions in the integrand of (\ref{Wsi}) one has the asymptotics%
\begin{eqnarray}
P_{x-1/2}^{-\mu }\left( u_{0}\right) &\sim &\frac{x^{-\mu -1/2}e^{r_{0}x}}{%
\sqrt{2\pi \sinh r_{0}}},  \notag \\
Q_{x-1/2}^{-\mu }\left( u_{0}\right) &\sim &\frac{\pi e^{i\mu \pi }x^{-\mu
-1/2}e^{-r_{0}x}}{\sqrt{2\pi \sinh r_{0}}}.  \label{PQas}
\end{eqnarray}%
From here it follows that in that limit, as expected, the part $G_{\mathrm{s}%
}\left( x,x^{\prime }\right) $ tends to zero. In the limit of large
curvature radius for the background spacetime, by making use of (\ref{PPJJp}%
), the sphere-induced contribution (\ref{Wsi}) is reduced to the
corresponding two-point function inside the sphere in the Milne universe,
given in \cite{Saha20}.

\section{VEV of the field squared}

\label{sec:phi2i}

We start the consideration of the local characteristics of the vacuum state
from the VEV of the field squared. By using (\ref{TikVev}) and (\ref{WFidec}%
), the VEV is presented in the decomposed form 
\begin{equation}
\left\langle \varphi ^{2}\right\rangle =\left\langle \varphi
^{2}\right\rangle _{0}+\left\langle \varphi ^{2}\right\rangle _{\mathrm{s}},
\label{phi2i}
\end{equation}%
where $\left\langle \varphi ^{2}\right\rangle _{0}$ is the VEV in the
boundary-free geometry and $\left\langle \varphi ^{2}\right\rangle _{\mathrm{%
s}}=\lim_{x^{\prime }\rightarrow x}G_{\mathrm{s}}\left( x,x^{\prime }\right)
/2$ is the contribution induced by the sphere. For the part depending on the
angular coordinates one has $\lim_{x^{\prime }\rightarrow x}2\mu
C_{l}^{n/2}\left( \cos \theta \right) =D_{l}$, where 
\begin{equation}
D_{l}=\frac{\left( 2l+n\right) \Gamma \left( l+n\right) }{l!\Gamma \left(
n+1\right) }  \label{Dl}
\end{equation}%
determines the degeneracy of the angular mode with fixed $l$. We consider
the properties of the VEVs outside and inside the sphere separately.

\subsection{Interior region}

For the region inside the sphere from (\ref{Wsi}) we get 
\begin{eqnarray}
\left\langle \varphi ^{2}\right\rangle _{\mathrm{s}} &=&-\frac{\alpha
^{1-D}\sinh ^{2-D}r}{S_{D}\sinh ^{D-1}\left( t/\alpha \right) }%
\sum_{l=0}^{\infty }D_{l}\int_{0}^{\infty }dx\,\frac{xe^{-i\mu \pi }}{\sin
\left( \pi x\right) }  \notag \\
&&\times \,\frac{\bar{Q}_{x-1/2}^{\mu }\left( u_{0}\right) }{\bar{P}%
_{x-1/2}^{-\mu }\left( u_{0}\right) }P_{\nu -1/2}^{x}\left( y\right) P_{\nu
-1/2}^{-x}\left( y\right) [P_{x-1/2}^{-\mu }\left( u\right) ]^{2}.
\label{phi2sin}
\end{eqnarray}%
As it has been already mentioned, for $r<r_{0}$ ($u<u_{0}$) the
renormalization is required for the part $\left\langle \varphi
^{2}\right\rangle _{0}$ only. The integral in (\ref{phi2sin}) (understood in
the sense of the principal value) can be presented in the form where the
integrand has no poles. That is done by using the formula (see \cite{Saha20})%
\begin{equation}
\int_{0}^{\infty }dx\,\frac{f(x)}{\sin \left( \pi x\right) }=\frac{2}{\pi }%
\sideset{}{'}{\sum}_{k=0}^{\infty }(-1)^{k}\int_{0}^{\infty }dx\,\frac{%
xf(x)-kf(k)}{x^{2}-k^{2}},  \label{PVint}
\end{equation}%
where the prime on the sign of summation means that the term $k=0$ is taken
with an additional coefficient 1/2. This replacement is convenient in the
numerical evaluations of the sphere-induced VEVs. In the flat spacetime
limit, corresponding to $\alpha \rightarrow \infty $, by using the relation (%
\ref{PPJJp}) with $t^{\prime }=t$, from (\ref{phi2sin}) the VEV of the field
squared is obtained inside a sphere in background of the Milne universe \cite%
{Saha20}.

For a conformally coupled massless field one has $\nu =1/2$ and, by using (%
\ref{P0}) with $z=ix$, we get%
\begin{equation}
\left\langle \varphi ^{2}\right\rangle _{\mathrm{s}}=\frac{\left\langle
\varphi ^{2}\right\rangle _{\mathrm{s}}^{\mathrm{(st)}}}{\sinh ^{D-1}\left(
t/\alpha \right) },  \label{phi2sconf}
\end{equation}%
where%
\begin{equation}
\left\langle \varphi ^{2}\right\rangle _{\mathrm{s}}^{\mathrm{(st)}}=-\frac{%
\alpha ^{1-D}}{\pi S_{D}}\sum_{l=0}^{\infty }e^{-i\mu \pi
}D_{l}\int_{0}^{\infty }dx\,\,\frac{\bar{Q}_{x-1/2}^{\mu }\left(
u_{0}\right) }{\bar{P}_{x-1/2}^{-\mu }\left( u_{0}\right) }\frac{%
[P_{x-1/2}^{-\mu }\left( u\right) ]^{2}}{\sinh ^{D-2}r},  \label{phi2st}
\end{equation}%
is the VEV for a massless conformally coupled scalar field induced by a
sphere with radius $r_{0}$ in a static negative constant curvature space
with the curvature radius $\alpha $ (see \cite{Saha14}). Equation (\ref%
{phi2sconf}) is the standard relation between two conformally related
problems.

The general formula (\ref{phi2sin}) is rather complicated and in order to
clarify the behavior of the sphere-induced VEV we consider asymptotic
regions of the parameters. We start with the region $t/\alpha \ll 1$. In
this region the argument of the functions $P_{\nu -1/2}^{\pm x}\left(
y\right) $ in the integrand of (\ref{phi2sin}) is close to 1 and for $\sigma
\neq 1,2,\ldots $ we use the asymptotic formula 
\begin{equation}
P_{\rho }^{\sigma }\left( y\right) \approx \frac{1}{\Gamma \left( 1-\sigma
\right) }\left( \frac{2}{y-1}\right) ^{\sigma /2},\;0<y-1\ll 1.  \label{Pas1}
\end{equation}%
For the time-dependent part in (\ref{phi2sin}) this gives%
\begin{equation}
\frac{P_{\nu -1/2}^{x}\left( y\right) P_{\nu -1/2}^{-x}\left( y\right) }{%
\alpha ^{D-1}\sinh ^{D-1}\left( t/\alpha \right) }\approx \frac{\sin \left(
\pi x\right) }{\pi xt^{D-1}}.  \label{PPas1}
\end{equation}%
Substituting this into (\ref{phi2sin}), to the leading order we get%
\begin{equation}
\left\langle \varphi ^{2}\right\rangle _{\mathrm{s}}\approx -\frac{t^{1-D}}{%
\pi S_{D}}\sum_{l=0}^{\infty }e^{-i\mu \pi }D_{l}\int_{0}^{\infty }dx\,\frac{%
\bar{Q}_{x-1/2}^{\mu }\left( u_{0}\right) }{\bar{P}_{x-1/2}^{-\mu }\left(
u_{0}\right) }\frac{[P_{x-1/2}^{-\mu }\left( u\right) ]^{2}}{\sinh ^{D-2}r}.
\label{phi2smt}
\end{equation}%
The expression in the right-hand side coincides with the VEV of the field
squared for the conformal vacuum of a massless scalar field in the Milne
universe \cite{Saha20}. Of course, this result is natural, because, for a
given $t$, the limit under consideration corresponds to large values of the
curvature radius $\alpha $ and the effects of gravity are weak. Comparing (%
\ref{phi2smt}) with (\ref{phi2st}), we see that in the limit $t/\alpha \ll 1$
one has the relation $\left\langle \varphi ^{2}\right\rangle _{\mathrm{s}%
}\approx \left( \alpha /t\right) ^{D-1}\left\langle \varphi
^{2}\right\rangle _{\mathrm{s}}^{\mathrm{(st)}}$ with $\left\langle \varphi
^{2}\right\rangle _{\mathrm{s}}^{\mathrm{(st)}}$ being the corresponding VEV
for a conformally coupled massless field in static spacetime with negative
constant curvature space.

The late stages of the expansion correspond to the opposite limit $t/\alpha
\gg 1$ and the argument of the functions $P_{\nu -1/2}^{\pm x}\left(
y\right) $ is large. In this case we use the asymptotic%
\begin{equation}
P_{\nu -1/2}^{\sigma }\left( y\right) \approx \frac{\Gamma \left( \nu
\right) \left( 2y\right) ^{\nu -1/2}}{\pi ^{1/2}\Gamma \left( \nu -\sigma
+1/2\right) },  \label{Pas2}
\end{equation}%
with $\nu >0$ and $\sigma -\nu +1/2\neq 1,2,\ldots $. For the sphere-induced
VEV this gives%
\begin{eqnarray}
\left\langle \varphi ^{2}\right\rangle _{\mathrm{s}} &\approx &-\frac{%
2^{D-1}\Gamma ^{2}\left( \nu \right) e^{-\left( D-2\nu \right) t/\alpha }}{%
\pi \alpha ^{D-1}S_{D}\sinh ^{D-2}r}\sum_{l=0}^{\infty
}D_{l}\int_{0}^{\infty }dx\,\frac{xe^{-i\mu \pi }}{\sin \left( \pi x\right) }
\notag \\
&&\times \,\frac{\bar{Q}_{x-1/2}^{\mu }\left( u_{0}\right) }{\bar{P}%
_{x-1/2}^{-\mu }\left( u_{0}\right) }\frac{[P_{x-1/2}^{-\mu }\left( u\right)
]^{2}}{\Gamma \left( \nu -x+1/2\right) \Gamma \left( \nu +x+1/2\right) }.
\label{phi2slt}
\end{eqnarray}%
For $\nu =0$ by using the asymptotic expression of the function $%
P_{-1/2}^{x}\left( y\right) $ for large $y$ one gets%
\begin{equation}
P_{-1/2}^{x}(y)P_{-1/2}^{-x}(y)\approx \frac{2\cos \left( \pi x\right) }{\pi
^{2}y}\ln ^{2}y.  \label{PPnu0}
\end{equation}%
With this result from (\ref{phi2sin}) we obtain%
\begin{equation}
\left\langle \varphi ^{2}\right\rangle _{\mathrm{s}}\approx -\frac{%
2^{D+1}t^{2}e^{-Dt/\alpha }}{\pi ^{2}\sinh ^{D-2}r}\sum_{l=0}^{\infty }\frac{%
e^{-i\mu \pi }D_{l}}{S_{D}\alpha ^{D+1}}\int_{0}^{\infty }dx\,x\cot \left(
\pi x\right) \,\frac{\bar{Q}_{x-1/2}^{\mu }\left( u_{0}\right) }{\bar{P}%
_{x-1/2}^{-\mu }\left( u_{0}\right) }[P_{x-1/2}^{-\mu }\left( u\right) ]^{2},
\label{phi2sltnu0}
\end{equation}%
for $t/\alpha \gg 1$. In the same limit and for imaginary values of $\nu $
we need the asymptotic of the function $P_{\rho }^{\sigma }\left( y\right) $
for large values of $y$ and for $\mathrm{Re}\,\rho =-1/2$, $\mathrm{Im}%
\,\rho >0$. In \cite{Nist10} the asymptotics are given for $\mathrm{Re}%
\,\rho =-1/2$. In order to find the required estimate we use the asymptotic
formula%
\begin{equation}
Q_{\rho }^{x}(y)\approx \frac{\sqrt{\pi }e^{ix\pi }\Gamma (\rho +x+1)}{%
\Gamma (\rho +3/2)(2y)^{\rho +1}}.  \label{Qly}
\end{equation}%
The asymptotic expression for the function $P_{\rho }^{x}(y)$ is obtained by
using the formula that relates this function with the functions $Q_{\rho
}^{x}(y)$ and $Q_{-\rho -1}^{x}(y)$. In this way we can see that for large $y
$%
\begin{equation}
P_{\rho }^{\pm x}(y)\approx \frac{2}{\sqrt{\pi }}\mathrm{Re}\left[ \frac{%
\Gamma \left( \rho +1/2\right) \left( 2y\right) ^{\rho }}{\Gamma \left( \rho
\mp x+1\right) }\right] .  \label{Ply}
\end{equation}%
By using this result in (\ref{phi2sin}), for $t/\alpha \gg 1$ and $\nu
=i|\nu |$ one gets%
\begin{eqnarray}
\left\langle \varphi ^{2}\right\rangle _{\mathrm{s}} &\approx &-\frac{%
2^{D}\alpha ^{1-D}e^{-Dt/\alpha }}{\pi S_{D}\sinh ^{D-2}r}\sum_{l=0}^{\infty
}D_{l}\int_{0}^{\infty }dx\,\frac{xe^{-i\mu \pi }}{\sin \left( \pi x\right) }%
\,\frac{\bar{Q}_{x-1/2}^{\mu }\left( u_{0}\right) }{\bar{P}_{x-1/2}^{-\mu
}\left( u_{0}\right) }  \notag \\
&&\times \left[ P_{x-1/2}^{-\mu }\left( u\right) \right] ^{2}\left\{ \frac{%
\coth \left( \pi |\nu |\right) }{|\nu |}\cos \left( \pi x\right) +B_{\nu
}(x)\cos \left[ \phi (t,x)\right] \right\} ,  \label{phi2slt3}
\end{eqnarray}%
where we have introduced the notation%
\begin{equation}
\phi (t,x)=2|\nu |t/\alpha +\phi _{\nu }(x).  \label{phitx}
\end{equation}%
The functions $B_{\nu }(x)>0$ and $\phi _{\nu }(x)$ are defined by the
relation%
\begin{equation}
B_{\nu }(x)e^{i\phi _{\nu }(x)}=\frac{\Gamma ^{2}\left( \nu \right) }{\Gamma
\left( 1/2+x+\nu \right) \Gamma \left( 1/2-x+\nu \right) }.  \label{Bdef}
\end{equation}%
In this case one has an oscillatory damping behavior.

Now let us consider the asymptotic regions with respect to the radial
coordinate. For points near the sphere center one has $r\ll 1$ and the
argument of the function $P_{x-1/2}^{-\mu }\left( u\right) $ is close to 1.
By using the asymptotic relation (\ref{Pas1}) we can see that the
contribution of term with a given $l$ to the VEV (\ref{phi2sin}) is of the
order $r^{l}$. The dominant contribution comes from the $l=0$ mode with the
leading term%
\begin{equation}
\left\langle \varphi ^{2}\right\rangle _{\mathrm{s}}\approx -\frac{e^{-i\pi
\left( D/2-1\right) }\left( 2\alpha \right) ^{1-D}}{\pi ^{D/2}\Gamma \left(
D/2\right) \sinh ^{D-1}\left( t/\alpha \right) }\int_{0}^{\infty }dx\frac{x}{%
\sin \left( \pi x\right) }\,\frac{\bar{Q}_{x-1/2}^{D/2-1}\left( u_{0}\right) 
}{\bar{P}_{x-1/2}^{1-D/2}\left( u_{0}\right) }P_{\nu -1/2}^{x}\left(
y\right) P_{\nu -1/2}^{-x}\left( y\right) .  \label{phi2sc}
\end{equation}%
Note that in the special case $D=3$, for a non-Dirichlet boundary condition
(i.e., $B\neq 0$), this expression is further simplified as%
\begin{equation}
\left\langle \varphi ^{2}\right\rangle _{\mathrm{s}}\approx -\frac{\sinh
^{-2}\left( t/\alpha \right) }{2\pi \alpha ^{2}}\int_{0}^{\infty }dx\,\frac{%
x^{2}P_{\nu _{0}-1/2}^{x}\left( y\right) P_{\nu _{0}-1/2}^{-x}\left(
y\right) }{\sin \left( \pi x\right) \left( \frac{\beta -x+\coth r_{0}}{\beta
+x+\coth r_{0}}e^{2xr_{0}}-1\right) },  \label{phi2scD3}
\end{equation}%
where we have used the notation%
\begin{equation}
\nu _{0}=\sqrt{\frac{9}{4}-m^{2}\alpha ^{2}-12\xi }.  \label{nu0}
\end{equation}%
In the same spatial dimension and for the Dirichlet boundary condition near
the sphere center one gets%
\begin{equation}
\left\langle \varphi ^{2}\right\rangle _{\mathrm{s}}\approx -\frac{\sinh
^{-2}\left( t/\alpha \right) }{2\pi \alpha ^{2}}\int_{0}^{\infty }dx\frac{%
x^{2}P_{\nu _{0}-1/2}^{x}\left( y\right) P_{\nu _{0}-1/2}^{-x}\left(
y\right) }{\sin \left( \pi x\right) \left( e^{2xr_{0}}-1\right) }\,.
\label{phi2scD3D}
\end{equation}

The boundary-induced contribution (\ref{phi2sin}) diverges on the sphere.
For points near the sphere the dominant contribution to the integral comes
from large values of $x$. By using the asymptotic expressions for the
functions $P_{\nu -1/2}^{\pm x}\left( y\right) $ \cite{Nist10} it can be seen%
\begin{equation}
P_{\nu -1/2}^{x}\left( y\right) P_{\nu -1/2}^{-x}\left( y\right) \approx 
\frac{\sin \left( \pi x\right) }{\pi x},\;x\gg 1.  \label{AsPP}
\end{equation}%
Substituting this into (\ref{phi2sin}), to the leading order we get%
\begin{equation}
\left\langle \varphi ^{2}\right\rangle _{\mathrm{s}}\approx \frac{%
\left\langle \varphi ^{2}\right\rangle _{\mathrm{s}}^{\mathrm{(st)}}}{\sinh
^{D-1}\left( t/\alpha \right) },  \label{phi2near}
\end{equation}%
where $\left\langle \varphi ^{2}\right\rangle _{\mathrm{s}}^{\mathrm{(st)}}$
is given by (\ref{phi2st}). Taking the near-sphere asymptotic for $%
\left\langle \varphi ^{2}\right\rangle _{\mathrm{s}}^{\mathrm{(st)}}$ from 
\cite{Saha14}, we find the leading order term in the corresponding
asymptotic expansion for (\ref{phi2sin}):%
\begin{equation}
\left\langle \varphi ^{2}\right\rangle _{\mathrm{s}}\approx \frac{\left(
1-2\delta _{0B}\right) \Gamma \left( \left( D-1\right) /2\right) }{\left(
4\pi \right) ^{\left( D+1\right) /2}\left[ \alpha \sinh \left( t/\alpha
\right) \left( r_{0}-r\right) \right] ^{D-1}}.  \label{phi2near2}
\end{equation}%
Note that $\alpha \sinh \left( t/\alpha \right) \left( r_{0}-r\right) $ is
the proper distance from the sphere. The leading term (\ref{phi2near2})
coincides with the corresponding term for a sphere in Minkowski spacetime
with the distance from the sphere replaced by the proper distance. As seen
from (\ref{phi2near}), near the sphere the boundary-induced VEV is negative
for Dirichlet boundary condition and positive for non-Dirichlet boundary
conditions. From the problem symmetry we expect the renormalized VEV $%
\left\langle \varphi ^{2}\right\rangle _{0}$ for the boundary-free geometry
will depend on the time coordinate only and near the sphere the total VEV is
dominated by the sphere-induced part.

\subsection{Exterior region}

Similar to the interior region, the VEV of the field squared outside the
sphere is decomposed into the boundary-free and sphere-induced contributions
(see (\ref{phi2i})). For points $r>r_{0}$ the latter is directly obtained
from (\ref{Wse2}) in the coincidence limit $x^{\prime }\rightarrow x$ and is
expressed as%
\begin{eqnarray}
\left\langle \varphi ^{2}\right\rangle _{\mathrm{s}} &=&-\frac{\alpha
^{1-D}\sinh ^{2-D}r}{S_{D}\sinh ^{D-1}\left( t/\alpha \right) }%
\sum_{l=0}^{\infty }D_{l}\int_{0}^{\infty }dx\,\frac{xe^{-i\mu \pi }}{\sin
\left( \pi x\right) }  \notag \\
&&\times \,\frac{\bar{P}_{x-1/2}^{-\mu }\left( u_{0}\right) }{\bar{Q}%
_{x-1/2}^{\mu }\left( u_{0}\right) }P_{\nu -1/2}^{x}\left( y\right) P_{\nu
-1/2}^{-x}\left( y\right) [Q_{x-1/2}^{\mu }\left( u\right) ]^{2}.
\label{phi2se}
\end{eqnarray}%
For a conformally coupled massless field this VEV is related to the
corresponding VEV outside a spherical boundary in static spacetime with a
negative constant curvature space by the formula (\ref{phi2sconf}), where
the expression for $\left\langle \varphi ^{2}\right\rangle _{\mathrm{s}}^{%
\mathrm{(st)}}$ is obtained from (\ref{phi2st}) by the replacements (\ref%
{ReplPQ}). The VEV outside a sphere in the Milne universe is obtained from (%
\ref{phi2se}) in the limit $\alpha \rightarrow \infty $. The latter limit is
reduced to the replacements (\ref{PPJJp}) and $\alpha \sinh \left( t/\alpha
\right) \rightarrow t$.

At early stages of the expansion, corresponding to $t/\alpha \ll 1$, the
leading order term in the expansion of (\ref{phi2se}) coincides with the
boundary-induced VEV for a massless scalar field in the conformal vacuum
outside the sphere in the Milne universe and the influence of the
gravitational field is weak. The corresponding expression is given by the
right-hand side of (\ref{phi2smt}) with the replacements (\ref{ReplPQ}). The
effect of gravity is essential at late stages of the expansion,
corresponding to $t/\alpha \gg 1$. The time dependence in the sphere-induced
VEVs for the interior and exterior regions appears through the same
functions and the investigation of the behavior of the VEV (\ref{phi2se}) is
similar to that presented in the previous subsection for the region inside
the sphere. The asymptotic behavior for $\left\langle \varphi
^{2}\right\rangle _{\mathrm{s}}$ in the cases $\nu >0$, $\nu =0$ and $\nu
=i|\nu |$ is given by the formulas (\ref{phi2slt}), (\ref{phi2sltnu0}) and (%
\ref{phi2slt3}), respectively, with the replacements (\ref{ReplPQ}). Note
that in the last case the fall-off of the sphere-induced VEV, as a function
of $t/\alpha $, is damped oscillatory.

The leading term in the asymptotic expansion of (\ref{phi2se}) with respect
to the distance from the sphere is given by (\ref{phi2near2}) with the
replacement $r_{0}-r\rightarrow r-r_{0}$. In the opposite limit of large
distances from the sphere, $r\gg 1$, we use the asymptotic 
\begin{equation}
Q_{x-1/2}^{\mu }\left( \cosh r\right) \approx \frac{\sqrt{\pi }e^{i\mu \pi
}\Gamma \left( x+\mu +1/2\right) }{\Gamma \left( x+1\right) e^{\left(
x+1/2\right) r}},  \label{Qasl}
\end{equation}%
for the associated Legendre function. With this asymptotic, the integral in (%
\ref{phi2se}) is dominated by the contribution coming from the region near
the lower limit and to the leading order we get 
\begin{equation}
\left\langle \varphi ^{2}\right\rangle _{\mathrm{s}}\approx -\frac{%
2^{D-3}\alpha ^{1-D}[P_{\nu -1/2}^{0}\left( y\right) ]^{2}}{S_{D}\sinh
^{D-1}\left( t/\alpha \right) re^{(D-1)r}}\sum_{l=0}^{\infty }\frac{D_{l}%
\bar{P}_{-1/2}^{-\mu }\left( u_{0}\right) }{e^{-i\mu \pi }\bar{Q}%
_{-1/2}^{\mu }\left( u_{0}\right) }\Gamma ^{2}(\mu +1/2).  \label{phi2selr}
\end{equation}%
Thus, for large values of $r$ the sphere-induced VEV is exponentially small.
Note that for large $r$ and for a fixed $t$, the geodesic distance from the
sphere is proportional to $\alpha r$ (see (\ref{dxx1})). The exponential
suppression takes place for both massive and massless fields. Note that for
a spherical boundary in flat spacetime the decay of the boundary-induced
VEVs at large distances from the sphere is power-law for a massless field.
At large distances, the total renormalized VEV $\left\langle \varphi
^{2}\right\rangle $ is dominated by the boundary-free contribution $%
\left\langle \varphi ^{2}\right\rangle _{0}$. By using the relations (\ref%
{rInf}), we can see that in the limit $r\rightarrow \infty $, for fixed $t$,
one has $r_{\mathrm{I}}\rightarrow \alpha $ and this corresponds to the
near-horizon limit for an observer located at $r=0$.

It is of interest to compare the sphere-induced VEVs in the problem under
consideration with the VEVs for a sphere having constant radius $r_{\mathrm{I%
}}=r_{\mathrm{I}}^{(0)}$ in inflationary coordinates (see Appendix \ref%
{sec:ApCoord}). The corresponding problem for the Bunch-Davies vacuum state
has been considered in \cite{Milt12}. Due to the maximal symmetry of the
Bunch-Davies vacuum, the VEVs in the latter problem depend on the sphere
radius and on the time and radial coordinates in the form of the ratios $r_{%
\mathrm{I}}^{(0)}/\eta _{\mathrm{I}}$ and $r_{\mathrm{I}}/\eta _{\mathrm{I}}$%
, where $\eta _{\mathrm{I}}=-\alpha e^{-t_{\mathrm{I}}/\alpha }$ is the
corresponding conformal time coordinate. Note that $r_{\mathrm{I}}/|\eta _{%
\mathrm{I}}|$ is the proper distance from the sphere, in units of the
curvature radius, measured by an observer with fixed $r_{\mathrm{I}}$. The
hyperbolic vacuum is not maximally symmetric and the mentioned feature does
not take place for the VEVs in the problem under consideration. An essential
difference between two problems is seen also in the behavior of the VEVs at
large distances from the sphere. For the problem in \cite{Milt12}, at large
distances the sphere-induced VEV of the field squared behaves as $(r_{%
\mathrm{I}}/|\eta _{\mathrm{I}}|)^{2-2\nu -2D}$ for real $\nu $ and like $%
(r_{\mathrm{I}}/|\eta _{\mathrm{I}}|)^{2-2D}\cos [2|\nu |\ln (r_{\mathrm{I}%
}/|\eta _{\mathrm{I}}|)+\phi _{0}]$ for imaginary $\nu $. In the second case
the decay of the VEV, as a function of the radial coordinate, is damping
oscillatory. In the problem we consider here the decay of the VEV is always
monotonic, as $e^{(1-D)r}/r$.

\subsection{Numerical analysis}

\begin{figure}[tbph]
\begin{center}
\begin{tabular}{cc}
\epsfig{figure=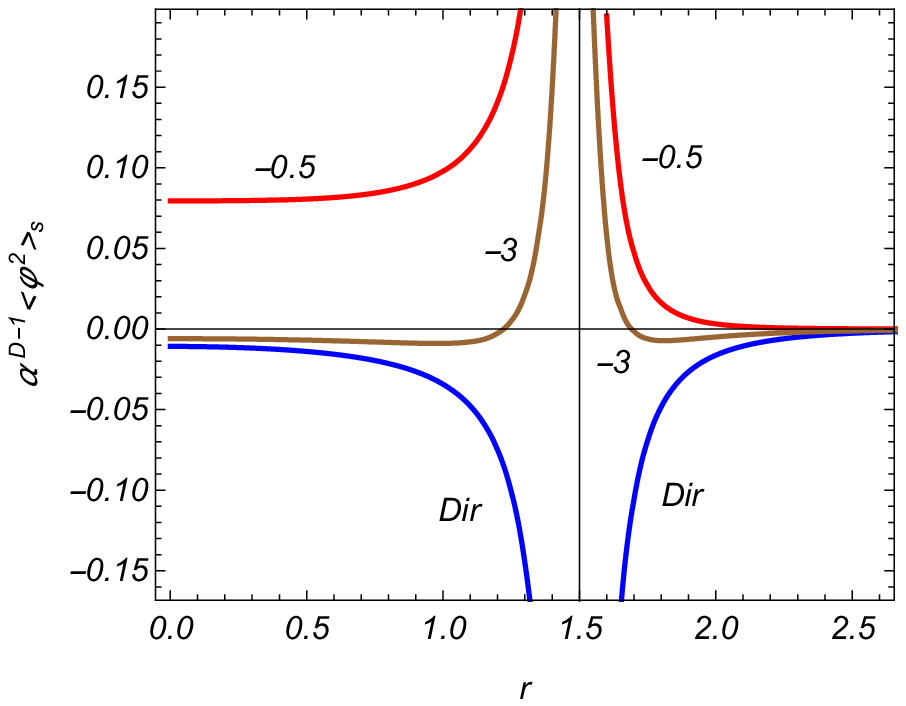,width=7.5cm,height=6.5cm} & \quad %
\epsfig{figure=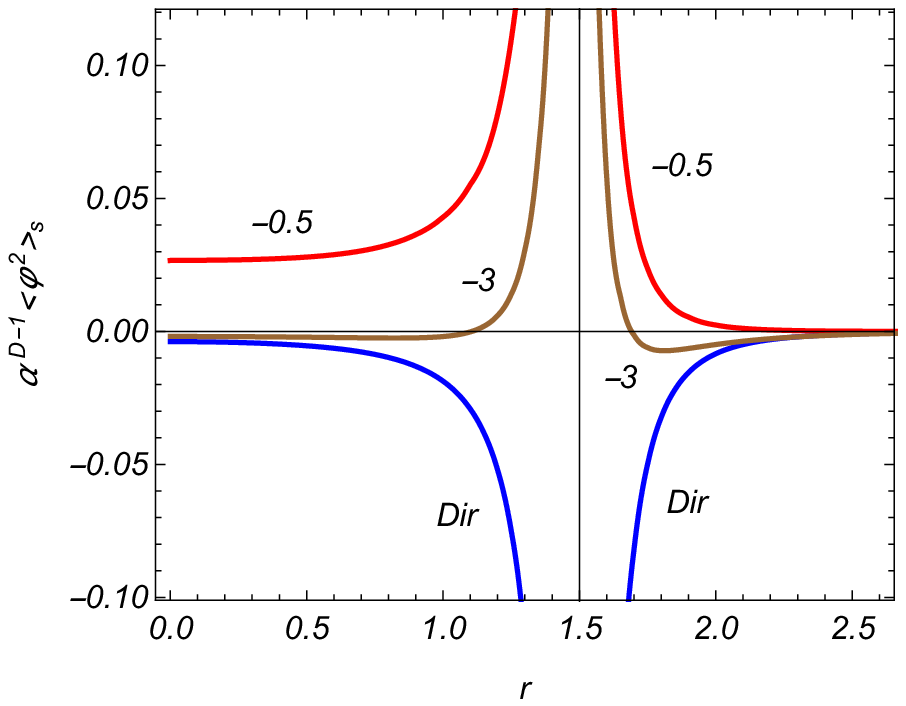,width=7.5cm,height=6.5cm}%
\end{tabular}%
\end{center}
\caption{The sphere-induced VEV of the field squared for $D=3$ scalar field
as a function of the radial coordinate in the cases of minimally (left
panel) and conformally (right panel) coupled fields for Dirichlet boundary
condition and for Robin conditions with $\protect\beta =-3,-0.5$. The graphs
are plotted for $r_{0}=1.5$, $m\protect\alpha =1$, $t/\protect\alpha =1$.}
\label{fig2}
\end{figure}

In the discussion below the numerical results will be presented for the most
important special cases of $D=3$ minimally and conformally coupled fields.
In Figure \ref{fig2}, we have plotted the sphere-induced contributions in
the VEV of the field squared inside and outside a spherical shell versus the
radial coordinate $r$ for Dirichlet boundary condition (curve Dir) and for
Robin boundary conditions with $\beta =-3,-0.5$ (the numbers near the
curves). The graphs are plotted for $r_{0}=1.5$, $m\alpha =1$, $t/\alpha =1$%
. The left and right panels correspond to minimal and conformal couplings,
respectively. In accordance with the asymptotic analysis given above, near
the sphere the boundary-induced VEV of the field squared behaves as $%
(r-r_{0})^{-2}$. It is negative for Dirichlet boundary condition and
positive for non-Dirichlet boundary conditions. At large distances from the
sphere, the VEV $\left\langle \varphi ^{2}\right\rangle _{\mathrm{s}}$ is
suppressed by the factor $e^{-2r}$ and is negative for all graphs in Figure %
\ref{fig2}. Near the sphere center the leading terms in the asymptotic are
given by (\ref{phi2scD3}) and (\ref{phi2scD3D}).

\begin{figure}[tbph]
\begin{center}
\begin{tabular}{cc}
\epsfig{figure=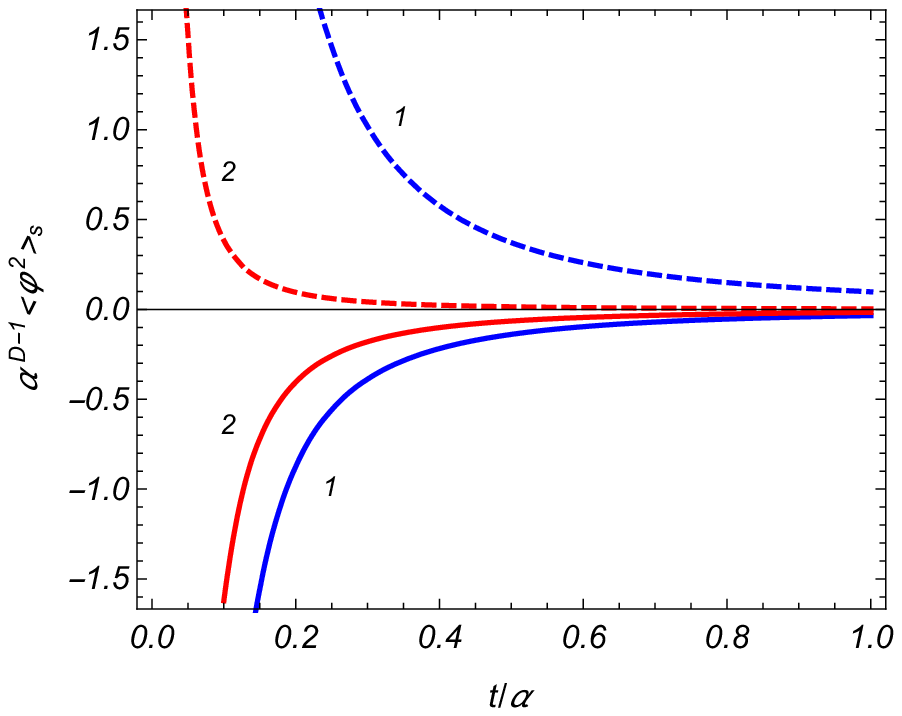,width=7.5cm,height=6.cm} & \quad %
\epsfig{figure=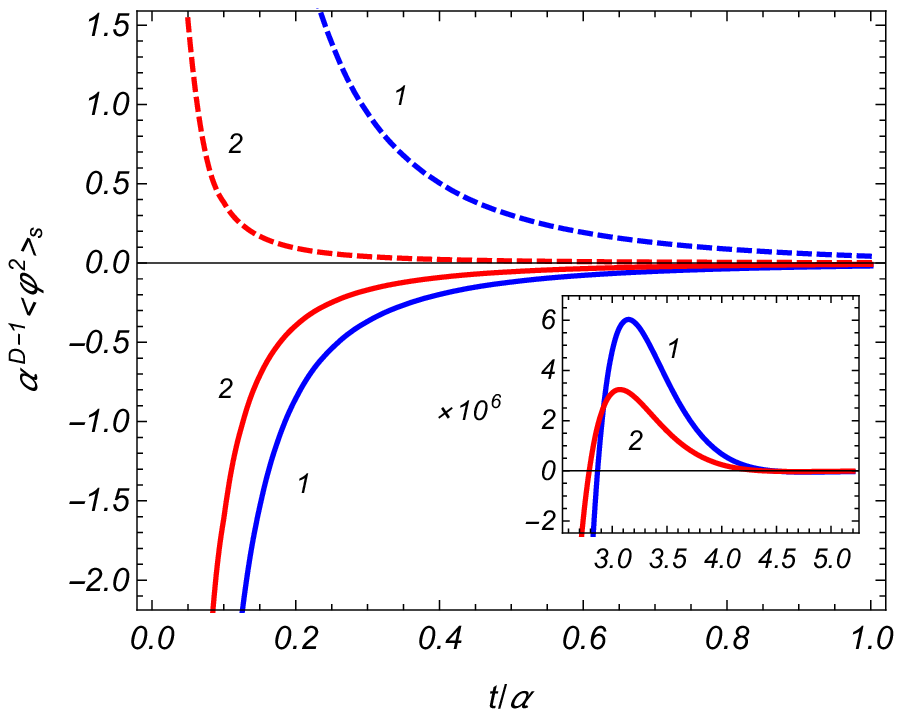,width=7.5cm,height=6.cm}%
\end{tabular}%
\end{center}
\caption{The sphere-induced VEV of the field squared for $D=3$ scalar field
as a function of the time coordinate in the cases of minimal (left panel)
and conformal (right panel) couplings. The graphs are plotted for $r_{0}=1.5$%
, $m\protect\alpha =1$ and the numbers near the curves correspond to the
values of the radial coordinate. The full and dashed curves present the
cases of Dirichlet and Robin (with $\protect\beta =-0.5$) boundary
conditions, respectively.}
\label{fig3}
\end{figure}

Figure \ref{fig3} displays the time-dependence of the sphere-induced
contribution in the VEV of the field squared for fixed $r$ (numbers near the
curves) and for $r_{0}=1.5$, $m\alpha =1$. The full curves correspond to
Dirichlet boundary condition and the dashed curves correspond to Robin
boundary condition with $\beta =-0.5$. The graphs for minimal and conformal
couplings are presented on the left and right panels, respectively.
According to the asymptotic analysis given above, for $D=3$ and in the
region $t/\alpha \ll 1$ the boundary-induced VEV of the field squared
behaves as $t^{-2}$. In the opposite limit $t/\alpha \gg 1$ the
corresponding approximation for minimal coupling (left panel) is obtained
from (\ref{phi2slt}), according to which $\left\langle \varphi
^{2}\right\rangle _{\mathrm{s}}$ behaves as nearly $e^{-(3-\sqrt{5})t/\alpha
}$. Contrary to this, in the case of conformal coupling (right panel) the
parameter $\nu $ is purely imaginary, $\nu =i\sqrt{3}/2$, and the late time
asymptotic is found from (\ref{phi2slt3}). In this case the field squared
decays oscillatory and this behavior is displayed on the right panel as
inset for $10^{6}\alpha ^{D-1}\left\langle \varphi ^{2}\right\rangle _{%
\mathrm{s}}$. 

\begin{figure}[tbph]
\begin{center}
\begin{tabular}{cc}
\epsfig{figure=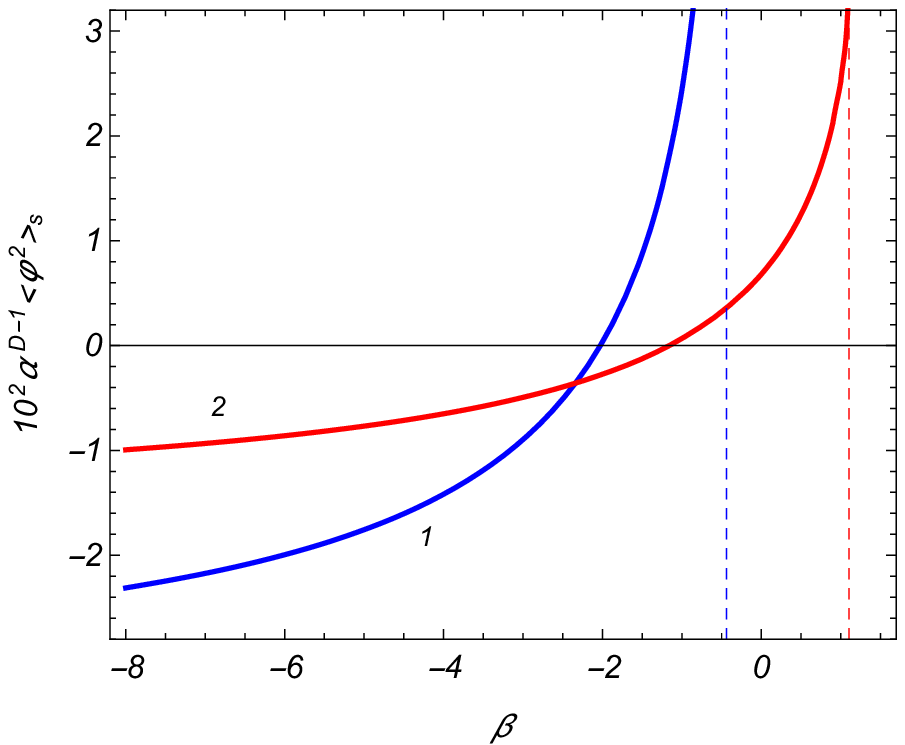,width=7.5cm,height=6.cm} & \quad %
\epsfig{figure=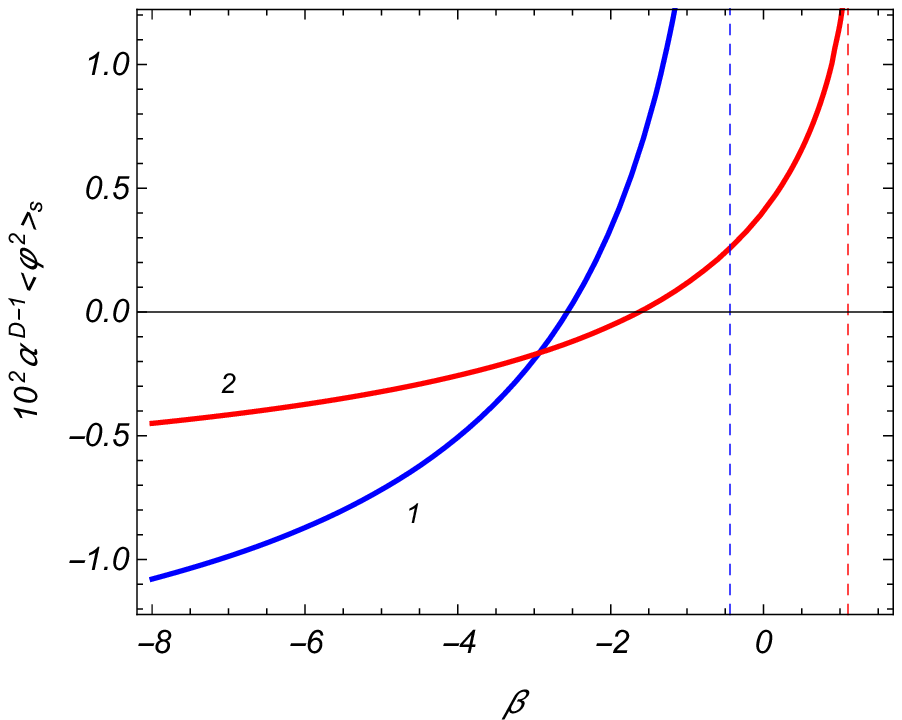,width=7.5cm,height=6.cm}%
\end{tabular}%
\end{center}
\caption{The sphere-induced contribution in the VEV of the field squared for 
$D=3$ scalar field versus the Robin coefficient in the cases of minimal
(left panel) and conformal (right panel) couplings. The graphs are plotted
for $m\protect\alpha =t/\protect\alpha =1$, $r_{0}=1.5$ and the numbers near
the curves are the values of the coordinate $r$.}
\label{fig4}
\end{figure}

The dependence of the sphere-induced VEV on the coefficient $\beta $ in
Robin boundary condition is displayed in Figure \ref{fig4} for minimally
(left panel) and conformally (right panel) coupled fields. The graphs are
plotted for $D=3$, $m\alpha =t/\alpha =1$, $r_{0}=1.5$ and the numbers near
the curves correspond to the values of the coordinate $r$. The vertical
dashed lines correspond to the critical values $\beta _{0}^{\mathrm{(i)}%
}(u_{0})$ and $\beta _{0}^{\mathrm{(e)}}(u_{0})$ for the Robin coefficient.
As seen, depending on the values of the Robin coefficient, the
boundary-induced VEV changes the sign. For $\beta \ll -1$ ($\beta =-\infty $
corresponds to Dirichlet boundary condition) $\left\langle \varphi
^{2}\right\rangle _{\mathrm{s}}$ is negative and it becomes positive with
increasing $\beta $. The VEV increases rapidly when $\beta $ approaches the
critical values $\beta _{0}^{(\mathrm{i})}(u_{0})$ and $\beta _{0}^{(\mathrm{%
e})}(u_{0})$ for the interior and exterior regions, respectively. For $\beta 
$ near the critical values of $\beta $ the main contribution to the VEV $%
\left\langle \varphi ^{2}\right\rangle _{\mathrm{s}}$ comes from the mode $%
l=0$. 

\section{VEV of the energy-momentum tensor}

\label{sec:EMT}

For the evaluation of the VEV of the energy-momentum tensor we use the
formula (\ref{TikVev}). On the base of (\ref{WFidec}) the VEV is decomposed
as%
\begin{equation}
\left\langle T_{ik}\right\rangle =\left\langle T_{ik}\right\rangle
_{0}+\left\langle T_{ik}\right\rangle _{\mathrm{s}},  \label{Tikdec}
\end{equation}%
with the boundary-free and sphere-induced contributions $\left\langle
T_{ik}\right\rangle _{0}$ and $\left\langle T_{ik}\right\rangle _{\mathrm{s}}
$. For points away from the sphere the sphere-induced part is finite and is
obtained directly by using the formula that is the analog of (\ref{TikVev})
for sphere-induced contributions. From the symmetry of the problem we expect
that the angular stresses are isotropic:%
\begin{equation}
\left\langle T_{2}^{2}\right\rangle _{\mathrm{s}}=\left\langle
T_{3}^{3}\right\rangle _{\mathrm{s}}=\cdots =\left\langle
T_{D}^{D}\right\rangle _{\mathrm{s}}.  \label{Tang}
\end{equation}%
In addition, we have the trace relation%
\begin{equation}
\left\langle T_{k}^{k}\right\rangle _{\mathrm{s}}=\left[ D\left( \xi -\xi
_{D}\right) \nabla _{k}\nabla ^{k}+m^{2}\right] \left\langle \varphi
^{2}\right\rangle _{\mathrm{s}}.  \label{Trel}
\end{equation}%
In the case of a conformally coupled massless field the sphere-induced
energy-momentum tensor is traceless. The trace anomaly is contained in the
boundary-free part $\left\langle T_{ik}\right\rangle _{0}$.

From the symmetry of the problem we expect that the renormalized VEV $%
\left\langle T_{ik}\right\rangle _{0}$ is diagonal with isotropic stresses, $%
\left\langle T_{1}^{1}\right\rangle _{0}=\left\langle T_{2}^{2}\right\rangle
_{0}=\cdots =\left\langle T_{D}^{D}\right\rangle _{0}$, and the components $%
\left\langle T_{i}^{k}\right\rangle _{0}$ are functions of the time
coordinate only. The continuity equation $\nabla _{k}\left\langle
T_{i}^{k}\right\rangle _{0}=0$ leads to the relation%
\begin{equation}
\left\langle T_{1}^{1}\right\rangle _{0}=\frac{\partial _{t/\alpha }\left[
\sinh ^{D}(t/\alpha )\left\langle T_{0}^{0}\right\rangle _{0}\right] }{%
D\sinh ^{D-1}(t/\alpha )\cosh \left( t/\alpha \right) },  \label{Tik0}
\end{equation}%
between the energy density and stress. For the Bunch-Davies vacuum state one
has $\left\langle T_{i}^{k}\right\rangle _{0}^{\mathrm{(BD)}}=\mathrm{const}%
\cdot \delta _{i}^{k}$. The divergences are determined by the local geometry
of the background spacetime and are the same in the unrenormalized VEVs $%
\left\langle T_{i}^{k}\right\rangle _{0}$ and $\left\langle
T_{i}^{k}\right\rangle _{0}^{\mathrm{(BD)}}$ for the hyperbolic and
Bunch-Davies vacua. From here it follows that the difference $\Delta
\left\langle T_{i}^{k}\right\rangle _{0}=\left\langle T_{i}^{k}\right\rangle
_{0}-\left\langle T_{i}^{k}\right\rangle _{0}^{\mathrm{(BD)}}$ needs no
renormalization and can be directly evaluated by applying the procedure
similar to (\ref{TikVev}) for $G(x,x^{\prime })-G_{\mathrm{BD}}(x,x^{\prime
})$, where $G_{\mathrm{BD}}(x,x^{\prime })$ is the Hadamard function for the
Bunch-Davies vacuum. In this way, the renormalization of the VEV $%
\left\langle T_{i}^{k}\right\rangle _{0}$ is reduced to the one for the
Bunch-Davies vacuum. The latter procedure has been widely discussed in the
literature. For a conformally coupled massless field the tensor $\Delta
\left\langle T_{i}^{k}\right\rangle _{0}$ is traceless and from (\ref{Tik0})
it follows that%
\begin{equation}
\Delta \left\langle T_{i}^{k}\right\rangle _{0}=\mathrm{const}\frac{\mathrm{%
diag}\left( 1,-1/D,\cdots ,-1/D\right) }{\alpha ^{D+1}\sinh ^{D+1}(t/\alpha )%
}.  \label{DelTik0}
\end{equation}%
A special case of (\ref{DelTik0}) for $D=3$, with $\mathrm{const}=-1/(480\pi
^{2})$, is considered in \cite{Pfau82}. Here, we are mainly interested in
the sphere-induced effects and they will be discussed for the interior and
exterior regions separately.

\subsection{Interior region}

By making use of the expression (\ref{Wsi}) for the sphere-induced Hadamard
function in (\ref{TikVev}), after long but straightforward calculations, for
the diagonal components of the sphere-induced vacuum energy-momentum tensor
in the interior region one finds (no summation over $k$)%
\begin{equation}
\left\langle T_{k}^{k}\right\rangle _{\mathrm{s}}=-\frac{\sinh ^{-2}\left(
t/\alpha \right) }{\alpha ^{D+1}S_{D}}\sum_{l=0}^{\infty
}D_{l}\int_{0}^{\infty }dx\frac{xe^{-i\mu \pi }}{\sin \left( \pi x\right) }\,%
\frac{\bar{Q}_{x-1/2}^{\mu }\left( u_{0}\right) }{\bar{P}_{x-1/2}^{-\mu
}\left( u_{0}\right) }\left[ \hat{F}_{k}^{(0)}(y)-\hat{F}_{k}^{(1)}(u)\right]
F^{\mathrm{(i)}}\left( x,y,u\right) ,  \label{Tkki}
\end{equation}%
where the notation%
\begin{equation}
F^{\mathrm{(i)}}\left( x,y,u\right) =\frac{P_{\nu -1/2}^{x}\left( y\right)
P_{\nu -1/2}^{-x}\left( y\right) }{\left( y^{2}-1\right) ^{(D-1)/2}}\frac{%
[P_{x-1/2}^{-\mu }\left( u\right) ]^{2}}{(u^{2}-1)^{D/2-1}}  \label{Fi}
\end{equation}%
is introduced. The operators $\hat{F}_{0}^{(0)}(y)$ and $\hat{F}_{k}^{(0)}(y)
$, $k=1,2,\ldots D$, are defined as 
\begin{eqnarray}
\hat{F}_{0}^{(0)}(y) &=&\left( y^{2}-1\right) \left\{ \frac{1}{4}\left(
y^{2}-1\right) \partial _{y}^{2}+\left[ D\left( \xi +\xi _{D}\right) +\frac{1%
}{2}\right] y\partial _{y}+m^{2}\alpha ^{2}+\xi D^{2}\right\}   \notag \\
&&+\frac{\left( D-1\right) ^{2}}{4}-x^{2},  \notag \\
\hat{F}_{k}^{(0)}(y) &=&\left( y^{2}-1\right) \left\{ \left( \xi -\frac{1}{4}%
\right) \left( y^{2}-1\right) \partial _{y}^{2}+\left[ D\left( \xi -\xi
_{D}\right) -\frac{1}{2}\right] y\partial _{y}-\xi D\right\}   \notag \\
&&+\delta _{1k}\left[ x^{2}-\frac{\left( D-1\right) ^{2}}{4}\right] .
\label{Fk0}
\end{eqnarray}%
The operators $\hat{F}_{k}^{(1)}(u)$ act on the functions of the argument $u$
and are given by the expressions 
\begin{eqnarray}
\hat{F}_{0}^{(1)}(u) &=&\left( \xi -\frac{1}{4}\right) \left[ \left(
u^{2}-1\right) \partial _{u}^{2}+Du\partial _{u}\right] ,  \notag \\
\hat{F}_{1}^{(1)}(u) &=&\frac{1}{4}\left( u^{2}-1\right) \partial _{u}^{2}+%
\left[ \xi \left( D-1\right) +\frac{D}{4}\right] u\partial _{u}-\frac{%
l\left( l+n\right) }{u^{2}-1},  \notag \\
\hat{F}_{k}^{(1)}(u) &=&\hat{F}_{0}^{(1)}-\xi u\partial _{u}+\frac{1}{D-1}%
\frac{l\left( l+n\right) }{u^{2}-1},  \label{Fk1}
\end{eqnarray}%
where $k=2,3,\ldots D$. As an additional check for the formula (\ref{Tkki})
we can see that the trace relation (\ref{Trel}) is obeyed.

The only nonzero off-diagonal component of the vacuum energy-momentum tensor
corresponds to $\left\langle T_{0}^{1}\right\rangle _{\mathrm{s}}$ that
describes energy flux directed along the radial direction. The corresponding
expression reads 
\begin{eqnarray}
\left\langle T_{0}^{1}\right\rangle _{\mathrm{s}} &=&\frac{\sinh ^{-3}\left(
t/\alpha \right) }{\alpha ^{D+2}S_{D}}\sum_{l=0}^{\infty
}D_{l}\int_{0}^{\infty }dx\frac{xe^{-i\mu \pi }}{\sin \left( \pi x\right) }%
\frac{\bar{Q}_{x-1/2}^{\mu }\left( u_{0}\right) }{\bar{P}_{x-1/2}^{-\mu
}\left( u_{0}\right) }  \notag \\
&&\times \,\left[ \left( 1/4-\xi \right) \left( y^{2}-1\right) \partial
_{y}+\xi y\right] \partial _{r}F^{\mathrm{(i)}}\left( x,y,u\right) .
\label{T10}
\end{eqnarray}%
With this result, it can be seen that the components given by (\ref{Tkki})
and (\ref{T10}) obey the covariant conservation equation $\nabla
_{k}\left\langle T_{i}^{k}\right\rangle _{\mathrm{s}}=0$. For the geometry
described by (\ref{LE}) the latter is reduced to the equations%
\begin{eqnarray}
\sum_{k=0,1}\partial _{k}\left\langle T_{0}^{k}\right\rangle _{\mathrm{s}%
}+\left( D-1\right) \left\langle T_{0}^{1}\right\rangle _{\mathrm{s}}\coth r+%
\frac{1}{\alpha }\left[ \left( D+1\right) \left\langle
T_{0}^{0}\right\rangle _{\mathrm{s}}-\left\langle T_{k}^{k}\right\rangle _{%
\mathrm{s}}\right] \coth \left( t/\alpha \right) &=&0,  \notag \\
\sum_{k=0,1}\partial _{k}\left\langle T_{1}^{k}\right\rangle _{\mathrm{s}}+%
\frac{D}{\alpha }\left\langle T_{1}^{0}\right\rangle _{\mathrm{s}}\coth
\left( t/\alpha \right) +\left( D-1\right) \left( \left\langle
T_{1}^{1}\right\rangle _{\mathrm{s}}-\left\langle T_{2}^{2}\right\rangle _{%
\mathrm{s}}\right) \coth r &=&0.  \label{Ceq}
\end{eqnarray}

The vacuum energy induced by the sphere in volume $V$ is expressed as $E_{%
\mathrm{(s)}V}=\int_{V}d^{D}x\,\sqrt{|g|}\left\langle T_{0}^{0}\right\rangle
_{\mathrm{s}}$. For the spherical layer $r_{1}\leq r\leq r_{2}$ it is
presented in the form 
\begin{equation*}
E_{\mathrm{(s)}V}=\alpha \sinh (t/\alpha )\int_{r_{1}}^{r_{2}}dr\,S_{\mathrm{%
p}}(r)\left\langle T_{0}^{0}\right\rangle _{\mathrm{s}}.
\end{equation*}%
where $S_{\mathrm{p}}(r)=S_{D}\left[ \alpha \sinh \left( t/\alpha \right)
\sinh r\right] ^{D-1}$ is the proper surface area of the sphere with radius $%
r$. From the first equation (\ref{Ceq}) it follows that%
\begin{equation}
\partial _{0}E_{\mathrm{(s)}V}=-\alpha \sinh \left( t/\alpha \right) S_{%
\mathrm{p}}(r)\left\langle T_{0}^{1}\right\rangle _{\mathrm{s}%
}|_{r=r_{1}}^{r=r_{2}}+\cosh \left( t/\alpha \right)
\int_{r_{1}}^{r_{2}}dr\,S_{\mathrm{p}}(r)\sum_{k=1}^{D}\left\langle
T_{k}^{k}\right\rangle _{\mathrm{s}}.  \label{Evar}
\end{equation}%
This relation shows that the quantity 
\begin{equation}
\left\langle \tilde{T}_{0}^{1}\right\rangle _{\mathrm{s}}=\alpha \sinh
\left( t/\alpha \right) \left\langle T_{0}^{1}\right\rangle _{\mathrm{s}}
\label{Flux}
\end{equation}%
is the energy flux density per unit proper surface area. The latter can be
written as $\left\langle \tilde{T}_{0}^{1}\right\rangle _{\mathrm{s}%
}=n_{k}\left\langle T_{0}^{k}\right\rangle _{\mathrm{s}}$, where $n_{k}$ is
the unit spatial vector normal to the sphere (external with respect to the
volume $V$). For the spherical layer corresponding to (\ref{Evar}) one has $%
n_{k}=\pm \delta _{k}^{1}\alpha \sinh \left( t/\alpha \right) $, where the
upper and lower signs stand for the spheres $r=r_{2}$ and $r=r_{1}$,
respectively.

Let us consider some limiting cases of the general result for the
sphere-induced VEV of the energy-momentum tensor. In the flat spacetime
limit $\alpha \rightarrow \infty $, by using the relation (\ref{PPJJp}) for
the product of the associated Legendre functions, it can be seen that from (%
\ref{Tkki}) and (\ref{T10}) the boundary-induced VEV for the conformal
vacuum inside a sphere in background of the Milne universe is obtained (see 
\cite{Saha20}).

Another special case corresponds to a conformally coupled massless scalar
field. In this case one has $\nu =1/2$ and the function (\ref{Fi}) is
simplified to 
\begin{equation}
F^{\mathrm{(i)}}\left( x,y,u\right) =\frac{\sin \left( \pi x\right) /(\pi x)%
}{(y^{2}-1)^{(D-1)/2}}\frac{[P_{x-1/2}^{-\mu }\left( u\right) ]^{2}}{%
(u^{2}-1)^{D/2-1}}.  \label{Ficc}
\end{equation}%
With this function, the off-diagonal component (\ref{T10}) vanishes for $\xi
=\xi _{D}$. For the diagonal components we find (no summation over $k$)%
\begin{equation}
\left\langle T_{k}^{k}\right\rangle _{\mathrm{s}}=\frac{\left\langle
T_{k}^{k}\right\rangle _{\mathrm{s}}^{\mathrm{(st)}}}{\sinh ^{D+1}(t/\alpha )%
},  \label{Tikc}
\end{equation}%
where 
\begin{equation}
\left\langle T_{k}^{k}\right\rangle _{\mathrm{s}}^{\mathrm{(st)}%
}=\sum_{l=0}^{\infty }\frac{e^{-i\mu \pi }D_{l}}{\pi \alpha ^{D+1}S_{D}}%
\int_{0}^{\infty }dx\,\frac{\bar{Q}_{x-1/2}^{\mu }\left( u_{0}\right) }{\bar{%
P}_{x-1/2}^{-\mu }\left( u_{0}\right) }\hat{F}_{k}^{\mathrm{(st)}}(u)\frac{%
[P_{x-1/2}^{-\mu }\left( u\right) ]^{2}}{(u^{2}-1)^{D/2-1}}.  \label{Tikstc}
\end{equation}%
with the operators%
\begin{eqnarray}
\hat{F}_{0}^{\mathrm{(st)}}(u) &=&-\frac{u^{2}-1}{4D}\partial _{u}^{2}-\frac{%
u}{4}\partial _{u}+x^{2},  \notag \\
\hat{F}_{1}^{\mathrm{(st)}}(u) &=&\frac{u^{2}-1}{4}\partial _{u}^{2}+\frac{%
D^{2}-D+1/2}{2D}u\partial _{u}+\frac{\left( D-1\right) ^{3}}{4D}-\frac{%
l\left( l+n\right) }{u^{2}-1}-x^{2},  \notag \\
\hat{F}_{2}^{\mathrm{(st)}}(u) &=&-\frac{u^{2}-1}{4D}\partial _{u}^{2}-\frac{%
2D-1}{4D}u\partial _{u}-\frac{\left( D-1\right) ^{2}}{4D}+\frac{l}{D-1}\frac{%
l+n}{u^{2}-1}.  \label{F2stc}
\end{eqnarray}%
The quantity (\ref{Tikstc}) is the boundary-induced VEV of the
energy-momentum tensor for a conformally coupled massless scalar field
inside a sphere with radius $r_{0}$ in background of a static negative
constant curvature space with the curvature radius $\alpha $. It is obtained
from the more general result in \cite{Saha14} in the special case $m=0$ and $%
\xi =\xi _{D}$ (see also \cite{Saha20}).

Let us consider the asymptotics with respect to the ratio $t/\alpha $. At
the early stages of the expansion, corresponding to $t/\alpha \ll 1$, by
using the relation (\ref{PPas1}) we find (no summation over $k$)%
\begin{equation}
\left\langle T_{k}^{k}\right\rangle _{\mathrm{s}}\approx \sum_{l=0}^{\infty }%
\frac{e^{-i\mu \pi }D_{l}}{\pi S_{D}t^{D+1}}\int_{0}^{\infty }dx\,\frac{\bar{%
Q}_{x-1/2}^{\mu }\left( u_{0}\right) }{\bar{P}_{x-1/2}^{-\mu }\left(
u_{0}\right) }\hat{R}_{k}(u)\frac{[P_{x-1/2}^{-\mu }\left( u\right) ]^{2}}{%
(u^{2}-1)^{D/2-1}},  \label{Tkksmt}
\end{equation}%
where the operators $\hat{R}_{k}(u)$ are defined by%
\begin{eqnarray}
\hat{R}_{0}(u) &=&\left( \xi -\frac{1}{4}\right) \left[ \left(
u^{2}-1\right) \partial _{u}^{2}+Du\partial _{u}\right] +x^{2}+D\left(
D-1\right) \left( \xi -\xi _{D}\right) ,  \notag \\
\hat{R}_{1}(u) &=&\frac{1}{4}\left( u^{2}-1\right) \partial _{u}^{2}+\left[
\left( D-1\right) \xi +\frac{D}{4}\right] u\partial _{u}-\frac{l\left(
l+n\right) }{u^{2}-1}-x^{2}-\frac{D-1}{4}\left( 4\xi -D+1\right) ,  \notag \\
\hat{R}_{2}(u) &=&\left( \xi -\frac{1}{4}\right) \left( u^{2}-1\right)
\partial _{u}^{2}+\left[ \left( D-1\right) \xi -\frac{D}{4}\right] u\partial
_{u}+\frac{l}{D-1}\frac{l+n}{u^{2}-1}-\xi \left( D-1\right) .  \label{R2}
\end{eqnarray}%
We note that the sphere-induced VEV of the energy-momentum tensor for a
scalar field in the static spacetime with negative constant curvature
spatial sections is expressed in terms of the operators (\ref{R2}) as (see 
\cite{Saha14}, no summation over $k$) 
\begin{equation}
\langle T_{i}^{k}\rangle _{\mathrm{s}}^{\mathrm{(st)}}=\delta
_{i}^{k}\sum_{l=0}^{\infty }\frac{e^{-i\mu \pi }D_{l}}{\pi S_{D}\alpha ^{D+1}%
}\int_{x_{m}}^{\infty }dx\,x\frac{\bar{Q}_{x-1/2}^{\mu }(u_{0})}{\bar{P}%
_{x-1/2}^{-\mu }(u_{0})}\frac{\hat{R}_{k}(u)-m^{2}a^{2}\delta _{k}^{0}}{%
\sqrt{x^{2}-x_{m}^{2}}}\frac{P_{x-1/2}^{-\mu }(u)}{(u^{2}-1)^{D/2-1}},
\label{Tikst}
\end{equation}%
where $x_{m}=\sqrt{m^{2}\alpha ^{2}-D(D-1)\left( \xi -\xi _{D}\right) }$.
For $\xi =\xi _{D}$ the operators (\ref{R2}) are reduced to the ones in (\ref%
{F2stc}): $\hat{R}_{k}(u)=\hat{F}_{k}^{\mathrm{(st)}}(u)$. In the same
limit, $t/\alpha \ll 1$, for the off-diagonal component we get%
\begin{equation}
\left\langle T_{0}^{1}\right\rangle _{\mathrm{s}}\approx D\left( \xi -\xi
_{D}\right) \sum_{l=0}^{\infty }\frac{e^{-i\mu \pi }D_{l}}{\pi S_{D}t^{D+2}}%
\int_{0}^{\infty }dx\,\frac{\bar{Q}_{x-1/2}^{\mu }\left( u_{0}\right) }{\bar{%
P}_{x-1/2}^{-\mu }\left( u_{0}\right) }\partial _{r}\frac{[P_{x-1/2}^{-\mu
}\left( u\right) ]^{2}}{(u^{2}-1)^{D/2-1}}.  \label{T01smt}
\end{equation}%
The energy flux density per unit proper surface area is given as $%
\left\langle \tilde{T}_{0}^{1}\right\rangle _{\mathrm{s}}\approx
t\left\langle T_{0}^{1}\right\rangle _{\mathrm{s}}$ and for non-conformally
coupled fields it is of the same order as the diagonal components. The
expressions in the right-hand sides of (\ref{Tkksmt}) and (\ref{T01smt})
coincide with the leading terms in the expansions of the corresponding VEVs
for a sphere in the Milne universe at early stages $mt\ll 1$. This is
related to the fact that the effects of curvature of dS spacetime are weak
in the range $t/\alpha \ll 1$.

At late stages of the expansion, $t/\alpha \gg 1$, we use the asymptotic (%
\ref{Pas2}). In the case $\nu >0$, to the leading order, for the diagonal
components we get (no summation over $k$) 
\begin{equation}
\left\langle T_{k}^{k}\right\rangle _{\mathrm{s}}\approx \frac{a_{k}}{\alpha
^{2}}\left\langle \varphi ^{2}\right\rangle _{\mathrm{s}}\,,  \label{Tkklti}
\end{equation}%
where the VEV of the field squared is estimated as (\ref{phi2slt}) and%
\begin{eqnarray}
a_{0} &=&\frac{D}{4}\left[ D-2\nu -4\xi \left( D+1-2\nu \right) \right] , 
\notag \\
a_{1} &=&a_{2}=\frac{2\nu }{D}a_{0}.  \label{a1}
\end{eqnarray}%
In this limit the sphere-induced VEV is suppressed by the factor $e^{-\left(
D-2\nu \right) t/\alpha }$. The corresponding asymptotic for the
off-diagonal component has the form%
\begin{equation}
\left\langle T_{0}^{1}\right\rangle _{\mathrm{s}}\approx \frac{4a_{0}}{%
D\alpha ^{3}}e^{-2t/\alpha }\partial _{r}\left\langle \varphi
^{2}\right\rangle _{\mathrm{s}},  \label{T10lt}
\end{equation}%
and the suppression is stronger, by the factor $e^{-\left( D+2-2\nu \right)
t/\alpha }$. Note that at late stages of the expansion we have the relation%
\begin{equation}
\left\langle \tilde{T}_{0}^{1}\right\rangle _{\mathrm{s}}=\frac{2}{D}%
e^{-t/\alpha }\partial _{r}\left\langle T_{0}^{0}\right\rangle _{\mathrm{s}},
\label{T10T00}
\end{equation}%
between the energy and the energy flux densities. The asymptotics (\ref%
{Tkklti}) and (\ref{T10lt}) are also valid for $\nu =0$ but now the behavior
of $\left\langle \varphi ^{2}\right\rangle _{\mathrm{s}}$ is described by (%
\ref{phi2sltnu0}). Note that in this case the leading terms in the vacuum
stresses vanish and we have (no summation over $k$) $|\left\langle
T_{k}^{k}\right\rangle _{\mathrm{s}}|\ll |\left\langle
T_{0}^{0}\right\rangle _{\mathrm{s}}|$, where $k=1,2,\ldots ,D$.

For $t/\alpha \gg 1$ and purely imaginary $\nu $, $\nu =i|\nu |$, we use the
asymptotic formula (\ref{Ply}). To the leading order this gives%
\begin{eqnarray}
\left\langle T_{0}^{0}\right\rangle _{\mathrm{s}} &\approx
&m^{2}\left\langle \varphi ^{2}\right\rangle _{\mathrm{s}}+\frac{2^{D}|\nu
|e^{-Dt/\alpha }}{\pi S_{D}\alpha ^{D+1}}\overset{\infty }{\sum_{l=0}}%
D_{l}\int_{0}^{\infty }dx\frac{xe^{-i\mu \pi }}{\sin \left( \pi x\right) }\,%
\frac{\bar{Q}_{x-1/2}^{\mu }\left( u_{0}\right) }{\bar{P}_{x-1/2}^{-\mu
}\left( u_{0}\right) }  \notag \\
&&\times \frac{\lbrack P_{x-1/2}^{-\mu }\left( u\right) ]^{2}}{\sinh ^{D-2}r}%
B_{\nu }(x)\left\{ 2D\left( \xi -\frac{1}{4}\right) \sin \left[ \phi (t,x)%
\right] +|\nu |\cos \left[ \phi (t,x)\right] \right\} ,  \label{T00lt3}
\end{eqnarray}%
for the energy density and 
\begin{eqnarray}
\left\langle T_{k}^{k}\right\rangle _{\mathrm{s}} &\approx &-\frac{2^{D}|\nu
|e^{-Dt/\alpha }}{\pi S_{D}\alpha ^{D+1}}\sum_{l=0}^{\infty
}D_{l}\int_{0}^{\infty }dx\frac{xe^{-i\mu \pi }}{\sin \left( \pi x\right) }\,%
\frac{\bar{Q}_{x-1/2}^{\mu }\left( u_{0}\right) }{\bar{P}_{x-1/2}^{-\mu
}\left( u_{0}\right) }\frac{[P_{x-1/2}^{-\mu }\left( u\right) ]^{2}}{\sinh
^{D-2}r}B_{\nu }(x)  \notag \\
&&\times \left\{ \left( 1-4\xi \right) |\nu |\cos \left[ \phi (t,x)\right] +2%
\left[ \left( D+1\right) \xi -\frac{D}{4}\right] \sin \left[ \phi (t,x)%
\right] \right\} ,  \label{T11lt3}
\end{eqnarray}%
for the stresses (no summation over $k$) $k=1,2,\ldots ,D$. Here, the phase $%
\phi (t,x)$ is defined in (\ref{phitx}). The asymptotic expression for the
energy flux density takes the form%
\begin{eqnarray}
\left\langle \tilde{T}_{0}^{1}\right\rangle _{\mathrm{s}} &\approx &\frac{%
D-4\left( D+1\right) \xi }{2\alpha ^{2}e^{t/\alpha }}\partial
_{r}\left\langle \varphi ^{2}\right\rangle _{\mathrm{s}}+\frac{2^{D}|\nu
|\left( 4\xi -1\right) }{\pi \alpha ^{D+1}S_{D}e^{(D+1)t/\alpha }}%
\sum_{l=0}^{\infty }D_{l}\int_{0}^{\infty }dx\frac{xe^{-i\mu \pi }}{\sin
\left( \pi x\right) }  \notag \\
&&\times \,\frac{\bar{Q}_{x-1/2}^{\mu }\left( u_{0}\right) }{\bar{P}%
_{x-1/2}^{-\mu }\left( u_{0}\right) }\partial _{r}\frac{[P_{x-1/2}^{-\mu
}\left( u\right) ]^{2}}{\sinh ^{D-2}r}B_{\nu }(x)\sin \left[ \phi (t,x)%
\right] .  \label{T10lt3}
\end{eqnarray}%
The expression for $\left\langle \varphi ^{2}\right\rangle _{\mathrm{s}}$ in
(\ref{T00lt3}) and (\ref{T10lt3}) is given by (\ref{phi2slt3}) and $B_{\nu
}(x)$, $\phi _{\nu }(x)$ are defined by (\ref{Bdef}). For purely imaginary $%
\nu $ the decay of the vacuum energy-momentum tensor at late stages is
damping oscillatory.

Now we turn to the asymptotics with respect to the radial coordinate. Near
the center the dominant contributions come from the terms with $l=0,1$ and
to the leading order we get (no summation over $k$)%
\begin{equation}
\left\langle T_{k}^{k}\right\rangle _{\mathrm{s}}\approx \frac{2\sinh
^{-2}\left( t/\alpha \right) }{\left( 4\pi \right) ^{D/2}\alpha ^{D+1}\Gamma
\left( D/2\right) }\int_{0}^{\infty }dx\sum_{l=0,1}\frac{xe^{-i\mu \pi }}{%
\sin \left( \pi x\right) }\frac{\bar{Q}_{x-1/2}^{\mu }\left( u_{0}\right) }{%
\bar{P}_{x-1/2}^{-\mu }\left( u_{0}\right) }\hat{F}_{\left( k\right) l}\frac{%
P_{\nu -1/2}^{x}\left( y\right) P_{\nu -1/2}^{-x}\left( y\right) }{%
(y^{2}-1)^{(D-1)/2}},  \label{Tkkr0}
\end{equation}%
where $\hat{F}_{\left( k\right) 0}=\hat{F}_{k}^{(0)}(y)$, with the operators
from (\ref{Fk0}), and 
\begin{eqnarray}
\hat{F}_{\left( 0\right) 1} &=&2\xi -\frac{1}{2},  \notag \\
\hat{F}_{\left( 1\right) 1} &=&\hat{F}_{\left( 2\right) 1}=\frac{2}{D}\left[
\left( D-1\right) \xi -\frac{D-2}{4}\right] .  \label{F11}
\end{eqnarray}%
The dominant contribution to the off-diagonal component comes from $l=1$
term and one finds%
\begin{eqnarray}
\left\langle T_{0}^{1}\right\rangle _{\mathrm{s}} &\approx &\frac{e^{-i\pi
D/2}r\sinh ^{-3}\left( t/\alpha \right) }{\left( 4\pi \right) ^{D/2}\alpha
^{D+2}D\Gamma \left( D/2\right) }\int_{0}^{\infty }dx\frac{x}{\sin \left(
\pi x\right) }\,\frac{\bar{Q}_{x-1/2}^{D/2}\left( u_{0}\right) }{\bar{P}%
_{x-1/2}^{-D/2}\left( u_{0}\right) }  \notag \\
&&\times \left[ \left( 1-4\xi \right) \left( y^{2}-1\right) \partial
_{y}+4\xi y\right] \frac{P_{\nu -1/2}^{x}\left( y\right) P_{\nu
-1/2}^{-x}\left( y\right) }{(y^{2}-1)^{(D-1)/2}},  \label{T01r0}
\end{eqnarray}%
and it linearly vanishes at the sphere center.

For points near the sphere the main contribution to the integral and series
in (\ref{Tkki}) comes from large $l$ and $x$. By using the large $x$
asymptotic (\ref{AsPP}) we see that the dependence on $y$ in the function (%
\ref{Fi}) appears in the form $(y^{2}-1)^{(1-D)/2}$ and the derivatives in (%
\ref{Fk0}) with respect to $y$ are easily evaluated. Keeping the leading
terms in $x$ we can see that for the components $\left\langle
T_{k}^{k}\right\rangle _{\mathrm{s}}$, $k=0,2,\ldots ,D$, the leading terms
in the asymptotic expansions over the distance from the sphere are expressed
in terms of the corresponding components $\left\langle
T_{k}^{k}\right\rangle _{\mathrm{s}}^{\mathrm{(st)}}$ for a sphere in static
spacetime with negative constant curvature space as $\left\langle
T_{k}^{k}\right\rangle _{\mathrm{s}}\approx \left\langle
T_{k}^{k}\right\rangle _{\mathrm{s}}^{\mathrm{(st)}}/\sinh ^{D+1}\left(
t/\alpha \right) $. By using the asymptotics for $\left\langle
T_{k}^{k}\right\rangle _{\mathrm{s}}^{\mathrm{(st)}}$ from \cite{Saha14} we
find (no summation over $k$)%
\begin{equation}
\left\langle T_{k}^{k}\right\rangle _{\mathrm{s}}\approx \frac{(2\delta
_{0B}-1)D\Gamma ((D+1)/2)\left( \xi -\xi _{D}\right) }{2^{D}\pi ^{(D+1)/2}%
\left[ \alpha \sinh (t/\alpha )(r_{0}-r)\right] ^{D+1}},  \label{Tkknear}
\end{equation}%
for $k=0,2,\ldots ,D$. In order to find the asymptotics for the radial
stress and the off-diagonal component it is more convenient to use the
covariant conservation equations (\ref{Ceq}). From the first equation it
follows that%
\begin{equation}
\left\langle T_{0}^{1}\right\rangle _{\mathrm{s}}\approx \frac{1}{\alpha }%
\coth \left( t/\alpha \right) \left( r_{0}-r\right) \left\langle
T_{0}^{0}\right\rangle _{\mathrm{s}}.  \label{T01near}
\end{equation}%
With this result, from the second equation in (\ref{Ceq}) we get%
\begin{equation}
\left\langle T_{1}^{1}\right\rangle _{\mathrm{s}}\approx \frac{D-1}{D}\coth
\left( r_{0}\right) \left( r_{0}-r\right) \left\langle
T_{0}^{0}\right\rangle _{\mathrm{s}}.  \label{T11near}
\end{equation}%
The leading terms do not depend on the mass. For a conformally coupled field
they vanish and one needs to keep the next-to-leading order terms. Note that
the leading terms in the VEVs of the field squared and of the diagonal
components, given by (\ref{phi2near2}) and (\ref{Tkknear}), are obtained
from the corresponding terms for a sphere in the Minkowski bulk (see \cite%
{Saha01SphM}) replacing the distance from the sphere by the proper distance $%
\alpha \sinh (t/\alpha )(r_{0}-r)$ for the geometry at hand.

\subsection{Exterior region}

The VEV of the energy-momentum tensor outside the sphere is decomposed as (%
\ref{Tikdec}), where the sphere-induced contribution is obtained from (\ref%
{TikVev}) and (\ref{Wse2}). The expression for the diagonal components reads
(no summation over $k$):%
\begin{equation}
\left\langle T_{k}^{k}\right\rangle _{\mathrm{s}}=-\frac{\sinh ^{-2}\left(
t/\alpha \right) }{\alpha ^{D+1}S_{D}}\sum_{l=0}^{\infty
}D_{l}\int_{0}^{\infty }dx\frac{e^{-i\mu \pi }x}{\sin \left( \pi x\right) }%
\frac{\bar{P}_{x-1/2}^{-\mu }\left( u_{0}\right) }{\bar{Q}_{x-1/2}^{\mu
}\left( u_{0}\right) }\,\left[ \hat{F}_{k}^{(0)}(y)-\hat{F}_{k}^{(1)}(u)%
\right] F^{\mathrm{(e)}}\left( x,y,u\right) ,  \label{Tkke}
\end{equation}%
with the function%
\begin{equation}
F^{\mathrm{(e)}}\left( x,y,u\right) =\frac{P_{\nu -1/2}^{x}\left( y\right)
P_{\nu -1/2}^{-x}\left( y\right) }{\sinh ^{D-1}\left( t/\alpha \right) }%
\frac{[Q_{x-1/2}^{\mu }\left( u\right) ]^{2}}{\sinh ^{D-2}r},  \label{Fe}
\end{equation}%
and the operators $\hat{F}_{k}^{(0)}(y)$ and $\hat{F}_{k}^{(1)}(u)$ are
defined by (\ref{Fk0}) and (\ref{Fk1}). The nonzero off-diagonal component
is expressed as%
\begin{eqnarray}
\left\langle T_{0}^{1}\right\rangle _{\mathrm{s}} &=&\frac{\sinh ^{-3}\left(
t/\alpha \right) }{\alpha ^{D+2}S_{D}}\sum_{l=0}^{\infty
}D_{l}\int_{0}^{\infty }dx\frac{xe^{-i\mu \pi }}{\sin \left( \pi x\right) }\,%
\frac{\bar{P}_{x-1/2}^{-\mu }\left( u_{0}\right) }{\bar{Q}_{x-1/2}^{\mu
}\left( u_{0}\right) }  \notag \\
&&\times \left[ \left( 1/4-\xi \right) \left( y^{2}-1\right) \partial
_{y}+\xi y\right] \partial _{r}F^{\mathrm{(e)}}\left( x,y,u\right) .
\label{T10e}
\end{eqnarray}%
Recall that the energy flux density per unit proper surface area is given by
(\ref{Flux}). One can check that the components (\ref{Tkke}) and (\ref{T10e}%
) obey the trace relation (\ref{Trel}) and covariant conservation equations (%
\ref{Ceq}).

For a conformally coupled massless field the off-diagonal component is zero
and for the diagonal components we have the relation (\ref{Tikc}), where the
VEV outside a sphere in static spacetime with a constant negative curvature
space is given by \cite{Saha14} 
\begin{equation}
\left\langle T_{i}^{k}\right\rangle _{\mathrm{s}}^{\mathrm{(st)}}=\delta
_{i}^{k}\sum_{l=0}^{\infty }\frac{e^{-i\mu \pi }D_{l}}{\pi \alpha ^{D+1}S_{D}%
}\int_{0}^{\infty }dx\,\frac{\bar{P}_{x-1/2}^{-\mu }\left( u_{0}\right) }{%
\bar{Q}_{x-1/2}^{\mu }\left( u_{0}\right) }\hat{F}_{k}^{\mathrm{(st)}}\frac{%
[Q_{x-1/2}^{\mu }\left( u\right) ]^{2}}{(u^{2}-1)^{D/2-1}},  \label{Tikstce}
\end{equation}%
with the operators $\hat{F}_{k}^{\mathrm{(st)}}$ defined in (\ref{F2stc}).
In the limit $\alpha \rightarrow \infty $ and for the case of a massive
field with general curvature coupling parameter, from (\ref{Tkke}) and (\ref%
{T10e}) we obtain the corresponding VEVs for the conformal vacuum outside a
spherical boundary in the Milne universe.

At early stages of the expansion, $t/\alpha \ll 1$, the leading terms of the
asymptotic expansion of the sphere-induced VEV $\left\langle
T_{i}^{k}\right\rangle _{\mathrm{s}}$ in the exterior region are obtained
from (\ref{Tkksmt}) and (\ref{T01smt}) by the replacements (\ref{ReplPQ}).
For a conformally coupled scalar field, to the leading order, we have the
relation $\left\langle T_{i}^{k}\right\rangle _{\mathrm{s}}\approx \left(
\alpha /t\right) ^{D+1}\left\langle T_{i}^{k}\right\rangle _{\mathrm{s}}^{%
\mathrm{(st)}}$, with $\left\langle T_{i}^{k}\right\rangle _{\mathrm{s}}^{%
\mathrm{(st)}}$ given by (\ref{Tikstce}). At late stages, $t/\alpha \gg 1$,
and for $\nu \geq 0$, the asymptotic expressions for the components of the
energy-momentum tensor are related to the corresponding asymptotic for the
VEV of the field squared by the formulas (\ref{Tkklti}) and (\ref{T10lt}).
For purely imaginary $\nu $, the behavior of the sphere-induced parts in the
VEV of the energy-momentum tensor is described by the formulas (\ref{T00lt3}%
)-(\ref{T10lt3}) with the replacements (\ref{ReplPQ}). In this limit, to the
leading order, the stresses are isotropic.

For points near the sphere the leading terms in the asymptotic expansions of
the energy density and stresses $\left\langle T_{k}^{k}\right\rangle _{%
\mathrm{s}}$, $k=0,2,\ldots ,D$, are given by (\ref{Tkknear}) with the
replacement $r_{0}-r\rightarrow r-r_{0}$. For points near the sphere these
components have the same sign in the interior and exterior regions. The
relations (\ref{T01near}) and (\ref{T11near}) for the off-diagonal component
and the radial stress remain the same and, hence, near the sphere these
components have opposite signs outside and inside the sphere. For large
distances from sphere the diagonal components of sphere-induced VEV of
energy-momentum tensor and the energy flux density are approximately given
by (no summation over $k$)%
\begin{eqnarray}
\left\langle T_{k}^{k}\right\rangle _{\mathrm{s}} &\approx &\frac{\hat{G}%
_{k}(y)\left\langle \varphi ^{2}\right\rangle _{\mathrm{s}}}{\alpha
^{2}\sinh ^{2}\left( t/\alpha \right) },\text{ }k=0,1,\ldots ,D,  \notag \\
\left\langle T_{0}^{1}\right\rangle _{\mathrm{s}} &\approx &\frac{D-1}{%
\alpha ^{3}\sinh ^{3}\left( t/\alpha \right) }\left[ \left( 1/4-\xi \right)
\left( y^{2}-1\right) \partial _{y}+\xi y\right] \left\langle \varphi
^{2}\right\rangle _{\mathrm{s}},  \label{Tkkelr}
\end{eqnarray}%
where $\left\langle \varphi ^{2}\right\rangle _{\mathrm{s}}$ is described by
the asymptotic expression (\ref{phi2selr}). The operators in (\ref{Tkkelr})
are defined as%
\begin{eqnarray}
\hat{G}_{0}(y) &=&\hat{F}_{0}^{(0)}(y)|_{x=0},\;\hat{G}_{1}(y)=\hat{F}%
_{2}^{(0)}(y)|_{x=0}+\left( D-1\right) ^{2}\left( \xi -1/4\right) ,  \notag
\\
\;\hat{G}_{k}(y) &=&\hat{F}_{2}^{(0)}(y)|_{x=0}-\left( D-1\right) \xi
,\;k=2,\ldots ,D.  \label{Gy}
\end{eqnarray}%
Hence, at large distances from the sphere we have an exponential suppression
of the sphere-induced VEVs by the factor $e^{-(D-1)r}/r$. For a conformally
coupled massless field the leading terms vanish. The corresponding behavior
is obtained by using the conformal relation (\ref{Tikc}) and the results
from \cite{Saha14} for static background. In this special case the energy
flux vanishes and the decay of the sphere-induced VEVs in the diagonal
components is stronger, like $e^{-(D-1)r}/r^{2}$.

\subsection{Numerical results}

\begin{figure}[tbph]
\begin{center}
\begin{tabular}{cc}
\epsfig{figure=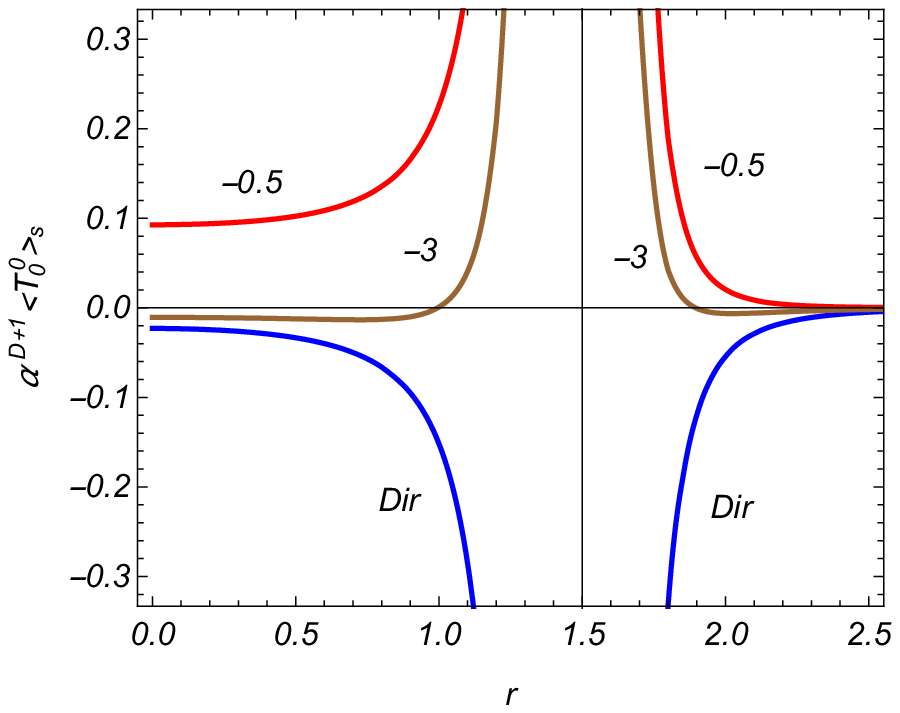,width=7.5cm,height=6.cm} & \quad %
\epsfig{figure=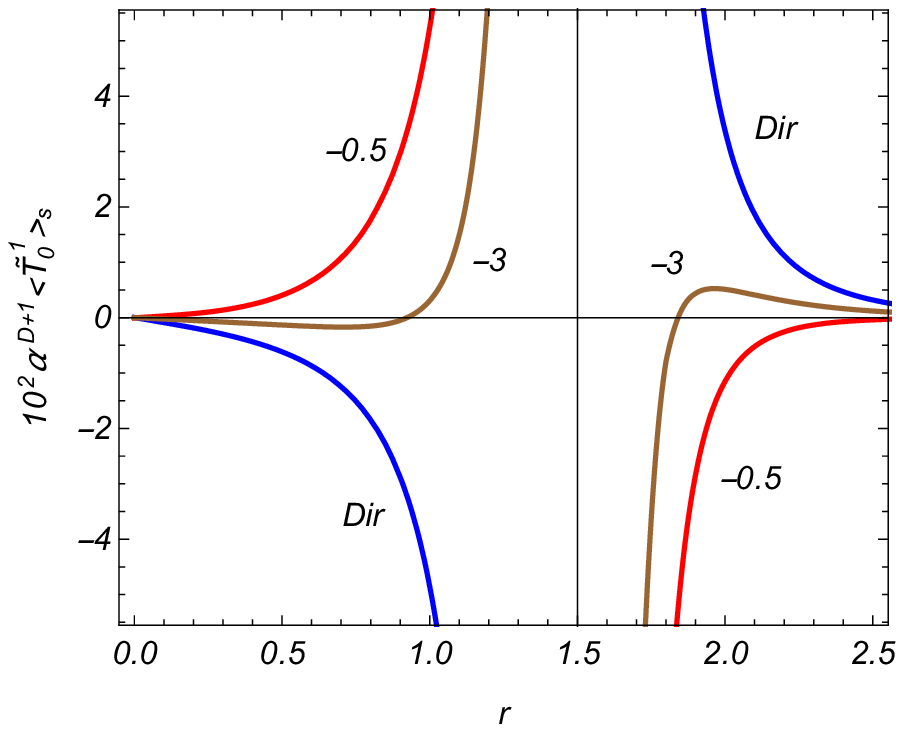,width=7.5cm,height=6.cm}%
\end{tabular}%
\end{center}
\caption{The sphere-induced energy density and the flux density as functions
of the radial coordinate for $D=3$ minimally coupled scalar field with
Dirichlet and Robin boundary conditions ($\protect\beta =-3,-0.5$). The
graphs are plotted for $m\protect\alpha =t/\protect\alpha =1$, $r_{0}=1.5$.}
\label{fig5}
\end{figure}

\begin{figure}[tbph]
\begin{center}
\begin{tabular}{cc}
\epsfig{figure=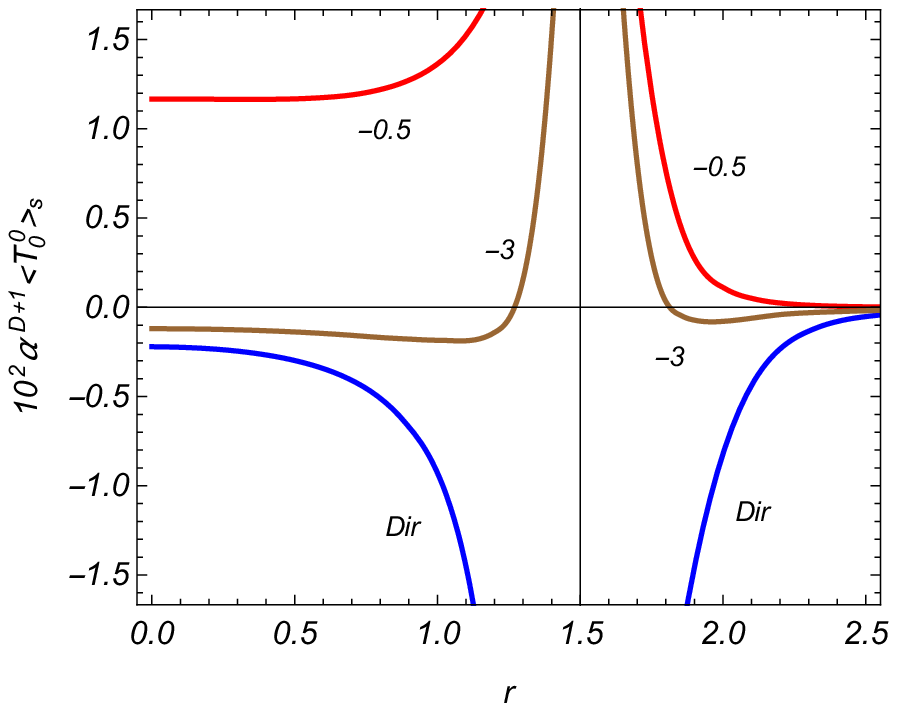,width=7.5cm,height=6.cm} & \quad %
\epsfig{figure=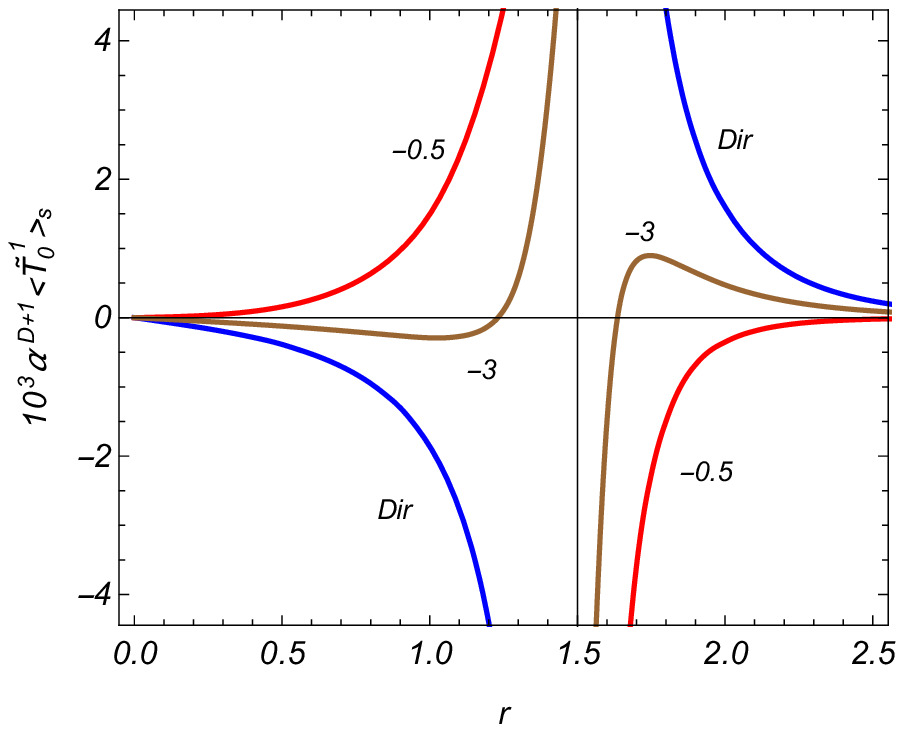,width=7.5cm,height=6.cm}%
\end{tabular}%
\end{center}
\caption{The same as in Figure \protect\ref{fig5} for a conformally coupled
field.}
\label{fig6}
\end{figure}

As before, the numerical results for the sphere-induced energy density and
energy flux will be presented for $D=3$ minimally and conformally coupled
scalar fields. In Figure \ref{fig5}, the boundary-induced energy (left
panel) and energy flux (right panel) densities are displayed as functions of
the radial coordinate for a minimally coupled scalar field. The graphs are
plotted for $m\alpha =t/\alpha =1$, $r_{0}=1.5$ in the cases of Dirichlet
boundary condition and for Robin conditions with $\beta =-3,-0.5$ (the
numbers near the curves). The same graphs for a conformally coupled scalar
field are presented in Figure \ref{fig6}. For both minimally and conformally
coupled fields, the energy flux in the interior and exterior egions is
directed from the boundary for Dirichlet boundary condition and towards the
boundary for Robin conditions.

The leading term for the energy density in the expansion near the sphere
center is given by (\ref{Tkkr0}). The energy flux linearly vanishes at the
center as a function of the radial coordinate (see (\ref{T01r0})). For a
minimally coupled field the leading term in the asymptotic expansion of the
energy density near the sphere is given by (\ref{Tkknear}) and the
sphere-induced VEV behaves as $(r-r_{0})^{-4}$. Near the sphere it has the
same sign for the interior and exterior regions. The leading term for the
energy flux is obtained from (\ref{T01near}) and it has opposite signs
inside and outside the sphere. The corresponding divergence on the sphere is
weaker, like $(r-r_{0})^{-3}$. The same is the case for the radial stress
(see (\ref{T11near})). For a conformally coupled field the leading terms in
the near-sphere expansions vanish and the energy density and energy flux
diverge as $(r-r_{0})^{-3}$ and $(r-r_{0})^{-2}$, respectively. At large
distances from the sphere, the boundary-induced contributions in both the
energy density and energy flux are suppressed by the factor $e^{-2r}$.

\begin{figure}[tbph]
\begin{center}
\begin{tabular}{cc}
\epsfig{figure=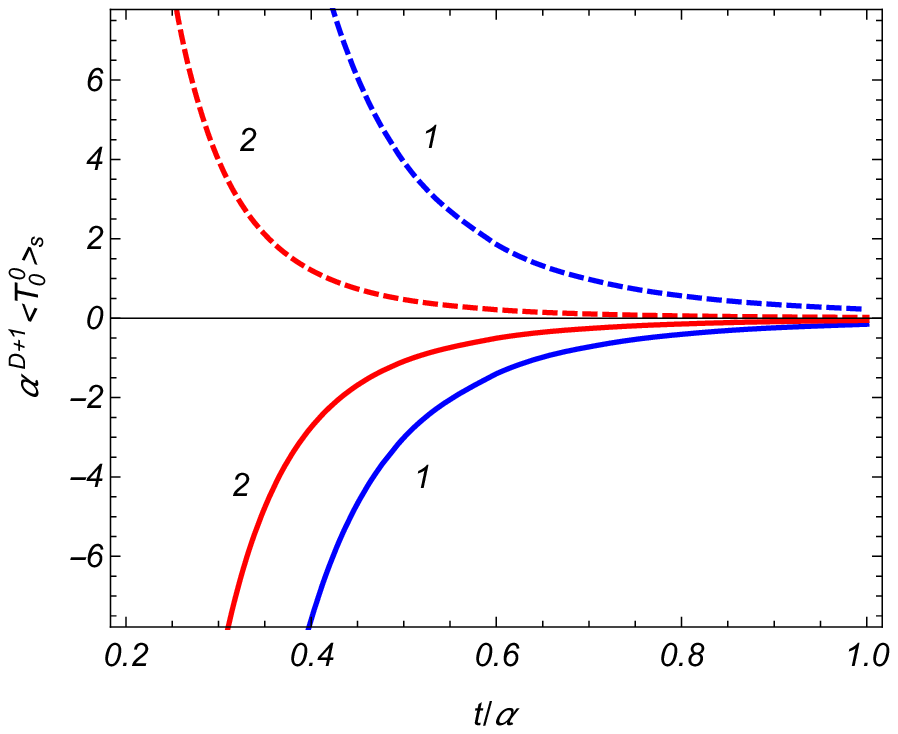,width=7.5cm,height=6.cm} & \quad %
\epsfig{figure=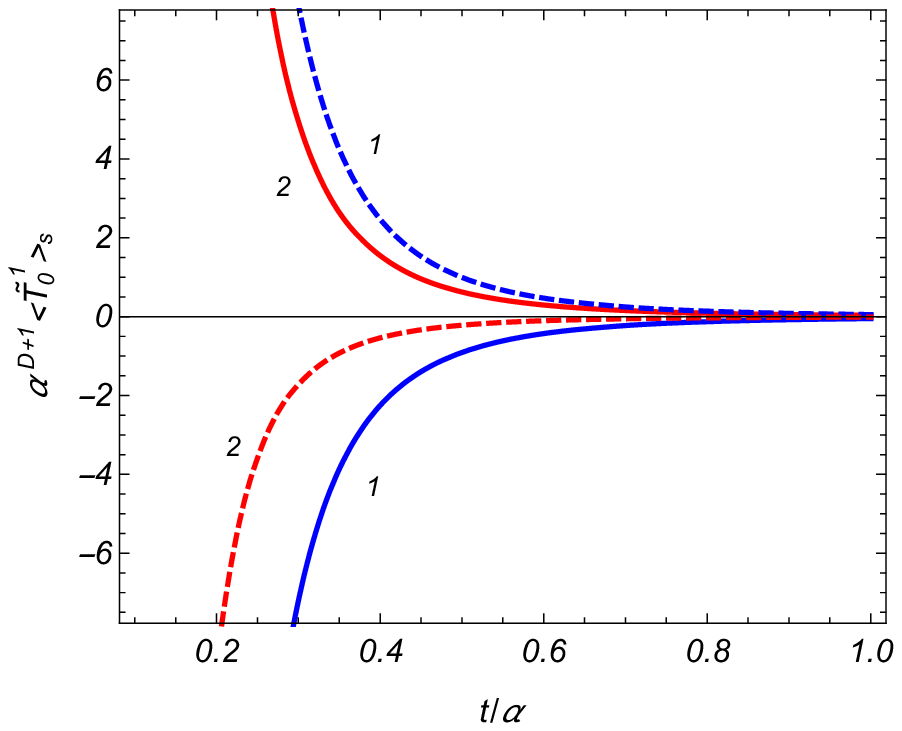,width=7.5cm,height=6.cm}%
\end{tabular}%
\end{center}
\caption{The sphere-induced energy density (left panel) and the energy flux
(right panel) for a minimally coupled field versus the time coordinate at
fixed values of the radial coordinate (the numbers near the curves). For the
sphere radius we have taken $r_{0}=1.5$ and for the field mass $m\protect%
\alpha =1$.}
\label{fig7}
\end{figure}

\begin{figure}[tbph]
\begin{center}
\begin{tabular}{cc}
\epsfig{figure=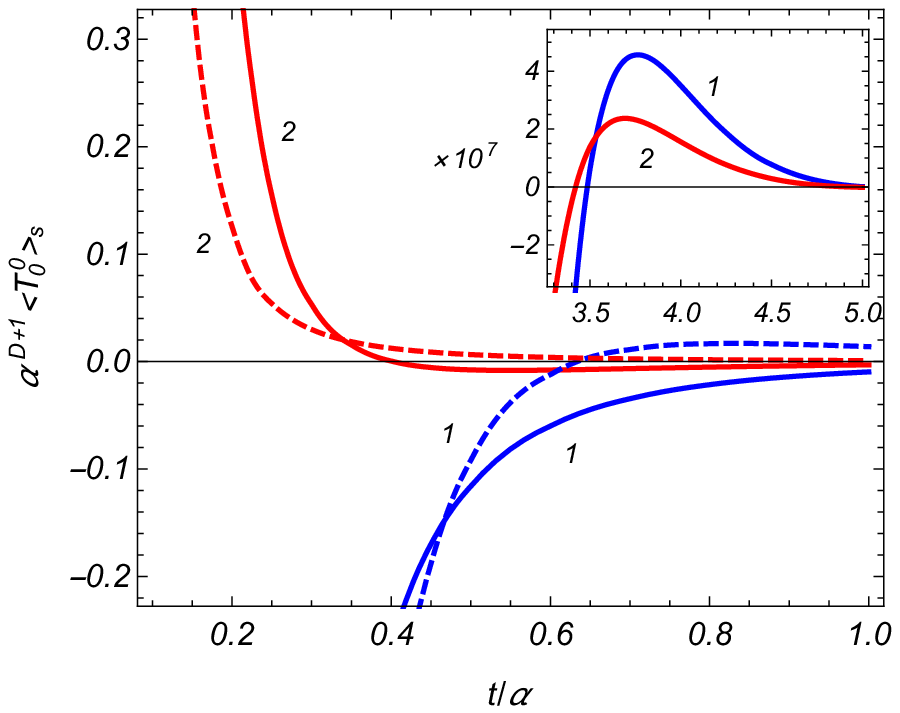,width=7.5cm,height=6.cm} & \quad %
\epsfig{figure=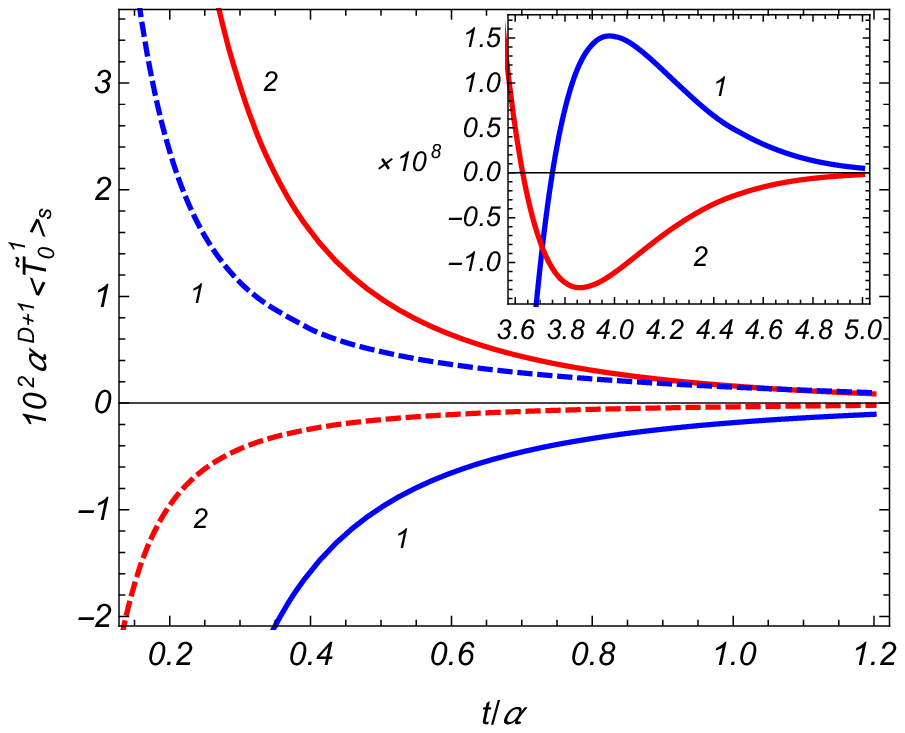,width=7.5cm,height=6.cm}%
\end{tabular}%
\end{center}
\caption{The same as in Figure \protect\ref{fig7} for a conformally coupled
scalar field.}
\label{fig8}
\end{figure}

Figure \ref{fig7} displays the time-dependence of the sphere-induced VEVs in
the energy density (left panel) and energy flux (right panel) for a
minimally coupled scalar field. The graphs are plotted for $r_{0}=1.5$, $%
m\alpha =1$ and the numbers near the curves are the values of the radial
coordinate $r$. The full curves correspond to Dirichlet boundary condition
and the dashed curves correspond to Robin boundary condition with $\beta
=-0.5$. The same graphs for a conformally coupled field are depicted in
Figure \ref{fig8}. According to (\ref{Tkksmt}), the boundary-induced
contribution in the energy density is nearly proportional to $1/t^{4}$ for $%
t/\alpha \ll 1$. For the energy flux density and for a minimally coupled
field one has the behavior $\left\langle \tilde{T}_{0}^{1}\right\rangle _{%
\mathrm{s}}\propto 1/t^{4}$. For a conformally coupled field the leading
term in the corresponding asymptotic expansion vanishes and $|\left\langle 
\tilde{T}_{0}^{1}\right\rangle _{\mathrm{s}}|\ll |\left\langle
T_{0}^{0}\right\rangle _{\mathrm{s}}|$ in the range $t/\alpha \ll 1$. This
is seen from Figure \ref{fig8}. In the opposite limit, $t/\alpha \gg 1$, the
corresponding approximation for minimal coupling is obtained from (\ref%
{Tkklti}), according to which $\left\langle T_{0}^{0}\right\rangle _{\mathrm{%
s}}$, as a function of $t$, behaves similar to the sphere-induced VEV of the
field squared. The corresponding approximation for a conformally coupled
scalar field is given by (\ref{T00lt3}). The oscillatory damping of the
sphere-induced VEVs in the case of a conformally coupled field is separately
displayed as insets (for $10^{7}\alpha ^{D+1}\left\langle
T_{0}^{0}\right\rangle _{\mathrm{s}}$ and $10^{8}\alpha ^{D+1}\left\langle 
\tilde{T}_{0}^{1}\right\rangle _{\mathrm{s}}$).

\begin{figure}[tbph]
\begin{center}
\begin{tabular}{cc}
\epsfig{figure=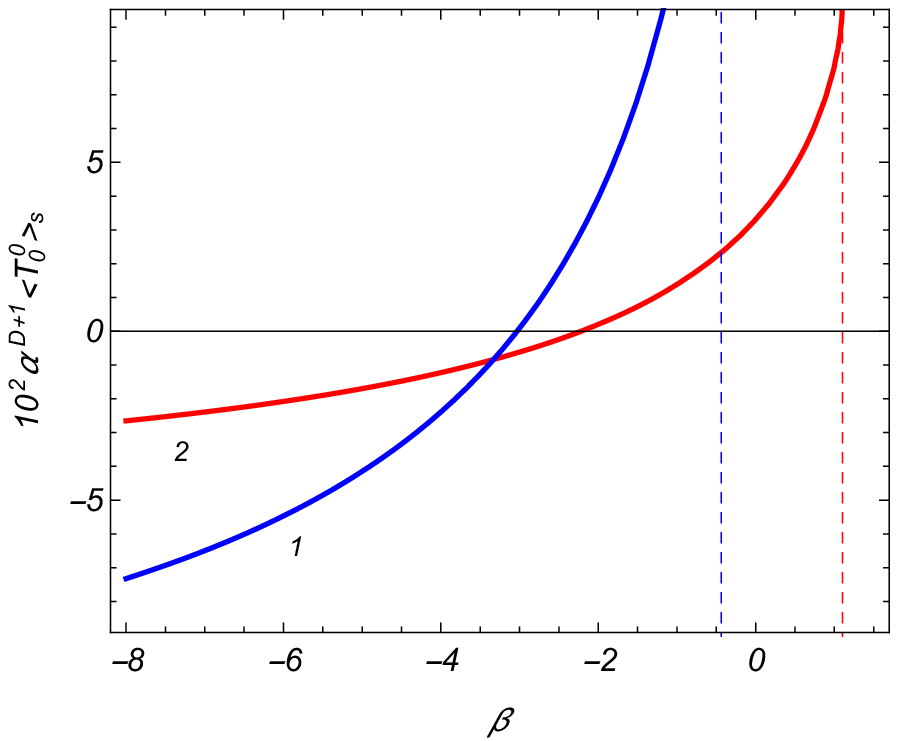,width=7.5cm,height=6.cm} & \quad %
\epsfig{figure=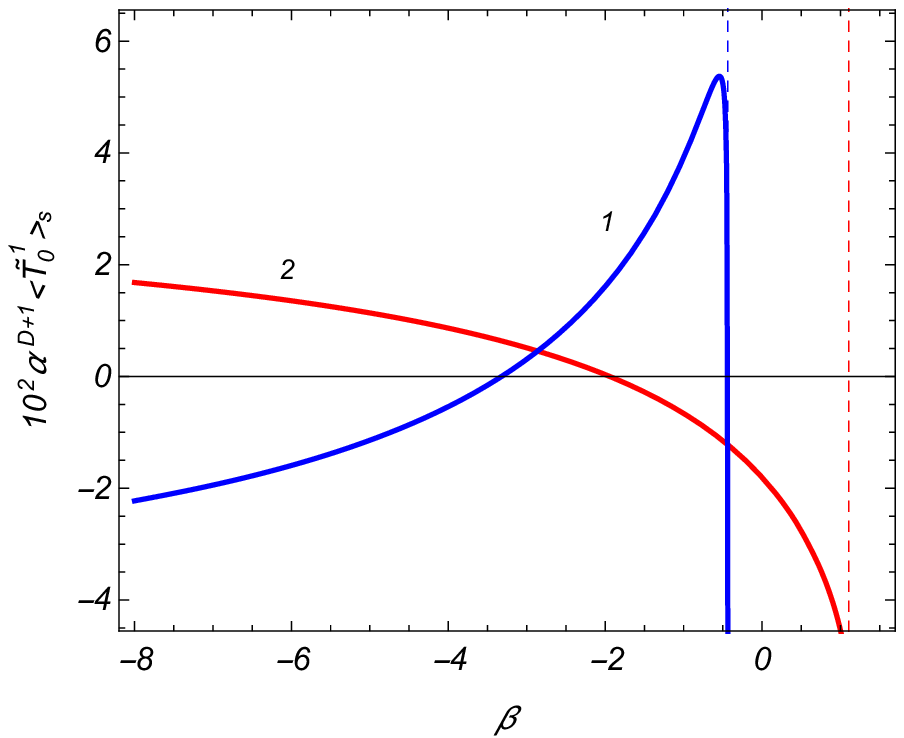,width=7.5cm,height=6.cm}%
\end{tabular}%
\end{center}
\caption{The boundary-induced energy density (left panel) and energy flux
(right panel) for $D=3$ minimally coupled scalar field as functions of the
coefficient $\protect\beta $ in Robin boundary condition. The graphs are
plotted for $m\protect\alpha =1$, $t/\protect\alpha =1$, $r_{0}=1.5$ and the
numbers near the curves correspond to the values of $r$.}
\label{fig9}
\end{figure}

\begin{figure}[tbph]
\begin{center}
\begin{tabular}{cc}
\epsfig{figure=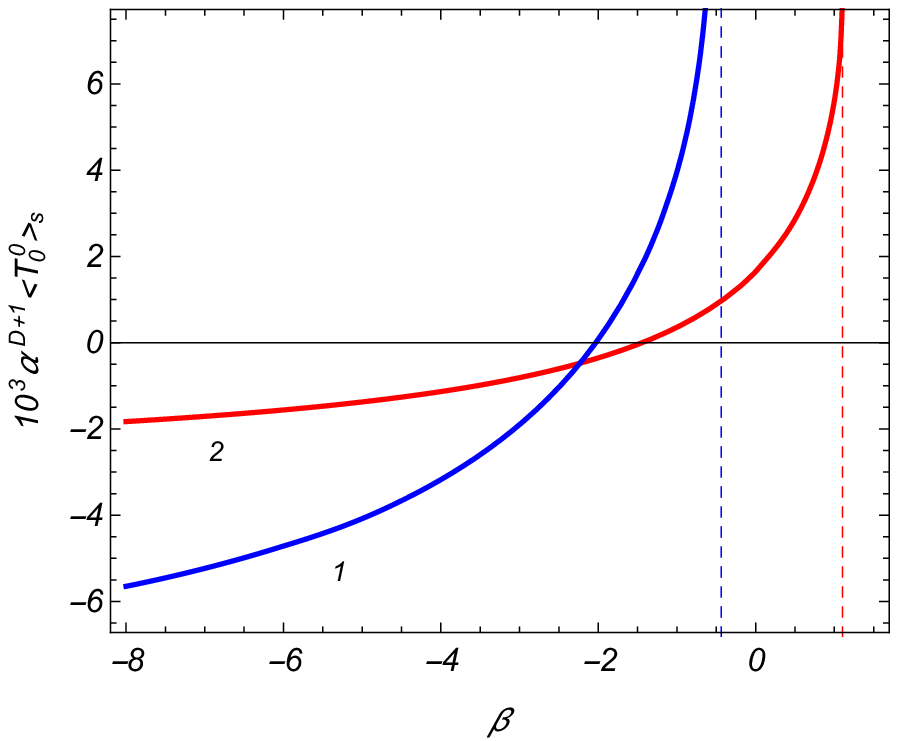,width=7.5cm,height=6.cm} & \quad %
\epsfig{figure=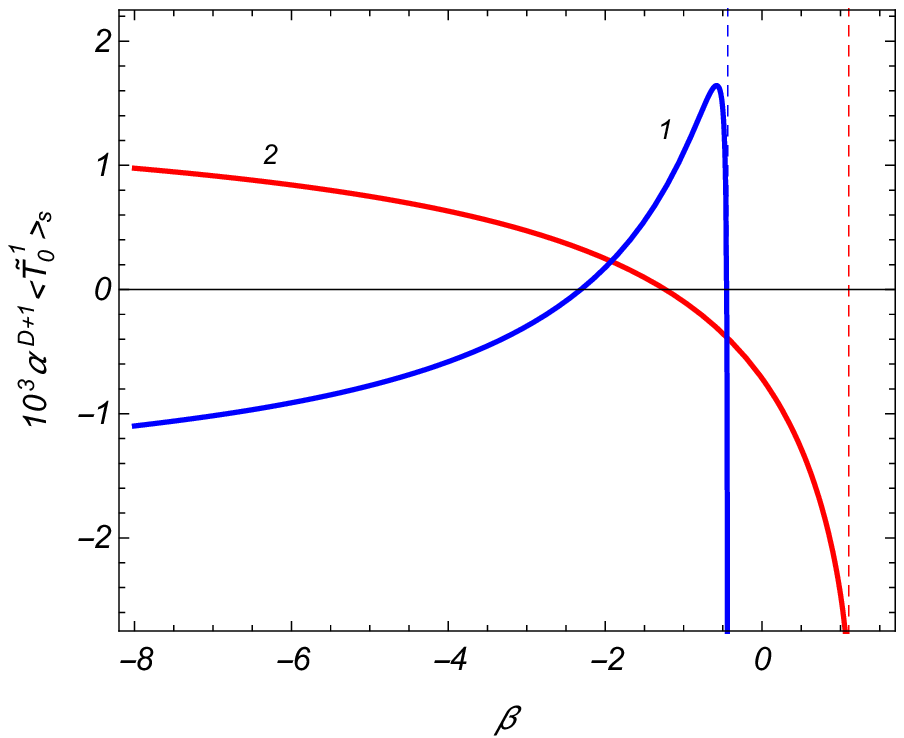,width=7.5cm,height=6.cm}%
\end{tabular}%
\end{center}
\caption{The same as in Figure \protect\ref{fig9} for a conformally coupled
field.}
\label{fig10}
\end{figure}

Figure \ref{fig9} presents the dependence of the sphere-induced VEV in the
energy density (left panel) and energy flux (right panel) on the coefficient 
$\beta $ in Robin boundary condition for a minimally coupled scalar fields.
The graphs are plotted for $D=3$, $m\alpha =t/\alpha =1$, $r_{0}=1.5$. The
numbers near the curves represent the values of the radial coordinate $r$.
For the interior region we have taken $r=1$ and for the exterior region $r=2$%
. The vertical dashed lines correspond to the critical values of the Robin
coefficient in the interior and exterior regions. The same graphs for a
conformally coupled field are presented in Figure \ref{fig10}. For the
values of the parameter $\beta $ close to the critical values the
sphere-induced energy density is positive. For large values of $-\beta $,
the VEVs tend to the values corresponding to Dirichlet boundary condition
and the energy density is negative. For some intermediate value of $\beta $
the sphere-induced contribution vanishes.

\section{Conclusion}

\label{sec:Conc}

For a given spacetime geometry, the quantum field theoretical vacuum is an
observer dependent notion. Among the interesting directions in the
investigations of the Casimir effect is the dependence of the physical
characteristics on the choice of the vacuum state. The previous studies of
the boundary and topology induced effects in dS spacetime mainly consider
the dS invariant Bunch-Davies vacuum state. The present paper concerns the
boundary induced effects on the local VEVs of a scalar field with general
curvature coupling for the hyperbolic vacuum in dS spacetime. As a boundary
we have considered a spherical shell with a constant comoving radius in
hyperbolic spatial coordinates. In inflationary coordinates this corresponds
to a spherical shell with time dependent radius, given by (\ref{rI0}). Note
that the Casimir effect for a spherical boundary with constant comoving
radius in inflationary coordinates and for the Bunch-Davies vacuum state has
been investigated in \cite{Milt12}.

As the first step in the investigation of the local VEVs we have constructed
the complete set of mode functions in hyperbolic coordinates without
specifying the vacuum state. Then, the mode functions are specified for the
conformal (hyperbolic) vacuum. It has been shown that the latter coincides
with the adiabatic vacuum. By using the complete set of mode functions, the
Hadamard functions are evaluated in the boundary-free geometry, outside the
spherical shell and inside the shell. For both regions in the problem with a
sphere, the contributions in the Hadamard function induced by the boundary
are separated explicitly. Inside the sphere, the eigenvalues of the quantum
number $z$ are given implicitly, as roots of the equation (\ref{Eigeq}), and
for the extraction of the sphere-induced part we have used the summation
formula (\ref{APF}). The corresponding contribution in the Hadamard function
is given by (\ref{Wsi}) and the explicit knowledge of the eigenvalues for $z$
is not required. Similar representations can be obtained for other two-point
functions (for example, for the Wightman function).

As a local characteristic of the hyperbolic vacuum, the VEV of the field
squared is considered. The latter is obtained taking the coincidence limit
of the arguments in the Hadamard function. In that limit divergences arise
and a renormalization is required. Having the decomposed representation of
the Hadamard function, for points away from the sphere the renormalization
is reduced to the one in the boundary-free geometry. The VEVs of the field
squared inside and outside the sphere are expressed as (\ref{phi2sin}) and (%
\ref{phi2se}). The corresponding expressions for the VEVs of the diagonal
components of the energy-momentum tensor are given by the expressions (\ref%
{Tkki}) and (\ref{Tkke}). Note that the expressions for the interior and
exterior regions are obtained from each other by the replacements (\ref%
{ReplPQ}). An interesting feature in the problem under consideration is the
presence of the vacuum energy flux along the radial direction. The latter is
described by the off-diagonal component of the energy-momentum tensor, given
by (\ref{T10}) and (\ref{T10e}). Depending on the value of the Robin
coefficient and also on the radial coordinate, that component may change the
sign. This shows that the energy flux can be directed either from the sphere
or towards the sphere.

The general formulas for the VEVs are complicated and in order to clarify
the qualitative features we have considered limiting cases and various
asymptotic regions of the parameters. In the flat spacetime limit,
corresponding to $\alpha \rightarrow \infty $, the line element (\ref{LE})
is reduced to the line element (\ref{ds2Milne}) for the Milne universe. It
is checked that, in this limit, from the results given above the
corresponding VEVs are obtained for a sphere in the Milne universe (see \cite%
{Saha20}), assuming that the scalar field is prepared in the conformal
vacuum. For a conformally coupled massless scalar field the problem under
consideration is conformally related to the problem with a spherical
boundary in static spacetime with constant negative curvature space. As
another check, we have shown that the VEVs in those problems are connected
by the standard conformal relation. Note that in this special case the
energy flux vanishes.

In early stages of the expansion, corresponding to $t/\alpha \ll 1$, the
effects of the spacetime curvature on the sphere-induced VEVs are weak and,
to the leading order, they coincide with the corresponding VEVs for a sphere
in the Milne universe. The effects of gravity are essential for $t/\alpha
\gtrsim 1$. In particular, at late stages, $t/\alpha \gg 1$, the behavior of
the VEVs is qualitatively different for positive and purely imaginary values
of the parameter $\nu $ in (\ref{nu}). For $\nu >0$ the decay of the
sphere-induced VEVs, as functions of the time coordinate, is monotonic, as $%
e^{-\left( D-2\nu \right) t/\alpha }$ for $\left\langle \varphi
^{2}\right\rangle _{\mathrm{s}}$, $\left\langle T_{k}^{k}\right\rangle _{%
\mathrm{s}}$, and like $e^{-\left( D+1-2\nu \right) t/\alpha }$ for the
energy flux density $\left\langle \tilde{T}_{0}^{1}\right\rangle _{\mathrm{s}%
}$. For imaginary $\nu $ the decay is oscillatory with the leading terms
given by (\ref{phi2slt3}), (\ref{T00lt3}), (\ref{T11lt3}), and (\ref{T10lt3}%
) in the interior region. The corresponding asymptotics outside the sphere
are obtained by the replacements (\ref{ReplPQ}).

For points near the sphere the dominant contribution to the VEVs comes from
the modes with large values of the angular momentum. The influence of the
gravitational field on those modes is weak and the leading terms in the
expansions of the VEVs for the field squared and for the energy density and
azimuthal stresses coincide with those for a spherical boundary in flat
spacetime with the distance from the sphere replaced by the proper distance $%
\alpha \sinh \left( t/\alpha \right) |r-r_{0}|$ in dS bulk. They behave as $%
|r-r_{0}|^{1-D}$ for the field squared and as $|r-r_{0}|^{-D-1}$ for the
energy density and azimuthal stresses. Near the sphere these VEVs have the
same sign in the exterior and interior regions. The leading terms for the
energy flux and radial stress are obtained by using the relations (\ref%
{T01near}) and (\ref{T11near}). These components behave like $|r-r_{0}|^{-D}$
and have opposite signs inside and outside the sphere. The leading terms do
not depend on the mass. In the case of the energy-momentum tensor they
vanish for a conformally coupled field. In the exterior region, at large
distances from the sphere, the sphere-induced VEVs are suppressed by the
factor $e^{-(D-1)r}/r$. For a conformally coupled massless field the leading
terms vanish and the suppression at large distances is stronger, like $%
e^{-(D-1)r}/r^{2}$.

\section*{Acknowledgments}

A.A.S. was supported by the grant No. 20RF-059 of the Committee of Science
of the Ministry of Education, Science, Culture and Sport RA. T.A.P. was
supported by the Committee of Science of the Ministry of Education, Science,
Culture and Sport RA in the frames of the research project No. 20AA-1C005.

\appendix

\section{Coordinate systems in dS spacetime}

\label{sec:ApCoord}

The dS spacetime is defined as a hyperboloid%
\begin{equation}
\eta _{MN}Z^{M}Z^{N}=-\alpha ^{2},\;M,N=0,1,\ldots ,D+1,  \label{Hyper}
\end{equation}%
in $(D+2)$-dimensional Minkowski spacetime with the line element $%
ds_{D+2}^{2}=\eta _{MN}dZ^{M}dZ^{N}$, where $\eta _{MN}=\mathrm{diag}%
(1,-1,\ldots ,-1)$. The global coordinates $(t_{g},\chi ,\theta _{1},\ldots
,\theta _{n},\phi )$ on the hyperboloid are defined by the relations%
\begin{eqnarray}
Z^{0} &=&\alpha \sinh (t_{g}/\alpha ),  \notag \\
Z^{1} &=&\alpha \cosh (t_{g}/\alpha )\cos \chi ,  \notag \\
Z^{i} &=&\alpha w^{i-1}\cosh (t_{g}/\alpha )\sin \chi ,  \label{Global}
\end{eqnarray}%
where $i=2,3,\ldots ,D+1$, $-\infty <t_{g}<+\infty $, $0<\chi <\pi $, $%
0\leqslant \theta _{k}\leqslant \pi $, $k=1,2,\ldots ,n$, $0\leqslant \phi
\leqslant 2\pi $, and 
\begin{eqnarray}
w^{1} &=&\cos \theta _{1},\;w^{2}=\sin \theta _{1}\cos \theta _{2},\ldots , 
\notag \\
w^{D-2} &=&\sin \theta _{1}\sin \theta _{2}\cdots \sin \theta _{D-3}\cos
\theta _{n},  \notag \\
w^{D-1} &=&\sin \theta _{1}\sin \theta _{2}\cdots \sin \theta _{n}\cos \phi ,
\notag \\
w^{D} &=&\sin \theta _{1}\sin \theta _{2}\cdots \sin \theta _{n}\sin \phi .
\label{wD}
\end{eqnarray}%
The line element on the hyperboloid takes the form%
\begin{equation}
ds^{2}=dt_{g}^{2}-\alpha ^{2}\cosh ^{2}(t_{g}/\alpha )\left( d\chi ^{2}+\sin
^{2}\chi d\Omega _{D-1}^{2}\right) .  \label{dsg}
\end{equation}%
The spatial sections with the global coordinates are spheres $S^{D-1}$.

Introducing the conformal time coordinate $\eta _{g}$ in accordance with%
\begin{equation}
\cosh (t_{g}/\alpha )=\frac{1}{\sin (\eta _{g}/\alpha )},\;0<\eta
_{g}/\alpha <\pi ,  \label{etag}
\end{equation}%
the line element is written in a conformally static form%
\begin{equation}
ds^{2}=\frac{d\eta _{g}^{2}-\alpha ^{2}\left( d\chi ^{2}+\sin ^{2}\chi
d\Omega _{D-1}^{2}\right) }{\sin ^{2}(\eta _{g}/\alpha )}.  \label{dsgc}
\end{equation}%
The Penrose diagram for the dS spacetime is presented by the square%
\begin{equation}
0\leqslant \eta _{g}/\alpha \leqslant \pi ,\;0\leqslant \chi \leqslant \pi ,
\label{PenrGlob}
\end{equation}%
in the plane $(\chi ,\eta _{g}/\alpha )$.

The coordinates $(t,r,\theta _{1},\ldots ,\theta _{n},\phi )$ corresponding
to the negative curvature spatial foliation are defined as%
\begin{eqnarray}
Z^{0} &=&\alpha \sinh (t/\alpha )\cosh r,  \notag \\
Z^{1} &=&\alpha \cosh (t/\alpha ),  \notag \\
Z^{i} &=&\alpha w^{i-1}\sinh (t/\alpha )\sinh r,  \label{Hyperb}
\end{eqnarray}%
with $i=2,3,\ldots ,D+1$. The corresponding line element is presented as (%
\ref{LE}) or in a conformally static form (\ref{ds2c}). In order to clarify
the region in the Penrose diagram corresponding to the coordinates (\ref%
{Hyperb}) it is useful to have the relations with conformal global
coordinates: 
\begin{eqnarray}
\cosh (t/\alpha ) &=&\frac{\cos \chi }{\sin (\eta _{g}/\alpha )},  \notag \\
\tanh r &=&-\frac{\sin \chi }{\cos \left( \eta _{g}/\alpha \right) }.
\label{RelGl1}
\end{eqnarray}%
Two separate regions are obtained. The region LI corresponds to $0<r<\infty $
and is given by%
\begin{equation}
\mathrm{LI}=\{(\chi ,\eta _{g}/\alpha ):\chi \in (0,\pi /2),\;\eta
_{g}/\alpha \in (\pi /2,\pi ),\;\eta _{g}/\alpha \geqslant \chi +\pi /2\}.
\label{LI}
\end{equation}%
From the relation%
\begin{equation}
\sinh (t/\alpha )=-\frac{\cot \left( \eta _{g}/\alpha \right) }{\cosh r},
\label{sht}
\end{equation}%
it follows that for this region $0<t<\infty $. The region LII, corresponding
to $-\infty <r<0$, is presented as%
\begin{equation}
\mathrm{LII}=\{(\chi ,\eta _{g}/\alpha ):\chi \in (0,\pi /2),\;\eta
_{g}/\alpha \in (0,\pi /2),\;\eta _{g}/\alpha \leqslant \pi /2-\chi \},
\label{LII}
\end{equation}%
and in this region $-\infty <t<0$. The other two triangular regions of the
Penrpose diagram, RI and RII, are covered by the coordinates $(t_{\mathrm{R}%
},r_{\mathrm{R}},\theta _{1},\ldots ,\theta _{n},\phi )$, defined in
accordance with%
\begin{eqnarray}
Z^{0} &=&\alpha \sinh (t_{\mathrm{R}}/\alpha )\cosh r_{\mathrm{R}},  \notag
\\
Z^{1} &=&-\alpha \cosh (t_{\mathrm{R}}/\alpha ),  \notag \\
Z^{i} &=&\alpha w^{i-1}\sinh (t_{\mathrm{R}}/\alpha )\sinh r_{\mathrm{R}},
\label{HyperbR}
\end{eqnarray}%
with $-\infty <t_{\mathrm{R}}<+\infty $, $-\infty <r_{\mathrm{R}}<+\infty $.
The relations to the global conformal coordinates are given as%
\begin{eqnarray}
\cosh (t_{\mathrm{R}}/\alpha ) &=&-\frac{\cos \chi }{\sin (\eta _{g}/\alpha )%
},  \notag \\
\tanh r_{\mathrm{R}} &=&-\frac{\sin \chi }{\cos \left( \eta _{g}/\alpha
\right) }.  \label{RelGL1R}
\end{eqnarray}%
The regions RI and RII in the Penrose diagram correspond to the ranges $0<r_{%
\mathrm{R}}<\infty $ and $-\infty <r_{\mathrm{R}}<0$, respectively and are
defined by 
\begin{eqnarray}
\mathrm{RI} &=&\{(\chi ,\eta _{g}/\alpha ):\chi \in (\pi /2,\pi ),\;\eta
_{g}/\alpha \in (\pi /2,\pi ),\;\eta _{g}/\alpha \geqslant 3\pi /2-\chi \}, 
\notag \\
\mathrm{RII} &=&\{(\chi ,\eta _{g}/\alpha ):\chi \in (\pi /2,\pi ),\;\eta
_{g}/\alpha \in (0,\pi /2),\;\eta _{g}/\alpha \leqslant \chi -\pi /2\}.
\label{RII}
\end{eqnarray}%
For the time coordinate in those regions we have the relation (\ref{sht})
with $t$ replaced by $t_{\mathrm{R}}$ and $r$ replaced by $r_{\mathrm{R}}$.
From here it follows that $0<t_{\mathrm{R}}<\infty $ and $-\infty <t_{%
\mathrm{R}}<0$ in the RI- and RII-regions, respectively.

The remaining region (C-region) of the Penrose diagram is covered by the
coordinates%
\begin{eqnarray*}
Z^{0} &=&\alpha \cos (t_{\mathrm{C}}/\alpha )\sinh r_{\mathrm{C}}, \\
Z^{1} &=&\alpha \sin (t_{\mathrm{C}}/\alpha ), \\
Z^{i} &=&\alpha w^{i-1}\cos (t_{\mathrm{C}}/\alpha )\cosh r_{\mathrm{C}},
\end{eqnarray*}%
with $i=2,3,\ldots ,D+1$ and $-\pi /2\leqslant t_{\mathrm{C}}/\alpha
\leqslant \pi /2$, $-\infty <r_{\mathrm{C}}<+\infty $. The corresponding
line element takes the form 
\begin{equation}
ds^{2}=-dt_{\mathrm{C}}^{2}+\alpha ^{2}\cos ^{2}\left( t_{\mathrm{C}}/\alpha
\right) (dr_{\mathrm{C}}^{2}-\cosh ^{2}r_{\mathrm{C}}d\Omega _{D-1}^{2}).
\label{ds2C}
\end{equation}%
We have the following relations with the global conformal coordinates:%
\begin{eqnarray}
\sin (t_{\mathrm{C}}/\alpha ) &=&\frac{\cos \chi }{\sin (\eta _{g}/\alpha )},
\notag \\
\tanh r_{\mathrm{C}} &=&-\frac{\cos (\eta _{g}/\alpha )}{\sin \chi }.
\label{CregGlob}
\end{eqnarray}%
The coordinate lines in all the regions of the Penrose diagram discussed
above are depicted in Figure \ref{fig1}.

It is also of interest to have the relations between the coordinates $%
(t,r,\vartheta ,\phi )$ and inflationary coordinates $(t_{\mathrm{I}},r_{%
\mathrm{I}},\vartheta ,\phi )$, with the line element%
\begin{equation}
ds^{2}=dt_{\mathrm{I}}^{2}-e^{2t_{\mathrm{I}}/\alpha }\left( dr_{\mathrm{I}%
}^{2}+r_{\mathrm{I}}^{2}d\Omega _{D-1}^{2}\right) .  \label{ds2Inf}
\end{equation}%
These relations are given by%
\begin{eqnarray}
\frac{t_{\mathrm{I}}}{\alpha } &=&\ln \left[ \cosh (t/\alpha )+\sinh
(t/\alpha )\cosh r\right] ,  \notag \\
\frac{r_{\mathrm{I}}}{\alpha } &=&e^{-t_{\mathrm{I}}/\alpha }\sinh (t/\alpha
)\sinh r.  \label{rInf}
\end{eqnarray}%
One has $t_{\mathrm{I}}=0$, $r_{\mathrm{I}}=0$ for $t=0$.

\section{Transformation of the Hadamard function in the boundary-free
geometry}

\label{sec:VEVbf}

In this section we will further transform the expression (\ref{W0g}) for the
Hadamard function in the boundary-free geometry. In \cite{Henr55} the
following addition theorem was proved for the associated Legendre functions
of the first kind (there is a misprint in formula (80) of \cite{Henr55}:
instead of $P_{\lambda }^{-\lambda -l}\left( \tau \right) $ should be $%
P_{\lambda }^{-\gamma -l}\left( \tau \right) $): 
\begin{eqnarray}
\frac{P_{\lambda }^{-\gamma }\left( \rho _{1}\right) }{\rho _{1}^{\prime
\gamma }} &=&\frac{2^{\gamma }\Gamma \left( \gamma \right) }{\rho ^{\prime
\gamma }\tau ^{\prime \gamma }}\overset{\infty }{\underset{l=0}{\sum }}%
\left( -1\right) ^{l}\left( \lambda +\gamma +1\right) _{l}\left( \gamma
-\lambda \right) _{l}  \notag \\
&&\times \left( l+\gamma \right) C_{l}^{\gamma }\left( \beta \right)
P_{\lambda }^{-\gamma -l}\left( \rho \right) P_{\lambda }^{-\gamma -l}\left(
\tau \right) ,  \label{AdP}
\end{eqnarray}%
where $(a)_{l}$ is Pochhammer's symbol, $\rho _{1}=\rho \tau +\rho ^{\prime
}\tau ^{\prime }\beta $ and $\chi ^{\prime }=(\chi ^{2}-1)^{1/2}$ for $\chi
=\rho ,\tau ,\rho _{1}$. Taking in this formula $\gamma =n/2$, $\lambda
=ix-1/2$, $\rho =u$, $\tau =u^{\prime }$, and $\beta =-\cos \theta $, it can
be rewritten in the form%
\begin{eqnarray}
&&\overset{\infty }{\underset{l=0}{\sum }}\left( l+\frac{n}{2}\right)
C_{l}^{n/2}\left( \cos \theta \right) \left\vert \Gamma \left( \frac{D-1}{2}%
+l+ix\right) \right\vert ^{2}P_{ix-1/2}^{-l-n/2}\left( u\right)
P_{ix-1/2}^{-l-n/2}\left( u^{\prime }\right)  \notag \\
&&\qquad =\frac{2^{-n/2}}{\Gamma \left( n/2\right) }\left[ \left(
u^{2}-1\right) \left( u^{\prime 2}-1\right) \right] ^{n/4}\left\vert \Gamma
\left( \frac{D-1}{2}+ix\right) \right\vert ^{2}\frac{P_{ix-1/2}^{-n/2}\left( 
\bar{u}\right) }{\left( \bar{u}^{2}-1\right) ^{n/4}},  \label{AdP2}
\end{eqnarray}%
where $\bar{u}$ is defined by (\ref{u}). The summation over $l$ in formula (%
\ref{W0g}) can be done by using the addition theorem (\ref{AdP2}) with $%
n=D-2 $ and $l+n/2=\mu $. This gives%
\begin{eqnarray}
G_{0}\left( x,x^{\prime }\right) &=&\frac{\alpha ^{1-D}}{2^{D/2}\pi ^{D/2+1}}%
\int_{0}^{\infty }dz\,z\sinh \left( \pi z\right) \left\vert \Gamma \left( 
\frac{D-1}{2}+iz\right) \right\vert ^{2}  \notag \\
&&\times \frac{X_{\nu }^{iz}\left( y\right) \left[ X_{\nu }^{iz}\left(
y^{\prime }\right) \right] ^{\ast }+X_{\nu }^{iz}\left( y^{\prime }\right) %
\left[ X_{\nu }^{iz}\left( y\right) \right] ^{\ast }}{\left[ \sinh (t/\alpha
)\sinh (t^{\prime }/\alpha )\right] ^{\frac{D-1}{2}}}\frac{%
P_{iz-1/2}^{1-D/2}\left( \bar{u}\right) }{\left( \bar{u}^{2}-1\right) ^{%
\frac{D-2}{4}}}.  \label{W0g1}
\end{eqnarray}%
This function depends on the spatial coordinates through the combination (%
\ref{u}). This property is a consequence of the maximal symmetry of the
spatial geometry. For the adiabatic vacuum one should take the function $%
X_{\nu }^{iz}\left( y\right) $ in the form (\ref{Xad}) and the Hadamard
function is expressed as (\ref{W01}).

Note that the geodesic distance $d(x,x^{\prime })$ between the points $x$
and $x^{\prime }$ is expressed in terms of $\bar{u}$. Considering the inner
product $\eta _{MN}Z^{M}Z^{\prime N}$ between the points $Z$ and $Z^{\prime }
$ in the embedding space, the geodesic distance is given by $\cosh \left[
d(x,x^{\prime })/\alpha \right] =\eta _{MN}Z^{M}Z^{\prime N}/\alpha ^{2}$ or
by $\cos \left[ d(x,x^{\prime })/\alpha \right] =\eta _{MN}Z^{M}Z^{\prime
N}/\alpha ^{2}$, depending on the separation between $x$ and $x^{\prime }$.
In the hyperbolic coordinates, by using (\ref{Hyperb}), we get%
\begin{equation}
\eta _{MN}Z^{M}Z^{\prime N}/\alpha ^{2}=\sinh (t/\alpha )\sinh (t^{\prime
}/\alpha )\bar{u}-\cosh (t/\alpha )\cosh (t^{\prime }/\alpha ).  \label{dxx}
\end{equation}%
For the special case of the maximally symmetric Bunch-Davies vacuum, the
function $G_{0}\left( x,x^{\prime }\right) $ depends on $x$ and $x^{\prime }$
through the geodesic distance (see, for example, the discussion in \cite%
{Sasa95} for $D=3$). In general, this is not the case for (\ref{W0g1}). For
points with $t=t^{\prime }$, $\theta =0$ we get%
\begin{equation}
\eta _{MN}Z^{M}Z^{\prime N}/\alpha ^{2}=2\sinh ^{2}(t/\alpha )\sinh ^{2}
\left[ (r-r^{\prime })/2\right] -1  \label{dxx1}
\end{equation}%
and for large radial separations $e^{d(x,x^{\prime })/\alpha }\approx \sinh
^{2}(t/\alpha )e^{r-r^{\prime }}$.

For a conformally coupled massless field one has $\nu =1/2$ and the
functions $P_{\nu -1/2}^{\pm iz}\left( \cosh \left( t/\alpha \right) \right) 
$ are given by (\ref{P0}). In the special case $D=3$, by taking into account
that%
\begin{equation}
P_{iz-1/2}^{-1/2}\left( \bar{u}\right) =\sqrt{\frac{2}{\pi }}\frac{\sin
\left( z\zeta \right) }{z\sqrt{\sinh \zeta }},  \label{P12}
\end{equation}%
with $\zeta $ defined by $\bar{u}=\cosh \zeta $, from (\ref{W01}) for the
Hadamard function one gets%
\begin{equation}
G_{0}\left( x,x^{\prime }\right) =\frac{\sinh (\eta /\alpha )\sinh (\eta
^{\prime }/\alpha )}{2\pi ^{2}\alpha ^{2}\sinh \zeta }\frac{\zeta }{\zeta
^{2}-\left( \eta -\eta ^{\prime }\right) ^{2}/\alpha ^{2}}.  \label{G0D3}
\end{equation}%
Note that in this expression $\zeta =\ln (\bar{u}+\sqrt{\bar{u}^{2}-1})$.
For points with $\theta =0$ we have $\zeta =r-r^{\prime }$ and the
expression (\ref{G0D3}) is specified as 
\begin{equation}
G_{0}\left( x,x^{\prime }\right) |_{\theta =0}=\frac{\sinh (\eta /\alpha
)\sinh (\eta ^{\prime }/\alpha )}{2\pi ^{2}\sinh \left( r-r^{\prime }\right) 
}\frac{r-r^{\prime }}{\alpha ^{2}\left( r-r^{\prime }\right) ^{2}-\left(
\eta -\eta ^{\prime }\right) ^{2}}.  \label{G0D3b}
\end{equation}%
This expression is conformally related (with the conformal factor $\sinh
(\eta /\alpha )\sinh (\eta ^{\prime }/\alpha )$) to the corresponding result
in static hyperbolic universes found in \cite{Bunc78}.


\addcontentsline{toc}{section}{References}

\begin{thebibliography}{99}
\bibitem{Hawk94} S. W. Hawking and G. F. R. Ellis, \textit{The Large Scale
Structure of Space-Time} (Cambridge University Press, Cambridge, England,
1994).

\bibitem{Grif09} J. B. Griffiths and J. Podolsky, \textit{Exact Space-Times
in Einstein's General Relativity} (Cambridge University Press, 2009).

\bibitem{Bass06} B. A. Bassett, S. Tsujikawa, and D. Wands, Inflation
dynamics and reheating, Rev. Mod. Phys. \textbf{78}, 537 (2006).

\bibitem{Mart14} J. Martin, C. Ringeval, and V. Vennin, Encyclop\ae dia
Inflationaris, Phys. Dark. Univ. \textbf{5-6}, 75 (2014).

\bibitem{Stro01} A. Strominger, The dS/CFT correspondence, JHEP \textbf{10}
(2001) 034.

\bibitem{Noji02} S. Nojiri and S. D. Odintsov, Quantum cosmology,
inflationary brane-world creation and dS/CFT correspondence, JHEP \textbf{12}
(2002) 033.

\bibitem{Anni17} D. Anninos, T. Hartman, and A. Strominger, Higher spin
realization of the dS/CFT correspondence, Class. Quantum Grav. \textbf{34},
015009 (2017).

\bibitem{Most97} V. M. Mostepanenko and N. N. Trunov, \textit{The Casimir
Effect and Its Applications} (Clarendon,Oxford, 1997).

\bibitem{Milt02} K. A. Milton, \textit{The Casimir Effect: Physical
Manifestation of Zero-Point Energy} (World Scientific, Singapore, 2002).

\bibitem{Pars05} V. A. Parsegian, \textit{Van der Waals Forces: A Handbook
for Biologists, Chemists, Engineers, and Physicists} (Cambridge University
Press, Cambridge, England, 2005).

\bibitem{Bord09} M. Bordag, G. L. Klimchitskaya, U. Mohideen, and V. M.
Mostepanenko, \textit{Advances in the Casimir Effect} (Oxford University
Press, New York, 2009).

\bibitem{Casi11} Casimir Physics, edited by D. Dalvit, P. Milonni, D.
Roberts, and F. da Rosa, \textit{Lecture Notes in Physics} Vol. 834
(Springer-Verlag, Berlin, 2011).

\bibitem{Saha20AdS} A. A. Saharian, Quantum vacuum effects in braneworlds on
AdS bulk, Universe \textbf{6}, 181 (2020).

\bibitem{Saha20AdSb} A. A. Saharian, A. S. Kotanjyan, and H. G. Sargsyan,
Electromagnetic field correlators and the Casimir effect for planar
boundaries in AdS spacetime with application in braneworlds, Phys. Rev. D 
\textbf{102}, 105014 (2020).

\bibitem{Eliz03dS} E. Elizalde, S. Nojiri, S.D. Odintsov, and S. Ogushi,
Casimir effect in de Sitter and Anti-de Sitter braneworlds, Phys. Rev. D 
\textbf{67}, 063515 (2003).

\bibitem{Saha09ds} A. A. Saharian and T. A. Vardanyan, Casimir densities for
a plate in de Sitter spacetime, Class. Quantum Grav. \textbf{26}, 195004
(2009).

\bibitem{Eliz10} E. Elizalde, A. A. Saharian, and T. A. Vardanyan, Casimir
effect for parallel plates in de Sitter spacetime, Phys. Rev. D \textbf{81},
124003 (2010).

\bibitem{Saha11ds} A. A. Saharian, Casimir effect in de Sitter spacetime,
Int. J. Mod. Phys. A \textbf{26}, 3833 (2011).

\bibitem{Burd11} P. Burda, Casimir effect for a massless minimally coupled
scalar field between parallel plates in de Sitter spacetime, JETP Lett. 
\textbf{93}, 632 (2011).

\bibitem{Espo15} G. Esposito and G. M. Napolitano, Towards obtaining Green
functions for a Casimir cavity in de Sitter spacetime, Phys. Scr. \textbf{90}%
, 074013 (2015).

\bibitem{Kota15} A. S. Kotanjyan, A. A. Saharian, and H. A. Nersisyan,
Electromagnetic Casimir effect for conducting plates in de Sitter spacetime,
Phys. Scr. \textbf{90}, 065304 (2015).

\bibitem{Saha15cyl} A. A. Saharian and V. F. Manukyan, Scalar Casimir
densities induced by a cylindrical shell in de Sitter spacetime, Class.
Quantum Grav. \textbf{32}, 025009 (2015).

\bibitem{Saha16cyl} A. A. Saharian, V. F. Manukyan, and N. A. Saharyan,
Electromagnetic Casimir densities for a cylindrical shell on de Sitter
space, Int. J. Mod. Phys. A \textbf{31}, 1650183 (2016).

\bibitem{Seta01a} M. R. Setare and R. Mansouri, Casimir effect for a
spherical shell in de Sitter space, Class. Quantum Grav. \textbf{18}, 2331
(2001).

\bibitem{Seta01b} M. R. Setare, Casimir stress for concentric spheres in de
Sitter space, Class. Quantum Grav. \textbf{18}, 4823 (2001).

\bibitem{Milt12} K. A. Milton and A. A. Saharian, Casimir densities for a
spherical boundary in de Sitter spacetime, Phys. Rev. D \textbf{85}, 064005
(2012).

\bibitem{Saha04dSbr} A. A. Saharian and M. R. Setare, Casimir
energy-momentum tensor for a brane in de Sitter spacetime, Phys. Lett. B 
\textbf{584}, 306 (2004). \ 

\bibitem{Bell14} S. Bellucci, A. A. Saharian, and A. H. Yeranyan, Casimir
densities from coexisting vacua, Phys. Rev. D \textbf{89}, 105006 (2014).

\bibitem{Saha08Compds} A. A. Saharian and M. R. Setare, Casimir effect in de
Sitter spacetime with compactified dimension, Phys. Lett. B \textbf{659},
367 (2008).

\bibitem{Saha08CompdsF} A. A. Saharian, The fermionic Casimir effect in
toroidally compactified de Sitter spacetime, Class. Quantum Grav. \textbf{25}%
, 165012 (2008).

\bibitem{Bell08Compds} S. Bellucci and A. A. Saharian, Wightman function and
vacuum densities in de Sitter spacetime with toroidally compactified
dimensions, Phys. Rev. D \textbf{77}, 124010 (2008).

\bibitem{Beze08Compds} E. R. Bezerra de Mello and A. A. Saharian, Fermionic
vacuum densities in higher-dimensional de Sitter spacetime, J. High Energy
Phys. 12 (2008) 081.

\bibitem{Saha09top} A. A. Saharian, Casimir effect in toroidally
compactified de Sitter spacetime, Int. J. Mod. Phys. A \textbf{24}, 1813
(2009).

\bibitem{Beze09CSds} E. R. Bezerra de Mello and A. A. Saharian, Vacuum
polarization by a cosmic string in de Sitter spacetime, J. High Energy Phys.
04 (2009) 046.

\bibitem{Moha15} A. Mohammadi, E. R. Bezerra de Mello, and A. A. Saharian,
Induced fermionic currents in de Sitter spacetime in the presence of a
compactified cosmic string, Class. Quantum Grav. \textbf{32}, 135002 (2015).

\bibitem{Saha17CSds} A. A. Saharian, V. F. Manukyan, and N. A. Saharyan,
Electromagnetic vacuum fluctuations around a cosmic string in de Sitter
spacetime, Eur. Phys. J. C \textbf{77}, 478 (2017).

\bibitem{Saha19Izv} A. A. Saharian, D. H. Simonyan, and A. S. Kotanjyan,
Vacuum polarization induced by a boundary in de Sitter space with compact
dimensions, J. Contemp. Phys. \textbf{54}, 1 (2019).

\bibitem{Mald13} J. Maldacena and G. L. Pimentel, Entanglement entropy in de
Sitter space, J. High Energy Phys. 02 (2013) 038.

\bibitem{Kann14} S. Kanno, J. Murugan, J. P. Shock, and J. Soda,
Entanglement entropy of $\alpha $-vacua in de Sitter space, J. High Energy
Phys. 07 (2014) 072.

\bibitem{Iizu16} N. Iizuka, T. Noumi, and N. Ogawa, Entanglement entropy of
de Sitter space $\alpha $-vacua, Nucl. Phys. B \textbf{910}, 23 (2016).

\bibitem{Kann17} S. Kanno, M. Sasaki, and T. Tanaka, Vacuum state of the
Dirac field in de Sitter space and entanglement entropy, J. High Energy
Phys. 03 (2017) 068.

\bibitem{Bhat19} S. Bhattacharya, S. Chakrabortty, and S. Goyal, Emergent $%
\alpha $-like fermionic vacuum structure and entanglement in the hyperbolic
de Sitter spacetime, Eur. Phys. J. C \textbf{79}, 799 (2019).

\bibitem{Sasa95} M. Sasaki, T. Tanaka, and K. Yamamoto, Euclidean vacuum
mode functions for a scalar field on open de Sitter space, Phys. Rev. D 
\textbf{51}, 2979 (1995).

\bibitem{Dimi15} F. V. Dimitrakopoulos, L. Kabir, B. Mosk, M. Parikh, and J.
P. van der Schaar, Vacua and correlators in hyperbolic de Sitter space, J.
High Energy Phys. 06 (2015) 095.

\bibitem{Abra72} \textit{Handbook of Mathematical Functions}, edited by M.
Abramowitz and I. A. Stegun (Dover, New York, 1972).

\bibitem{Nist10} F.W. Olver et al, \textit{NIST Handbook of Mathematical
Functions} (Cambridge University Press, USA, 2010).

\bibitem{Erd53V2} A. Erd\'{e}lyi et al, \textit{Higher Transcendental
Functions} (McGraw Hill, New York, 1953), Vol.2.

\bibitem{Saha20} A. A. Saharian and T. A. Petrosyan, The Casimir densities
for a sphere in the Milne universe, Symmetry \textbf{12}, 619 (2020).

\bibitem{Saha14} S. Bellucci, A. A. Saharian, and N. A. Saharyan, Wightman
function and the Casimir effect for a Robin sphere in a constant curvature
space, Eur. Phys. J. C \textbf{74}, 3047 (2014).

\bibitem{Saha08a} A. A. Saharian, A summation formula over the zeros of the
associated Legendre function with a physical application, J. Phys. A Math.
Theor. \textbf{41}, 415203 (2008).

\bibitem{Saha08b} A. A. Saharian, \textit{The Generalized Abel-Plana Formula
with Applications to Bessel Functions and Casimir Effect} (Yerevan State
University Publishing House, Yerevan, 2008); Report No. ICTP/2007/082;
arXiv:0708.1187.

\bibitem{Pfau82} J. D. Pfautsch, A new vacuum state in de Sitter space,
Phys. Lett. B \textbf{117}, 283 (1982).

\bibitem{Saha01SphM} A. A. Saharian, Scalar Casimir effect for D-dimensional
spherically symmetric Robin boundaries, Phys. Rev. D \textbf{63}, 125007
(2001).

\bibitem{Henr55} P. Henrici, Addition theorems for teneral Legendre and
Gegenbauer tunctions, Journal of Rational Mechanics and Analysis \textbf{4},
983 (1955).

\bibitem{Bunc78} T. S. Bunch, Stress tensor of massless conformal quantum
fields in hyperbolic universes, Phys. Rev. D \textbf{18}, 1844 (1978).
\end{thebibliography}
\end{document}